%% file: main.tex
\documentclass{article}
 
\usepackage{arxiv}

\usepackage[utf8]{inputenc}
\usepackage{amsmath,amssymb,amsfonts,amsthm}
\usepackage{ifthen}
\usepackage{comment}

\usepackage{url}            
\usepackage{booktabs}       
\usepackage{nicefrac}       
\usepackage{microtype}      
\usepackage{xcolor}         
\usepackage{enumerate}
\usepackage{pgf,pgfplots,graphicx,subcaption,siunitx}     
\usepackage[export]{adjustbox}
\usepackage[normalem]{ulem}
\usepackage{multirow, multicol}
\usepackage{doi}

\newtheorem{definition}{Definition}
\newtheorem{theorem}{Theorem}
\newtheorem{lemma}{Lemma}
\newtheorem{corollary}{Corollary}

\newtheorem{remark}{Remark}

\usepackage{tikz}
    \usetikzlibrary{matrix,positioning}
    \usetikzlibrary{decorations.pathreplacing}
    \usetikzlibrary{through,calc,intersections}
    \usetikzlibrary{shapes,shapes.geometric}

\DeclareMathOperator*{\argmax}{arg\,max}

\RequirePackage{tikz}

\usepackage{amsmath,commath}
\usepackage{amssymb}        
\usepackage{amsfonts}       
\usepackage{graphicx}       
\usepackage{subcaption}     
\usepackage{multirow,mathrsfs}
\usepackage{array,adjustbox,bbm,mathdots,program}
\usepackage{picture,xcolor,color} 
\usetikzlibrary{arrows,patterns}
\usetikzlibrary{automata}
\usetikzlibrary{matrix,calc}
\usetikzlibrary{arrows.meta,graphs,quotes,shadows,positioning}
 
    \usetikzlibrary{tikzmark,intersections,positioning}
    \usetikzlibrary{automata,shapes,calc,through}
    \usetikzlibrary{matrix,shapes.geometric,fit,shapes.symbols,chains}
    \usetikzlibrary{arrows.meta,graphs,quotes,shadows,backgrounds}
    \usetikzlibrary{decorations.pathreplacing,decorations.pathmorphing}
    \usetikzlibrary{decorations.text,fadings}

\newcolumntype{M}[1]{>{\centering\arraybackslash}m{#1}}

\newenvironment{proofOf}[1]	{\noindent {\textbf{Proof of #1}}\\}		{\hfill$\Box$\\\medskip}

\newcommand{\noop}[1]{}

\newcounter{equationthm}

\newcommand{\tuple}[1]{\ensuremath{\langle #1 \rangle}}

\newcommand{\CASE}[1]{

\ \\
\noindent
\textbf{#1}\\

}

\newcommand{\mGdefo}{\ensuremath{\tuple{G^{0}, G^{1}, \ldots, G^{n}}}}
\newcommand{\mbGdefo}{\ensuremath{\tuple{G^{0}, G^{r}, G^{c}}}}

\newcommand{\close}[2][\varepsilon]{\ensuremath{Cl^{#1}(#2)}}

\newcommand{\tbl}[1]    {\ensuremath{\boldsymbol{[#1]}}}  

\newcommand{\ug}[3][]{\ensuremath{\text{\textsc{ugain}}^{#1}(#2, #3)}}

\newcommand{\ugr}[3]{\ensuremath{U(#1,#2,#3)}}

\newcommand{\OP}[3]     {\ensuremath{\text{\textsc{op}}^{#1}_{#2}(#3)}}
\newcommand{\OM}[3]     {\ensuremath{\text{\textsc{om}}^{#1}_{#2}(#3)}}
\newcommand{\PM}[3]     {\ensuremath{\text{\textsc{pm}}^{#1}_{#2}(#3)}}
\newcommand{\ROM}[4]    {\ensuremath{\text{\textsc{rom}}^{#1}_{#2}(#3,#4)}}
\newcommand{\RPM}[4]    {\ensuremath{\text{\textsc{rpm}}^{#1}_{#2}(#3,#4)}}
\newcommand{\IN}[2]     {\ensuremath{\text{\textsc{in}}^{#1}_{#2}}}

\newcommand{\prob}[2][]{\ensuremath{{\mathcal P}_{#1} \boldsymbol{\left[\right.} #2 \boldsymbol{\left.\right]}}}

\newcommand{\devi}{\ensuremath{d}}
\newcommand{\Devi}{\ensuremath{D}}

\newcommand{\gndist}[1]{\ensuremath{{\mathcal N} \left(\ensuremath{#1}\right)}}

\definecolor{seagreen}{HTML}{34A853}
\definecolor{cinnabar}{HTML}{EA4335}
\definecolor{cyanblueazure}{HTML}{4B8BBE}
\definecolor{sunglow}{HTML}{FFD43B}

\tikzset{
diagonal fill/.style 2 args={fill=#2, path picture={
\fill[#1, sharp corners] (path picture bounding box.south west) -|
                         (path picture bounding box.north east) -- cycle;}},
reversed diagonal fill/.style 2 args={fill=#2, path picture={
\fill[#1, sharp corners] (path picture bounding box.north west) |- 
                         (path picture bounding box.south east) -- cycle;}}
}

\newcommand{\createDepiction}[4]{
$
\begin{adjustbox}{width=0.14\textwidth}
\centering
\begin{tikzpicture}
\node [rectangle,draw,text width=.8cm,minimum height=.8cm,
        text centered,rounded corners,name = re] {};
       \filldraw[sunglow,drop shadow][] (re.south west) 
        [rounded corners=4pt] -- (re.south east)
        [rounded corners=4pt] -- (re.north east)--cycle
        ;
        \filldraw[cyanblueazure][] (re.south west) 
        [rounded corners=4pt] -- (re.north west)
        [rounded corners=4pt] -- (re.north east)--cycle
        ;
\node [rectangle,draw,thick, text width=.8cm,minimum height=.8cm, 
        text centered,rounded corners,name = re] (A) {$(1,1)$}; 
\node[diagonal fill={sunglow}{cyanblueazure},
      text width=.8cm, minimum height=.8cm,
      text centered, rounded corners, draw, drop shadow] (B) [right=2cm of A] {$(1,2)$}; 
\node[diagonal fill={sunglow}{cyanblueazure},
      text width=.8cm, minimum height=.8cm,
      text centered, rounded corners, draw, drop shadow] (C) [below=2cm of A] {$(2,1)$}; 
\node[diagonal fill={sunglow}{cyanblueazure},
      text width=.8cm, minimum height=.8cm,
      text centered, rounded corners, draw, drop shadow] (D) [below=2cm of B] {$(2,2)$}; 
      
\draw[line width=.5mm,cyanblueazure] (A.west) .. controls +(left:7mm) and +(up:7mm) .. (C.north) node[midway,left,label={[xshift=-.5cm, yshift=-1cm,text=black,scale=2.5]$#1$}] {} ;
\draw[line width=.5mm,cyanblueazure] (B.west) .. controls +(left:7mm) and +(up:7mm) .. (D.north) node[midway,left,label={[xshift=1cm, yshift=-1cm,text=black,scale=2.5]$\large #2$}] {} ;

\draw[line width=.5mm,sunglow] (A.east) .. controls +(right:7mm) and +(down:7mm) .. (B.south) node[midway,above,label={[xshift=-0.5cm, yshift=0.05cm,text=black,scale=2.5]$\large #3$}] {} ;
\draw[line width=.5mm,sunglow] (C.east) .. controls +(right:7mm) and +(down:7mm) .. (D.south) node[midway,above,label={[xshift=-0.5cm, yshift=0.05cm,text=black,scale=2.5]$\large #4$}] {} ;
\end{tikzpicture}
\end{adjustbox}
$
}

\newcommand{\addLineTable}[7]{

({#7}) &
\ensuremath{ {#1} 0} &
\ensuremath{ {#2} 0} &
\ensuremath{ {#3} 0} &
\ensuremath{ {#4} 0} & 
\ensuremath{ \{ {#5} \} } 
\ensuremath{ {#6} } &
\createDepiction{#1}{#2}{#3}{#4} \\ 
\hline
}

\usepackage{etoolbox}
\usepackage{xparse}
\usepackage{cleveref}
\usepackage{crossreftools}

\makeatletter
\@ifundefined{crtcrefcountervalue}{%
\newcommand{\crt@crefundefinedcountervalue}{1977}
\newcommand{\crtrefundefinedcountervalue}[1]{\renewcommand{\crt@refundefinedcountervalue}{#1}}

\newcommand{\crtcrefcountervalue}[1]{%
  \crtcrefifundefinedlabel{#1}{%
    \crt@crefundefinedcountervalue%
  }{%
    \crtcrefnumber{#1}%
  }%
}
}{}

\makeatother

\newcounter{proofcntr}
\crefname{proofcntr}{proof}{proofs}
\Crefname{proofcntr}{Proof}{Proofs}

\newenvironment{ex1}
  {\ex}
  {\endex}

\newsavebox\IBoxA \newsavebox\IBoxB \newlength\IHeight
\newcommand\TwoFig[6]{
  \sbox\IBoxA{\includegraphics[width=0.45\textwidth]{#1}}
  \sbox\IBoxB{\includegraphics[width=0.45\textwidth]{#4}}%
  \ifdim\ht\IBoxA>\ht\IBoxB
    \setlength\IHeight{\ht\IBoxB}%
  \else\setlength\IHeight{\ht\IBoxA}\fi
  \begin{figure}[!htb]
  \minipage[t]{0.45\textwidth}\centering
  \includegraphics[height=\IHeight]{#1}
  \caption{#2}\label{#3}
  \endminipage\hfill
  \minipage[t]{0.45\textwidth}\centering
  \includegraphics[height=\IHeight]{#4}
  \caption{#5}\label{#6}
  \endminipage 
  \end{figure}%
}

\newtheorem{ex}{Example}
\newtheorem{example}{Example}%

\newtheorem{proposition}[theorem]{Proposition}
\theoremstyle{thmstyleone}%

\title{Noisy Games: A Study on the Effect of Noise on Game Specifications}

\date{} 					

\author{   
  Constantinos Varsos \\
  Centrum Wiskunde \& Informatica \\
  (CWI) \\
  \texttt{Konstantinos.Varsos@cwi.nl} \\
  \And
  Giorgos Flouris \\
  Foundation for Research and Technology-Hellas \\ (FORTH) \\
  \texttt{fgeo@ics.forth.gr} \\
  \AND
  Marina Bitsaki \\
  Institute of Computer Science, University of Crete \\
  \texttt{ecbitsaki@gmail.com} \\
}

\hypersetup{
pdftitle={Noisy Games: A Study on the Effect of Noise on Game Specifications},
pdfsubject={cs.GT,cs.LG},
pdfauthor={Constantinos~Varsos,Giorgos~Flouris,Marina~Bitsaki},
pdfkeywords={Noisy games, Misinformation games, noise, natural misinformed equilibrium},
}

\begin{document}
\maketitle

\begin{abstract}
We consider misinformation games, i.e., multi-agent interactions where the players are misinformed with regards to the game that they play, essentially having an \emph{incorrect} understanding of the game setting, without being aware of their misinformation. In this paper, we introduce and study a new family of misinformation games, called Noisy games, where misinformation is due to structured (white) noise that affects additively the payoff values of players. 
We analyse the general properties of Noisy games and derive theoretical formulas related to ``behavioural consistency'', i.e., the probability that the players behaviour will not be significantly affected by the noise.
We show several properties of these formulas, and present an experimental evaluation that validates and visualises these results.
\end{abstract}

\section{Introduction}
\label{sec:intro}

A common assumption in game theory \cite{Algorithmic_Game_Theory_book,Osborne_Rubinstein_book} is that the abstract formulation of the game (number of players, strategies available to the players and payoffs depending on the chosen strategies) are publicly available to all players. Even for games with incomplete information \cite{Zamir2009}, the fact that knowledge is incomplete, and the exact form of incompleteness, is embedded in the game specification.

\begin{figure}
    \centering
    \includegraphics[width = .45\textwidth]{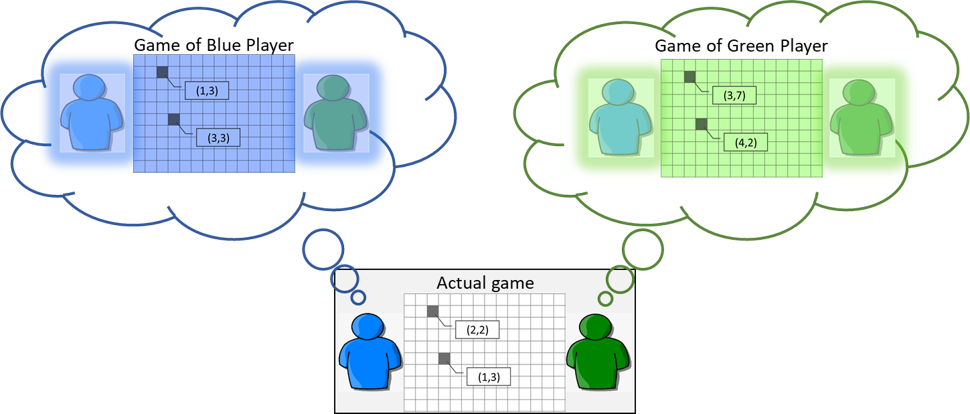}
    \caption{Visualisation of a misinformation/noisy game}
    \label{fig:mG}
\end{figure}

However, in several scenarios, it could be the case that the players may have wrong information regarding the game setup, and at the same time being unaware of the fact that their information is wrong, thus being \emph{misinformed}. The agents, being unaware of their misinformation, may make choices that are unexpected and seem irrational from the external viewpoint, leading to unexpected results. 
These games are called \emph{misinformation games}~\cite{VFBM} (see also Figure \ref{fig:mG}).

The main defining characteristic of misinformation games is that the players have no reason to believe that they have the wrong payoff information, and will play the game under the misconceived definition (payoffs) that they have. Nevertheless, the payoff that they will get is the one provisioned by the actual game. This makes the concept of misinformation games quite different from other types of games that have been defined in the literature, in particular games that incorporate uncertainty in their payoffs (e.g., Bayesian games \cite{Zamir2009}); in games with incomplete information, although uncertainty makes the players unsure as to their actual payoff for each different strategy, the players are well-aware of that, and they accommodate their strategies accordingly, in order to make the best out of the uncertainty that they have. On the contrary, in misinformation games the players believe the information that was given, and do not consider mitigation measures ``just in case'' the information is wrong.

In previous works \cite{VFBM,MVF}, various different causes of misinformation are identified, including deception and misleading reports, human errors, deliberate attempts by the game designer to channel players into different behaviours, erroneous sensor readings and random effects. 
In this paper, we focus on a special case of misinformation, attributed to noise and signal errors, a situation often occurring in distributed multiagent systems. This class of misinformation games will be called \emph{noisy games}.

Specifically, in distributed multiagent systems, agents\footnote{Note that we use the terms ``agent'' and ``player'' interchangeably throughout the paper.} are equipped with an internal logic that allows them to autonomously solve problems of a given nature. However, at deployment time, the precise specification of these problems is often unknown; instead, the details are communicated as needed at operation time, during the so-called ``online phase'' \cite{DBLP:journals/corr/abs-1710-08500}.
In such cases, unexpected communication errors, malfunctions in the communication module or noise may cause the agents to operate under a distorted problem specification, leading to unexpected behaviour.

For example, consider the scenario where we have two autonomous self-interested agents, already deployed in an unfriendly environment. At some point in time, the human controller asks each of the agents to choose among two actions, also specifying the payoffs for each combination of choices. 
If the communication goes through as expected, then the behaviour of the agents is predictable by the well-known results of game theory. 
However, if one (or both) of the agents' communication module malfunctions, or if there is unexpected noise in the communication channel, the signal may arrive distorted. This could lead agents to receive an erroneous payoff matrix, essentially causing them to believe that they play a game different than the one communicated to them, with unpredictable results (Figure \ref{fig:mG}).

Note that, if, at deployment time, the designer had foreseen the possibility for the agents to receive an erroneous game specification, then the agents would have been programmed to treat all signals as uncertain (i.e., true under a certain probability). In this case, the possibility of error is integrated in the agents' logic (even when no communication error occurs), and their behaviour can be modelled using the rich results on Bayesian games and games with incomplete information \cite{Zamir2009}.
On the other hand, if such a scenario had not been foreseen at deployment time, then the agents will operate under the payoff matrices received, without considering the possibility that the payoff matrices are not the correct ones. This is quite different, as the agents' decisions will be totally misled by the erroneous setting, and will not consider mitigation measures ``just in case'' the specification that they received is wrong.

The aim of this paper is to provide the theoretical machinery necessary to study scenarios of this kind. 
In particular, the main research question to be addressed is: \emph{given a game and a specific noise pattern affecting the players' perceived payoff matrices, compute the probability that players' behaviour (i.e., chosen strategies) will be as close as possible (in a manner to be formally defined later) to the behaviour that they would have in the absence of noise}. 

In summary, the main contributions of this paper are the following:
\begin{enumerate}
    \item The provision of motivation for the need to define misinformation in the context of noisy games, by positioning our work with respect to other similar efforts in the literature, in particular related to games with uncertainty, and games where the players have some kind of misconception related to the game's payoffs (Section~\ref{sec:related}).
    \item The definition of a formal model for the description of misinformation in noisy games (Section~\ref{sec:noisy_definitions}).
    \item The computation of the probability that the players' behaviour is not significantly affected by random noise, a feature that we call \emph{behavioural consistency} (Section~\ref{sec:noisy_probabilities}).
    \item A thorough analysis of the properties of these probabilities (Section~\ref{sec:results_noisy}). 
    \item Experimentation to visualize and validate our results (Section~\ref{sec:discuss}).
\end{enumerate}

\section{Related Work}
\label{sec:related}

The works most related to the concept of noisy games are those of games with misspecified views (e.g., \cite{BENNETT1980489}, \cite{RaiffanLuce}, \cite{DBLP:journals/mss/Schipper14}). Despite the fact that Bayesian techniques are very popular in this stream of works, there is no consideration about the structure of misinformation that results in these views. As opposed to Bayesian games, where there is a rich literature that studies the influence of the structure of uncertainty in the knowledge of the players to their strategic behaviour. 
Thus, conceptually we are closer to the first group of works. On the other hand though, we study too the effect of distributions in the knowledge of the players as to their strategic behaviour, thus our results can be related to the latter stream of works.

Bayesian games and games with incomplete information \cite{Zamir2009} have been introduced to handle uncertainty about players' payoffs. This uncertainty is represented by probabilities over the alternative payoff matrices. Also, there are cases where some payoffs are simply unknown.
In all cases, the players are aware of the fact that the information they have is incomplete, and their strategies are adapted to cater for this incompleteness. This is the main difference with regards to misinformation games, where information is complete, but incorrect. That is, players are unaware of the fact that the information they have is incorrect, and thus, their strategic choices are entirely based on it.

The works \cite{DBLP:conf/aaai/MeirP15,DBLP:journals/sigmetrics/MeirP15,DBLP:conf/atal/MeirP18} are also relevant to ours. 
In these studies, the authors consider non-atomic routing games, and suggest that the players experience their own cost functions, which are potentially different from the actual ones (e.g., to model player-specific biases).
This setting is similar to a misinformation game, except that, in our methodology, each player has a potentially different view of the entire game (including the payoffs of the other players), not just her own payoffs, and plays according to that view. 
In addition, the two approaches have a quite different motivation: in \cite{DBLP:conf/aaai/MeirP15} authors assume that players modify their payoffs from the objective ones, by themselves, for personal reasons (bias or some kind of personal preference); in our case, the modifications are accidental, caused by communication errors.

In \cite{Balcan:2009:PU:1566374.1566416}, the authors consider the impact of small fluctuations in the cost functions or in players’ perceptions of the cost structure in congestion and load balancing games, and study its effect on players' behaviour.
A fluctuation is a departure from the classical viewpoint that treats payoffs as a number; under \cite{Balcan:2009:PU:1566374.1566416}, the payoff is a range of values ``close'' to the actual payoff.
An extension of \cite{Balcan:2009:PU:1566374.1566416} considered normal-form games, aiming to define a new notion of equilibrium that maximizes the worst case outcome over possible actions by other players~\cite{DBLP:journals/mp/AghassiB06} in the presence of fluctuations, whereas a further extension (\cite{DBLP:journals/toc/BalcanB17}) studied the robustness of this equilibrium solution, utilizing the notion of approximations of payoffs using a fuzziness to the values of payoff matrices. 
There are several differences of the concept of fluctuation as compared to misinformation. First, the players are aware of the fluctuations and, thus, take them into account while deciding on their strategic choices. Second, fluctuations affect all players and all payoffs uniformly. Third, fluctuations have a limited effect, whereas the noise considered in our work may have unlimited effect (subject to a certain probability function). 

Further, in \cite{DBLP:journals/corr/abs-1710-08500} the authors study how resilient is the strategic behaviour of players when an unexpected communication loss occurs, and explore game settings in which communication failures can/cannot cause harm in the strategic behaviour of players. They introduce the notion of \emph{proxy payoffs} in order to funnel communication failures and show that, in several settings, loss of information may cause arbitrary strategic behaviours. 
Our work has a similar contribution as both works prove that in the presence of communication inefficiencies any strategic behaviour is possible. Though, authors in \cite{DBLP:journals/corr/abs-1710-08500} focus on how the agents choose policies so as to cope with communication failures, in this study we analyse the impact of disorder in the strategic behaviour of the players.
Also, we model communication failure using probabilities, and provide formulas that quantify the probability of arbitrariness in strategic behaviours, when information is degraded due to noise.

Moreover, there is another stream of works considering random payoff matrices, (i.e. \cite{DBLP:journals/corr/abs-math-0703902,DBLP:journals/geb/Takahashi08,DBLP:journals/mor/Stanford96,DBLP:journals/geb/RinottS00,Brny2007NashEI,Dresher1970PROBABILITYOA}) where the focus is on the distribution of pure Nash equilibria. A tweak of this methodology is presented in \cite{DBLP:journals/corr/abs-2007-08518} where authors study the distribution of players' average social utility. 

\section{Preliminaries}
\label{sec:preliminaries}

\subsection{A brief refresher on probability theory}
\label{subsec:probabilities}

We provide here some basic knowledge on probability theory that will be useful in the following sections. The interested reader is referred to \cite{Shiri} for further details.

A random variable $X$ is characterised by its \emph{probability density function (pdf)}, denoted by $f_X$, which represents the ``intensity'' of the probability in each given point. The pdf can be used to compute the probability that $X$ falls within a given range, say $[a,b]$, for any $a \leq b$. Formally, $f_X$ is such that: 

\begin{equation*}
    \prob{a \leq X \leq b} = \int^b_a f_X(x) dx
\end{equation*}

We denote by $F_X$ the \emph{cumulative distribution function (cdf)} of a random variable $X$, which equals the probability that the value of $X$ is at most $x$. Formally:
\begin{equation*}
    F_X(x) = \int^x_{-\infty} f_X(t) dt = \prob{X \leq x}
\end{equation*}

In this paper, we focus on random variables $X$ following the normal distribution, denoted by $X \sim \gndist{\mu,\devi^2}$ (for some mean $\mu \in \mathbb{R}$ and standard deviation\footnote{In probability theory, standard deviation is typically denoted by $\sigma$. To avoid confusion with the strategies of normal form games which use the same symbol (see Subsection \ref{subsec:normal-form-no-mis}), we use $\devi$ as a symbol for standard deviation in this paper.} $\devi > 0$). 
For the special case where $\mu = 0, \devi = 1$ (i.e., when $X\sim \gndist{0,1}$), we get the \emph{standard normal distribution}, with the following pdf ($\phi$) and cdf ($\Phi$):

\begin{equation} \label{eq:normal-std-pdf-cdf}
\begin{aligned}
    \phi(x) = \tfrac{1}{\sqrt{2\pi}} e^{- \tfrac{x^2}{2}}, \quad \Phi(x) = \tfrac{1}{\sqrt{2\pi}} \int^{x}_{-\infty} e^{-\frac{t^2}{2}} dt \\
\end{aligned}
\end{equation}

For the general case, where $X \sim \gndist{\mu,\devi^2}$, the pdf and cdf are:
\begin{equation}\label{eq:normal-gen-pdf-cdf}
\begin{aligned}    
    f_X(x) = \frac{1}{\devi} \phi \left(\frac{x-\mu}{\devi}\right) =
        \tfrac{1}{\devi \sqrt{2\pi}} e^{- \left( \tfrac{x - \mu}{\devi \sqrt{2}}\right)^2} \\
    F_X(x) = \Phi \left(\frac{x-\mu}{\devi}\right) = 
        \tfrac{1}{\sqrt{2\pi}} \int^{\tfrac{x-\mu}{\devi}}_{-\infty} e^{-\frac{t^2}{2}} dt \\
\end{aligned}
\end{equation}

It has been shown that, if $X_i \sim \gndist{\mu_i, \devi_i^2}$, $c_0,c_i \in \mathbb{R}$, then:
\begin{equation}
\label{eq:normal-dist-sum}
    c_0 + \sum c_i X_i \sim \gndist{c_0 + \sum c_i\mu_i, \sum c_i^2 \devi_i^2}    
\end{equation}

Given two events $A,B$, the symbol $\prob{A \vert B}$ denotes the \emph{conditional probability of $A$ given $B$}, which amounts to the probability that $A$ is true under the condition that $B$ is true.

When combining events, the following are true: 
\begin{equation}
\label{eq:prob-conj-disj}
\begin{split}
    & \text{General Conjunction Rule: }  \\
    & \hspace{5em} \prob{A \wedge B} = \prob{B} \cdot \prob{A \vert B} = \prob{A} \cdot \prob{B \vert A} \\
    & \text{Restricted Conjunction Rule: } \\
    & \hspace{5em} \prob{A \wedge B} = \prob{A} \cdot \prob{B} \quad \text{(when } A, B \text{ are mutually exclusive)} \\
    & \text{General Disjunction Rule: } \\
    & \hspace{5em} \prob{A \vee B} = \prob{A} + \prob{B} - \prob{A \wedge B} \\ 
    & \text{Restricted Disjunction Rule: } \\
    & \hspace{5em} \prob{A \vee B} = \prob{A} + \prob{B} \quad \text{(when } A, B \text{ are mutually exclusive)} \\
\end{split}
\end{equation}

\subsection{Normal-form games}
\label{subsec:normal-form-no-mis}

Normal-form games \cite{Algorithmic_Game_Theory_book,Osborne_Rubinstein_book} is the most commonly-studied class of games. A game in normal-form is represented by a \emph{payoff matrix} that defines the payoffs of all players for all possible combinations of pure strategies. Formally:

\begin{definition}[Normal-form games]
\label{def:Strategic_game}
A normal-form game $G$ is a tuple $G = \tuple{N, S, P}$, where:
\begin{itemize}
    \item $N =\{1,2,\dots,n\}$ is the set of players.
    \item $S = S_1\times \dots \times S_{n}$, $S_i$ is the set of pure strategies of player $i\in N$.
    \item $P = (P_1; \dots; P_n)$, $P_i \in \mathbb{R}^{\vert S_1\vert \times \ldots \times \vert S_n\vert }$ is the payoff matrix of player $i \in \{1,2,\dots,n\}$.
\end{itemize}
\end{definition}

In this paper we focus on \emph{$2 \times 2$ bimatrix games}, a popular class of games defined as follows:

\begin{definition}[$2 \times 2$ bimatrix games]
\label{def:Bimatrix_game}
A $2\times 2$ bimatrix game $G$ is a normal-form game $G= \tuple{N, S, P}$, such that:
\begin{itemize}
    \item $N = \{r, c\}$ is the set of players
    \item $S = S_r \times S_c$, where $S_r = S_c = \{s_1,s_2\}$
    \item $P = (P_r; P_c)$, $P_r,P_c \in \mathbb{R}^{2\times 2}$
\end{itemize}
\end{definition}

Let us now fix some player $x\in \{r,c\}$. 
A \emph{strategy} of $x$ is a pair $\sigma_x = (\sigma_{x,1}, \sigma_{x,2})$, where $(\sigma_{x,1}, \sigma_{x,2})$ form a discrete probability distribution over $S_x$ (i.e., $\sigma_{x,1},\sigma_{x,2} \in [0,1], \sigma_{x,1} + \sigma_{x,2} = 1$).
When $\sigma_{x,1}, \sigma_{x,2} \in (0,1)$ the strategy is called a \emph{mixed strategy}; otherwise, it is called a \emph{pure strategy}.
The \emph{support} of a strategy $\sigma_x$, denoted by $supp(\sigma_x)$, is the set of pure strategies (from $S_x$) that are played with positive probability on $\sigma_x$ (thus, $supp(\sigma_x) \subseteq S_x$).
We denote by $\Sigma_x$ the set of all possible strategies of player $x$.
Apparently, for $2 \times 2$ bimatrix games, $\Sigma_x = \{(p,1-p) \mid 0 \leq p \leq 1 \}$. 

A \emph{strategy profile} is a pair $\sigma = (\sigma_r, \sigma_c)$, for $\sigma_r \in \Sigma_r, \sigma_c \in \Sigma_c$. We denote by $\Sigma$ the set of all strategy profiles, i.e., $\Sigma = \Sigma_r \times \Sigma_c$. A strategy profile is called \emph{pure} if it consists of pure strategies only, \emph{mixed} if it consists of mixed strategies only, and \emph{hybrid} if it consists of a pure and a mixed strategy.

The \emph{payoff function} of player $x$, under a given strategy profile $\sigma = (\sigma_r, \sigma_c)$, $u_x: \Sigma \to \mathbb{R}$, is defined as: 
\begin{equation*}
    u_x(\sigma_r,\sigma_c) = \sigma_r^T P_x \sigma_{c} 
\end{equation*}
where $\sigma_r^T$ represents the transposition of vector $\sigma_r$.

For $x \in \{r,c\}$, we denote by $\bar x$ the other player, i.e., $\bar r = c, \bar c = r$.
Given a strategy $\sigma_x$ of $x \in \{r,c\}$, the \emph{best response} of $\bar x$ is the strategy $\sigma_{\bar x}$ that maximizes her payoff, given $\sigma_x$.
A \emph{Nash equilibrium} is a strategy profile for which any unilateral change in the strategy of any given player would not produce a better payoff for that player. 
In other words, a Nash equilibrium is a strategy profile where each player plays her best response, given the other player's strategic choice.
For bimatrix games, this notion can be formalised as follows:

\begin{definition}[Nash equilibrium \cite{Nash51}]
\label{def:Nash}
A strategy profile $\sigma^* = (\sigma^*_r, \sigma^*_c)$ is a \emph{Nash equilibrium} if and only if, for any $\hat{\sigma}_r \in \Sigma_r$, $\hat{\sigma}_c \in \Sigma_c$,
\begin{equation*}
    {\sigma^*_r}^T P_r \sigma^*_{ c} \geq \hat{\sigma}_r^T P_r \sigma^*_{c} 
        \quad \text{ and } \quad 
    {\sigma^*_r}^T P_c \sigma^*_c \geq {\sigma^*_r}^T P_c \hat{\sigma}_c
\end{equation*}
\end{definition}

It has been shown that all games possess at least one Nash equilibrium \cite{Nash51}.
If $\sigma^* = (\sigma^*_r, \sigma^*_c)$ is a Nash equilibrium, then $\sigma^*_r, \sigma^*_c$ are called \emph{Nash equilibrium strategies}.
We denote by $NE(G)$ the set of all Nash equilibria for a game $G$, and by $NE_x(G)$ the Nash equilibrium strategies of player $x$ in $G$.

A $2 \times 2$ bimatrix game is called \emph{degenerate} if and only if there is a pure strategy that has two pure best responses.

In the seminal work of \cite{koutsoupias-papadimitriou:kts-pap}, the authors defined a metric, the {\em Price of Anarchy (PoA)}, that measures the efficiency of a system with non-cooperative players. 
Let $SW(\sigma)$ be the social welfare function defined as the sum of players' payoffs for the strategy profile $\sigma$, and $opt$ the socially optimal strategy profile, i.e., $opt = \argmax_{\sigma} SW(\sigma)$. Then, $PoA$ is defined as follows: 

\begin{definition}\label{def:PoA}
Given a normal-form game $G$, the \emph{Price of Anarchy (PoA)} is defined as:
\begin{equation}\label{eq:PoA_max}
    PoA = \frac{SW(opt)}{\min_{\scriptscriptstyle \sigma \in ne} SW(\sigma)}
\end{equation}
\end{definition}


\subsection{Notational conventions and shorthands}
\label{subsec:notation}

To avoid confusion caused by the use of multiple indices in subsequent sections, we will use the notation $A[i,j]$ to refer to the element in the $i^{th}$ row and $j^{th}$ column of a matrix $A$, i.e., if $A=(a_{ij})$, then $A[i,j] = a_{ij}$. 

We will use boldface to indicate tables whose elements are all equal to a certain value. For example $\tbl{b}_{n\times m}$ represents the $n\times m$ table $B$, such that $B[i,j] = b$ for all $i,j$. The $n\times m$ subscript will be omitted when obvious from the context.

For three tables $A, M, \Devi$ of the same dimensions, we write $A \sim \gndist{M,\Devi}$ to indicate that $A[i,j] \sim \gndist{M[i,j],\Devi[i,j]}$ for all $i,j$.

Analogously, we extend the notation used in limits ($x \to a$) for tables. In particular, for two tables $X,A$ of the same dimension, we write $X \to A$ to denote $X[i,j] \to A[i,j] \; \forall i,j$. We extend the notation to include infinity, e.g., $X \to \tbl{+\infty}$ is equivalent to $X[i,j] \to +\infty$ for all $i,j$.

We define operators on payoff matrices as follows. Consider a $2\times 2$ bimatrix game $G = \tuple{N,S,P}$, where $P=(P_r;P_c)$. Then:

\begin{itemize}
    \item For $2 \times 2$ tables $M_r, M_c, D_r, D_c$, the expression
    $G \sim \gndist{(M_r;M_c),(\Devi_r;\Devi_c)}$ indicates that $P_r \sim \gndist{M_r, \Devi_r}, P_c \sim \gndist{M_c,\Devi_c}$
    \item For a $2\times 2$ bimatrix $A=(A_r;A_c)$ and $\lambda \in \mathbb{R}$, the result of the operation
    $\lambda G + A$ is the $2\times 2$ bimatrix game $G' = \tuple{N',S',P'}$, where $N'=N, S'=S, P' = \lambda P + A$
\end{itemize}

\section{Misinformation Games and Noisy Games}
\label{sec:noisy_definitions}

\subsection{Basic Definitions}
\label{subsec:mis-noisy-games}

\emph{Misinformation games} have been originally defined in \cite{VFBM}. In this section, we extend the main definitions and concepts for the case of noisy games.

Misinformation games have been defined to capture the idea that different players may have a different view of the game they play (see Figure \ref{fig:mG}). This leads to the following definition:

\begin{definition}[Misinformation game]
\label{def:mis_normal_basic}
    A \emph{misinformation normal-form game} (or simply \emph{misinformation game}) is a tuple $mG = \tuple{G^0, G^1, \ldots, G^n}$, where all $G^i$ are normal-form games and $G^0$ contains $n$ players.
\end{definition}

In the above definition, $G^0$ is called the \emph{actual game} (corresponding to the game actually being played), whereas $G^i$ is the \emph{game of player $i$} (corresponding to the game that player $i$ thinks that it is being played). 

In \cite{VFBM}, it was shown that, without loss of generality, we only need to concentrate ourselves in the special class of \emph{canonical misinformation games}.
A misinformation game $\tuple{G^0, G^1, \ldots, G^n}$ is called \emph{canonical} if and only if:
\begin{itemize}
    \item In $G^0$, all players have an equal number of pure strategies
    \item For any $x \in \{1,\dots,n\}$, $G^0, G^x$ differ only in their payoff matrices
\end{itemize}

\emph{Noisy games} are a special class of misinformation games, where misinformation is due to a random distortion in the original payoff matrix. Formally:

\begin{definition}[Noisy game]
A \emph{noisy game} is a canonical misinformation game $mG = \mGdefo$, where $G^i = G^0+\Delta^i$ for some matrix $\Delta^i$ whose elements follow a certain probability distribution.
\end{definition}

Note that the restriction of a noisy game being canonical implies that noise affects only the payoff matrix. In a more general scenario, noise could also affect the number of players and/or the strategies that a player understands (knows) regarding a game. However, as shown in \cite{VFBM}, we can restrict ourselves to canonical games for simplicity.

In this paper, we concentrate on noisy games whose actual game is a $2 \times 2$ bimatrix game, and where each element of $\Delta^i$ follows the normal distribution. We call such games \emph{normal noisy games}. Therefore:

\begin{definition}[Normal noisy game]\label{def:normal_noisy_games}
A \emph{normal noisy game} is a tuple $mG = \mbGdefo$, where: 
\begin{itemize}
    \item $G^0, G^r, G^c$ are $2 \times 2$ bimatrix games
    \item For $x\in \{r,c\}$, $G^x = G^0 + \Delta^x$, where $\Delta^x$ is a bimatrix whose elements follow the normal distribution (possibly for a different mean and standard deviation)
\end{itemize}
\end{definition}

For $M = (M_r;M_c)$, $\Devi = (\Devi_r;\Devi_c)$, we write $mG \sim G^0 + \gndist{M,\Devi}$ to indicate a normal noisy game $mG = \tuple{G^0, G^r, G^c}$, where $G^x = G^0 + \Delta^x$, $\Delta^x \sim \gndist{M_x,\Devi_x}$. Formula \ref{eq:normal-dist-sum} implies that, when $mG \sim G^0 + \gndist{M,\Devi}$, then $G^x \sim \gndist{G^0 + M_x, \Devi_x}$.

\subsection{Strategies, strategy profiles and equilibria in misinformation games}

We define strategies and equilibrium concepts for the case of normal noisy games of two players. Our formulation can be extended to apply to arbitrary misinformation games.

Consider a canonical normal noisy game $mG$.
A \emph{misinformed strategy} $\sigma_x$ for player $x$ in $mG$ is a strategy in $G^x$. A \emph{misinformed strategy profile} results by the agglomeration of misinformed strategies for the individual game, and is defined as a pair $\sigma = (\sigma_r,\sigma_c)$, where $\sigma_x$ is a misinformed strategy of $x \in \{r,c\}$.
Pure/mixed misinformed strategies, and pure/mixed/hybrid strategy profiles are defined analogously to their standard counterparts (see Subsection \ref{subsec:normal-form-no-mis}). 
Since $mG$ is canonical, a misinformed strategy (and misinformed strategy profile) is also a strategy (strategy profile) in $G^0$.
Thus, we simply use $\Sigma_x$ to denote the set of misinformed strategies of player $x$ in $mG$, and $\Sigma$ to denote the set of misinformed strategy profiles of $mG$.

It is important to note that, although the decisions of a player are made based on his own payoff matrix (the one in $G^x$), payoffs are computed on the basis of the actual payoff matrix (the one in $G^0$). This is reflected in the definition of payoffs and equilibria below.

Let $P^0=(P_r^0;P_c^0), P^r=(P_r^r;P_c^r), P^c=(P_r^c;P_c^c)$ be the payoff matrices of $G^0, G^r, G^c$ respectively. Then:
\begin{itemize}
    \item The \emph{actual payoff function} of player $x$, under a given strategy profile $\sigma = (\sigma_r, \sigma_c)$, $u_x: \Sigma \to \mathbb{R}$, is defined as: 
    \begin{equation*}
        u_x(\sigma_r,\sigma_c) = \sigma_r^T P_x^0 \sigma_{c} 
    \end{equation*}
    \item The \emph{misinformed payoff function} of player $x$, under the viewpoint of player $y$ and the strategy profile $\sigma = (\sigma_r, \sigma_c)$, $u^y_x: \Sigma \to \mathbb{R}$, is defined as: 
    \begin{equation*}
        u^y_x(\sigma_r,\sigma_c) = \sigma_r^T P_x^y \sigma_{c} 
    \end{equation*}
\end{itemize}

Note that the actual payoff function represents the payoff that player $x$ will receive as a response to her strategic choices. On the contrary, the misinformed payoff function represents the payoff that player $x$ believes that she will receive, under the (erroneous) view of the game that player $y$ has.

A notion of equilibrium, defined in \cite{VFBM}, considers misinformed equilibrium simply as the agglomeration of the Nash equilibrium strategies of each player in her own game:

\begin{definition}[Natural misinformed equilibrium (nme)]\label{def:natural_misinformed_eq}
A misinformed strategy, $\sigma^*_x$, of player $x$, is a \emph{natural misinformed equilibrium strategy}, if and only if it is a Nash equilibrium strategy for $x$ in $G^x$. A misinformed strategy profile $\sigma^*$ is called a \emph{natural misinformed equilibrium (nme)} if it consists of natural misinformed equilibrium strategies.
\end{definition}

The natural misinformed equilibrium will occur in one-off settings, i.e., when each player just picks a (seemingly optimal) strategy based on his own viewpoint. It is easy to see that at least one natural misinformed equilibrium exists in any misinformation game.

Inspired by \cite{koutsoupias-papadimitriou:kts-pap}, the authors of \cite{VFBM} defined a metric to measure the effect of misinformation compared to the social optimum, based on a social welfare function $SW$. This metric is called \emph{Price of Misinformation (PoM)}, and is defined as follows:

\begin{definition}\label{def:PoM}
Given a misinformation game $mG$, the \emph{Price of Misinformation (PoM)} is defined as:
\begin{equation}\label{eq:PoM_max}
    PoM = \frac{SW(opt)}{\min_{\scriptscriptstyle \sigma \in nme} SW(\sigma)}
\end{equation}
\end{definition}

Apparently, if $PoM = 1$, the players adopt optimal behaviour, due to misinformation. Moreover, interesting results can be derived by comparing the $PoA$ of $G^0$ with the PoM of $mG$: if $PoM < PoA$, then misinformation has a beneficial effect on social welfare, as the players are inclined (due to their misinformation) to choose socially better strategies; on the other hand, if $PoM > PoA$, then misinformation leads to a worse outcome, from the perspective of social welfare.

\subsection{Behavioural Consistency and \texorpdfstring{$\varepsilon$-}{}closeness}
\label{subsec:distance-closeness}

The misinformed equilibria of a normal noisy game may be different than the Nash equilibria of the actual game. We define a metric to quantify the distance among these equilibria and their respective strategies, essentially measuring the effect of noise on the behaviour of the players. For the definition, we use the infinite norm distance for vectors; formally, for a vector $\vec v = (v_1,v_2)$, the infinite norm distance $\| \vec v \|_{\infty} = \max\{v_1,v_2\}$. The notion of $\varepsilon$-closeness is now defined as follows:

\begin{definition}[$\varepsilon$-closeness]\label{def:strat_close}
Let 
$\sigma = (\sigma_1, \sigma_2)$, $\sigma' = (\sigma_1', \sigma_2')$
be two strategies and $\varepsilon\geq 0$.
Then we say that $\sigma, \sigma'$ are \emph{$\varepsilon$-close} if and only if 
$supp(\sigma) = supp(\sigma')$ and $\| \sigma-\sigma' \|_{\infty} \leq \varepsilon$.
For strategy $\sigma$,
the set of strategies that are $\varepsilon$-close to it, is denoted by $\close{\sigma}$.
\end{definition}

Intuitively, the definition states that two strategies are $\varepsilon$-close if and only if they have identical supports and the allocation imposed by the players' strategies does not differ by more than $\varepsilon$ in any dimension. 
The fact that $\varepsilon$-closeness requires identical supports is based on the idea that adding (or removing) a pure strategy to (from) the support of a strategy is considered a major change in the player's behaviour. 

Note also that the above definition applies on strategies in general, and, thus, allows us to apply it also to check $\varepsilon$-closeness among strategies and/or misinformed strategies, as long as in each strategy profile each player has the same number of pure strategies.

We extend Definition \ref{def:strat_close} to (misinformed and non-misinformed) strategy profiles (and equilibria), in the obvious manner: 
$\sigma = (\sigma_r,\sigma_c)$ is $\varepsilon$-close to
$\sigma' = (\sigma_r',\sigma_c')$ if and only if
$\sigma_r$ is $\varepsilon$-close to $\sigma_r'$ and
$\sigma_c$ is $\varepsilon$-close to $\sigma_c'$.
We denote by $\close{\sigma}$ the strategy profiles that are $\varepsilon$-close to $\sigma$.
For a set of strategy profiles $\Sigma^*$, we set $\close{\Sigma^*} = \bigcup_{\sigma \in \Sigma^*} \close{\sigma}$, i.e., the strategy profiles that are $\varepsilon$-close to at least one of the strategy profiles in $\Sigma^*$.

The definition of $\varepsilon$-closeness gives formal substance to the idea of the behaviour of the players (expressed as an equilibrium) being ``similar'': two equilibria that are $\varepsilon$-close are ``similar'' (and vice-versa).
This notion allows us to formally define the behavioural consistency of players in the presence of noise, which amounts to checking whether the equilibria of the noisy game are similar (i.e., $\varepsilon$-close) to the ``expected'' ones under the actual game. Formally:

\begin{definition} \label{def:epsilon-misinformed}
Consider a normal noisy game $mG$ and some \emph{tolerance} $\varepsilon \geq 0$. Then,
\begin{itemize}
    \item $mG$ is \emph{$\varepsilon$-misinformed}
    iff for every natural misinformed equilibrium $\sigma^*$ of $mG$,
    there is a Nash equilibrium $\sigma^0$ of $G^0$, 
    such that $\sigma^* \in \close{\sigma^0}$.
    \item $mG$ is \emph{inverse-$\varepsilon$-misinformed}
    iff for every Nash equilibrium $\sigma^0$ of $G^0$,
    there is a natural misinformed equilibrium $\sigma^*$ of $mG$, 
    such that $\sigma^* \in \close{\sigma^0}$.
\end{itemize}
\end{definition}

\subsection{Running example}
\label{subsec:motivating}

The following example will be used as a running example for the rest of the paper to illustrate our results. 

\begin{ex1}[Running example]\label{ex:mot-def}
We consider two autonomous robotic agents $r, c$, deployed in a remote environment. At some point in time, the human controller asks each of the agents to choose among two actions $s_1, s_2$, also specifying the payoffs for each combination of choices, as shown in matrix $P^0$ below. 
\begin{equation*}
P^0 = \left( 
\begin{array}{cc}
     (3,2) & (0,0) \\
     (0,0) & (2,3) \\
\end{array}
\right)
\end{equation*}
The above payoff matrix corresponds to the well-known Battle of the Sexes (BoS) game\footnote{See \url{https://en.wikipedia.org/wiki/Battle_of_the_sexes_(game_theory)}.}, which has three Nash equilibria, namely $\sigma^0_1 = ((1,0),(1,0))$, $\sigma^0_2 = ((0,1),(0,1))$, $\sigma^0_3 = ((0.6,0.4),(0.4,0.6))$.

However, one of the components of the central communication module has received damage, unknowingly to the agents or the human controller, causing it to introduce a random noise ($\delta \sim \gndist{0,1}$) to each of the values in $P^0$ during transmission. The above setting can be modelled as a normal noisy game $mG = \tuple{G^0, G^r, G^c} \sim G^0 + \gndist{M^x_y,\Devi^x_y}$, where the payoff matrix of $G^0$ is $P^0$, and $M^x_y = \tbl{0}_{2\times 2}, \Devi^x_y = \tbl{1}_{2\times 2}$ for all $x,y\in \{r,c\}$.

Our objective here is to compute the probability that the robotic agents will exhibit behavioural consistency (Definition \ref{def:epsilon-misinformed}), despite the noise caused by the malfunction.
This question will be addressed in Section~\ref{sec:noisy_probabilities} below.
\end{ex1}

\section{Probabilities for behavioural consistency}
\label{sec:noisy_probabilities}

In this section, we will compute the probabilities for a normal noisy game being (inverse-)$\varepsilon$-misinformed. For better readability, we split our analysis in 3 subsections.
In Subsection \ref{subsec:game-theory-results-for-prob}, we recast some known results from game theory in a way that is more suitable for our analysis, whereas in Subsection \ref{subsec:mis-games-results-for-prob}, we develop some results that determine necessary and sufficient conditions for a misinformation game to be (inverse-)$\varepsilon$-misinformed. These results are then employed in Subsection \ref{subsec:main-results-prob-emis-invemis} to compute the required probabilities. The respective results are summarized in Table \ref{tab:game-theory-basics-depiction} (for Subsection \ref{subsec:game-theory-results-for-prob}), Table \ref{tab:misinformed-cases} (for Subsection \ref{subsec:mis-games-results-for-prob}) and Tables \ref{tab:prob-formulas-U}, \ref{tab:prob-formulas-NE} and \ref{tab:prob-emis-invemis} (for Subsection \ref{subsec:main-results-prob-emis-invemis}).

\subsection{Determining Equilibrium Strategies}
\label{subsec:game-theory-results-for-prob}

For a $2 \times 2$ bimatrix game $G$, we denote by \ug[G]{x}{i} the \emph{utility gain} of strategy $s_1$ (compared to $s_2$) for player $x \in \{r,c\}$ when her opponent plays $s_i$, in game $G$. The reference to $G$ will be omitted when obvious from the context.
Note that \ug{x}{i} is determined by the elements of the payoff matrix of $G$ (say $P= (P_r;P_c)$) as follows:
\begin{itemize}
    \item For $x = r$, $\ug{r}{i} = P_r[1,i] - P_r[2,i]$
    \item For $x = c$, $\ug{c}{i} = P_c[i,1] - P_c[i,2]$
\end{itemize}

Intuitively, $\ug{x}{i} > 0$ would mean that player $x$ would play $s_1$, if her opponent chooses to play $s_i$, i.e., that $s_1$ is the best response (for $x$) to $s_i$.
Similarly, $\ug{x}{i} < 0$ would mean that player $x$ would play $s_2$, if her opponent chooses to play $s_i$, i.e., that $s_2$ is the best response (for $x$) to $s_i$.
Finally, when $\ug{x}{i} = 0$, then player $x$ is indifferent as to whether to play $s_1$ or $s_2$, i.e., it has two pure best responses for her opponent's pure strategy $s_i$, indicating that the game is degenerate.

\begin{ex1}[Running example]\label{ex:mot-def-computeUGAIN}
From Example~\ref{ex:mot-def} we have that, for $G^0$, the following hold: $\ug{r}{1} = 3$, $\ug{r}{2} = -2$, $\ug{c}{1} = 2$, and $\ug{c}{2} = -3$.
\end{ex1}

Some well-known results from game theory for bimatrix games can be recast using the concept of \ug{x}{i}. 
For example, the following proposition gives an equivalent formulation of the degeneracy criterion for $2 \times 2$ bimatrix games\footnote{Proofs for all results appear in the Appendix.}:

\begin{proposition}
\label{prop:game-theory-degenerate}
A $2\times 2$ bimatrix game $G$ is degenerate if and only if $\ug{x}{i} = 0$ for some $x \in \{r,c\}, i \in \{1,2\}$.
\end{proposition}

When a non-degenerate $2 \times 2$ bimatrix game has a mixed Nash equilibrium, then its value is determined by $\ug{x}{i}$:

\begin{proposition}
\label{prop:game-theory-basics-mixed-value}
Consider a non-degenerate $2\times 2$ bimatrix game $G = \tuple{N,S,P}$, for $P = (P_r;P_c)$. 
If $(p,1-p) \in NE_x(G)$ for some $0 < p < 1$, $x\in \{r,c\}$, then:
\begin{equation*}
    p = \frac {\ug{\bar x}{2}} {\ug{\bar x}{2} - \ug{\bar x}{1}}
\end{equation*}
\end{proposition}

Now consider a non-degenerate $2 \times 2$ bimatrix game $G$ and some player $x\in \{r,c\}$. From classical results in game theory, we know that there are 4 possible cases for $NE_x(G)$, namely $NE_x(G) = \{(1,0)\}$, $NE_x(G) = \{(0,1)\}$, $NE_x(G) = \{(p,1-p)\}$ for some $0 < p < 1$ and $NE_x(G) = \{(1,0), (0,1), (p,1-p)\}$ for some $0 < p < 1$. If the game is degenerate, then there is one additional possibility, namely that $NE_x(G) = \{(p,1-p) \mid 0 \leq p \leq 1\} = \Sigma_x$.

For non-degenerate games, the value of $NE_x(G)$ can be determined using the following:
\begin{itemize}
    \item $NE_x(G) = \{(1,0)\}$ if and only if $s_1$ is dominant for $x$, or $s_i$ is dominant for $\bar x$ and $s_1$ is the best response for $x$ on $s_i$.
    \item $NE_x(G) = \{(0,1)\}$ if and only if $s_2$ is dominant for $x$, or $s_i$ is dominant for $\bar x$ and $s_2$ is the best response for $x$ on $s_i$.
    \item $NE_x(G) = \{(p,1-p)\}$ for some $0 < p < 1$ if and only if no strategy is dominant for either player and no pure Nash equilibrium exists.
    \item $NE_x(G) = \{(1,0), (0,1), (p,1-p)\}$ for some $0 < p < 1$ if and only if no strategy is dominant for either player and two pure Nash equilibria exist.
\end{itemize}

The above conditions can also be expressed in terms of \ug{x}{i}, as shown in Table \ref{tab:game-theory-basics-depiction}. 
In the table, the various (mutually exclusive) cases are visualised for player $r$ and for a non-degenerate game. The small figure in the rightmost column shows the depicted condition in terms of the relative order among the elements of $P_r$ (blue lines) or $P_c$ (yellow lines), which is determined by the sign (positive or negative) of \ug{x}{i}. The first column provides a reference to the formulation of Proposition \ref{prop:game-theory-basics}, where the above are formally stated and proved.

Before showing Proposition \ref{prop:game-theory-basics}, for brevity, we introduce the following predicates to refer to the different cases with regards to the value of $NE_x(G)$:

\begin{itemize}

    \item \emph{Only-pure}: \OP{G}{x}{i}, which is true if and only if the only equilibrium strategy for player $x$ in game $G$ is to play $s_i$, i.e.:
    \begin{equation*}
    \begin{split}
        & \OP{G}{x}{1} \text{ iff } NE_x(G) = \{(1,0)\} \text{ and } \OP{G}{x}{2} \text{ iff } NE_x(G) = \{(0,1)\}
    \end{split}
    \end{equation*}
    
    \item \emph{Only-mixed}: \OM{G}{x}{p}, which is true if and only if the only equilibrium strategy for player $x$ in game $G$ is $(p, 1-p)$ (where $0 < p < 1$), i.e.:
    \begin{equation*}
        \OM{G}{x}{p} \text{ iff } NE_x(G) = \{(p,1-p)\}
    \end{equation*}
    
    \item \emph{Pure-and-mixed}: $\PM{G}{x}{p}$, which is true if and only if player $x$ has 3 equilibrium strategies in game $G$, two pure and one mixed, and the mixed one is $(p,1-p)$ (where $0 < p < 1$), i.e.:
    \begin{equation*}
        \PM{G}{x}{p} \text{ iff } NE_x(G) = \{(1,0),(0,1),(p,1-p)\}
    \end{equation*}
    
    \item \emph{Ranged-only-mixed}: $\ROM{G}{x}{\omega_1}{\omega_2}$, which is true if and only if \OM{G}{x}{p} is true for some $\omega_1 < p < \omega_2$, i.e.:
    \begin{equation*}
        \ROM{G}{x}{\omega_1}{\omega_2} \text{ iff } NE_x(G) = \{(p,1-p)\} \text{ for some } p \text{ such that } \omega_1 < p < \omega_2
    \end{equation*}
    
    \item \emph{Ranged-pure-and-mixed}: $\RPM{G}{x}{\omega_1}{\omega_2}$, which is true if and only if \PM{G}{x}{p} is true for some $\omega_1 < p < \omega_2$, i.e.:
    \begin{equation*}
    \begin{split}
        & \RPM{G}{x}{\omega_1}{\omega_2} \text{ iff } NE_x(G) = \{(1,0),(0,1),(p,1-p)\} \text{ for some }  p \text{ such that } \omega_1 < p < \omega_2
    \end{split}
    \end{equation*}
    
    \item \emph{Infinite-Nash}: $\IN{G}{x}$, which is true if and only if player $x$ has an infinite number of equilibrium strategies, namely the entire $\Sigma_x$ (note that this is possible only for degenerate games), i.e.:
    \begin{equation*} 
        \IN{G}{x} \text{ iff } NE_x(G) = \Sigma_x
    \end{equation*}
\end{itemize}

When the game $G$ is obvious from the context, we will omit the superscript $G$ from the above.
Now we can formally state Proposition \ref{prop:game-theory-basics}, which formalises the intuition of Table \ref{tab:game-theory-basics-depiction}:

\input{Tables/game-theory-basics-depiction}

\begin{proposition}
\label{prop:game-theory-basics}
For any non-degenerate $2 \times 2$ bimatrix game the following hold:
\begin{enumerate}
    \item $\OP{}{x}{1}$ if and only if either one of the following is true:
    \begin{enumerate}
        \item $(\ug{x}{1} > 0) \bigwedge (\ug{x}{2} > 0)$
        \item $(\ug{x}{1} > 0) \bigwedge (\ug{x}{2} < 0) \bigwedge (\ug{\bar x}{1} > 0) \bigwedge (\ug{\bar x}{2} > 0)$
        \item $(\ug{x}{1} < 0) \bigwedge (\ug{x}{2} > 0) \bigwedge (\ug{\bar x}{1} < 0) \bigwedge (\ug{\bar x}{2} < 0)$
    \end{enumerate}
    
    \item $\OP{}{x}{2}$ if and only if either one of the following is true:
    \begin{enumerate}
        \item $(\ug{x}{1} < 0) \bigwedge (\ug{x}{2} < 0)$
        \item $(\ug{x}{1} < 0) \bigwedge (\ug{x}{2} > 0) \bigwedge (\ug{\bar x}{1} > 0) \bigwedge (\ug{\bar x}{2} > 0)$
        \item $(\ug{x}{1} > 0) \bigwedge (\ug{x}{2} < 0) \bigwedge (\ug{\bar x}{1} < 0) \bigwedge (\ug{\bar x}{2} < 0)$
    \end{enumerate}
    
    \item $\OM{}{x}{p}$ if and only if $p = \frac{\ug{\bar x}{2}}{\ug{\bar x}{2} - \ug{\bar x}{1}}$ and either one of the following is true:
    \begin{enumerate}
        \item $(\ug{x}{1} > 0) \bigwedge (\ug{x}{2} < 0) \bigwedge (\ug{\bar x}{1} < 0) \bigwedge (\ug{\bar x}{2} > 0)$
        \item $(\ug{x}{1} < 0) \bigwedge (\ug{x}{2} > 0) \bigwedge (\ug{\bar x}{1} > 0) \bigwedge (\ug{\bar x}{2} < 0)$
    \end{enumerate}
    
    \item $\PM{}{x}{p}$ if and only if $p = \frac{\ug{\bar x}{2}}{\ug{\bar x}{2} - \ug{\bar x}{1}}$ and either one of the following is true:
    \begin{enumerate}
        \item $(\ug{x}{1} > 0) \bigwedge (\ug{x}{2} < 0) \bigwedge (\ug{\bar x}{1} > 0) \bigwedge (\ug{\bar x}{2} < 0)$
        \item $(\ug{x}{1} < 0) \bigwedge (\ug{x}{2} > 0) \bigwedge \ug{\bar x}{1} < 0) \bigwedge (\ug{\bar x}{2} > 0)$
    \end{enumerate}
\end{enumerate}
\end{proposition}

An analogous set of conditions determines whether the ``ranged'' versions of the above predicates are true:

\begin{corollary}
\label{cor:game-theory-basics}
Given a non-degenerate $2\times 2$ bimatrix game $G$, the following hold:
\begin{enumerate}

    \item $\ROM{}{x}{\omega_1}{\omega_2}$ if and only if $\omega_1 < \frac{\ug{\bar x}{2}}{\ug{\bar x}{2} - \ug{\bar x}{1}} < \omega_2$ and either one of the following is true:
    \begin{enumerate}
        \item $(\ug{x}{1} > 0) \bigwedge (\ug{x}{2} < 0) \bigwedge (\ug{\bar x}{1} < 0) \bigwedge (\ug{\bar x}{2} > 0)$
        \item $(\ug{x}{1} < 0) \bigwedge (\ug{x}{2} > 0) \bigwedge (\ug{\bar x}{1} > 0) \bigwedge (\ug{\bar x}{2} < 0)$
    \end{enumerate}
    
    \item $\RPM{}{x}{\omega_1}{\omega_2}$ if and only if $\omega_1 < \frac{\ug{\bar x}{2}}{\ug{\bar x}{2} - \ug{\bar x}{1}} < \omega_2$ and either one of the following is true:
    \begin{enumerate}
        \item $(\ug{x}{1} > 0) \bigwedge (\ug{x}{2} < 0) \bigwedge (\ug{\bar x}{1} > 0) \bigwedge (\ug{\bar x}{2} < 0)$
        \item $(\ug{x}{1} < 0) \bigwedge (\ug{x}{2} > 0) \bigwedge (\ug{\bar x}{1} < 0) \bigwedge (\ug{\bar x}{2} > 0)$
    \end{enumerate}
\end{enumerate}
\end{corollary}

\begin{ex1}[Running example]\label{ex:mot-def-computeNash}
Continuing Example~\ref{ex:mot-def}, we note that the signs of the various $\ug x i$ (as computed in Example \ref{ex:mot-def-computeUGAIN}) indicate that case (4a) of Table \ref{tab:game-theory-basics-depiction} holds. Thus, the actual game $G^0$ has both pure and mixed Nash equilibria.
Given that $\frac { \ug c 2} {\ug c2 - \ug c1} = 0.6$, it follows that \PM{}{r}{0.6} is true.
Analogously, since $\frac { \ug r 2} {\ug r2 - \ug r1} = 0.4$, it follows that \PM{}{c}{0.4} is true.
\end{ex1}


\subsection{Misinformation Games}
\label{subsec:mis-games-results-for-prob}

In this subsection, we provide necessary and sufficient conditions for a misinformation game to be \emph{(inverse-)$\varepsilon$-misinformed}. These are given in Propositions \ref{prop:e-mis-no-prob}, \ref{prop:inv-e-mis-no-prob}, and use the notation previously introduced. Note that the propositions apply for all canonical misinformation games, not just noisy games. The results of this subsection are summarized in Table \ref{tab:misinformed-cases}.

\begin{proposition}
\label{prop:e-mis-no-prob}
Consider a canonical misinformation game $mG = \tuple{G^0,G^r,G^c}$, where $G^0$ is a $2 \times 2$ bimatrix game and $G^r, G^c$ are non-degenerate.
Then, $mG$ is $\varepsilon$-misinformed if and only if, for all $x \in \{r,c\}$, one of the following is true:
\begin{enumerate}
    \item $\OP{G^0}{x}{i}$ and $\OP{G^x}{x}{i}$ for some $i \in \{1,2\}$
    \item $\OM{G^0}{x}{p^0}$ for some $0 < p^0 < 1$ and $\ROM{G^x}{x}{\omega_1}{\omega_2}$, 
    where $\omega_1 = \max\{0, p^0 - \varepsilon\}$, $\omega_2 = \min\{1, p^0 + \varepsilon\}$
    \item $\PM{G^0}{x}{p^0}$ for some $0 < p^0 < 1$ and 
    $\OP{G^x}{x}{1} \bigvee \OP{G^x}{x}{2} \bigvee \ROM{G^x}{x}{\omega_1}{\omega_2} \bigvee \RPM{G^x}{x}{\omega_1}{\omega_2}$,
    where $\omega_1 = \max\{0, p^0 - \varepsilon\}$, $\omega_2 = \min\{1, p^0 + \varepsilon\}$
    \item $\IN{G^0}{x}$
\end{enumerate}
\end{proposition}

\begin{proposition}
\label{prop:inv-e-mis-no-prob}
Consider a canonical misinformation game $mG = \tuple{G^0,G^r,G^c}$, where $G^0$ is a $2 \times 2$ bimatrix game and $G^r, G^c$ are non-degenerate.
Then, $mG$ is inverse-$\varepsilon$-misinformed if and only if, for all $x \in \{r,c\}$, one of the following is true:
\begin{enumerate}
    \item $\OP{G^0}{x}{i}$ and $\OP{G^x}{x}{i} \bigvee \RPM{G^x}{x}{0}{1}$ for some $i \in \{1,2\}$
    \item $\OM{G^0}{x}{p^0}$ for some $0 < p^0 < 1$ and $\ROM{G^x}{x}{\omega_1}{\omega_2} \bigvee \RPM{G^x}{x}{\omega_1}{\omega_2}$,  
    where $\omega_1 = \max\{0, p^0 - \varepsilon\}$, $\omega_2 = \min\{1, p^0 + \varepsilon\}$
    \item $\PM{G^0}{x}{p^0}$ for some $0 < p^0 < 1$ and
    $\RPM{G^x}{x}{\omega_1}{\omega_2}$,  
    where $\omega_1 = \max\{0, p^0 - \varepsilon\}$, $\omega_2 = \min\{1, p^0 + \varepsilon\}$
    \item $\IN{G^0}{x}$ and $\varepsilon > 0.5$ and 
    $\RPM{G^x}{x}{\omega_1'}{\omega_2'}$, 
    where $\omega_1' = \max\{0, 1 - \varepsilon\}$, $\omega_2' = \min\{1, \varepsilon\}$
\end{enumerate}
\end{proposition}

\input{Tables/misinformed-cases}

\subsection{Probabilities}
\label{subsec:main-results-prob-emis-invemis}

We will now exploit the results of the previous subsections, in order to compute the probabilities associated to various events, eventually leading up to the computation that a given normal noisy game $mG$ is (inverse-)$\varepsilon$-misinformed. 
The results are summarized in Table \ref{tab:prob-emis-invemis}, whereas intermediate results necessary to compute the above probabilities appear in Tables \ref{tab:prob-formulas-U} and \ref{tab:prob-formulas-NE}.

For a normal noisy game $mG \sim G^0 + \gndist{M, \Devi}$, we define the family of random variables $\ugr{y}{x}{i}$, such that, for any $x,y \in \{r,c\}, i \in \{1,2\}$:
$$
\ugr{y}{x}{i} = \ug[G^y]{x}{i}
$$
Applying formula (\ref{eq:normal-dist-sum}) from Subsection \ref{subsec:probabilities}, we observe that $\ugr{y}{x}{i} \sim \gndist{\mu_{\ugr{y}{x}{i}}, \devi_{\ugr{y}{x}{i}}}$ for $\mu_{\ugr{y}{x}{i}}, \devi_{\ugr{y}{x}{i}}$ as shown in Table \ref{tab:prob-formulas-U}.
The cdf and pdf of \ugr{y}{x}{i} (as resulting from formula (\ref{eq:normal-gen-pdf-cdf}) in Subsection \ref{subsec:probabilities}), as well as the probabilities for $\ugr{y}{x}{i}$ taking certain values are also shown in the same table.

\begin{ex1}[Running example]\label{ex:mot-def-computeCDFPDF}
Continuing our running example (Example \ref{ex:mot-def}), we can now compute the distribution followed by the random variables $\ugr yxi$ for $x,y \in \{r,c\}$, $i \in \{1,2\}$, using Table \ref{tab:prob-formulas-U}.
Indeed, by Table \ref{tab:prob-formulas-U}, 
$\mu_{\ugr yxi} = \ug xi$ and
$\devi_{\ugr yxi} = \sqrt 2$ for all $x,y \in \{r,c\}$, $i \in \{1,2\}$.\\
As a more specific example, let us consider $\ugr r r 1$, which corresponds to the random variable representing the utility gain of strategy $s_1$ (as opposed to $s_2$), for player $r$ under the viewpoint of player $r$. By Table \ref{tab:prob-formulas-U}, and the values for $M, \Devi$ (Example \ref{ex:mot-def}), we conclude that
$\mu_{\ugr rr1} = 3$, $\devi_{\ugr rr1} = \sqrt 2$. Analogously, we can compute the rest.
From this, and formulas \ref{eq:normal-gen-pdf-cdf} in subsection \ref{subsec:probabilities}, we get that the pdf and cdf of $\ugr rr1$ are:
\begin{equation*}
    \begin{aligned}    
    f_{\ugr rr1}(x) = 
        \tfrac{1}{ 2 \sqrt{\pi}} e^{- \left( \tfrac{x - 3}{2}\right)^2} \quad \text{   and   } \quad F_{\ugr rr1}(x) = 
        \tfrac{1}{\sqrt{2\pi}} \int^{\tfrac{x-3}{\sqrt 2}}_{-\infty} e^{-\frac{t^2}{2}} dt \\
\end{aligned}
\end{equation*}
\end{ex1}

\input{Tables/prob-formulas-U}

Propositions \ref{prop:game-theory-degenerate} and \ref{prop:game-theory-basics} can now be employed to determine the probability that $NE_x(G^x)$ takes a certain value, based on the probabilities that $\ugr{y}{x}{i}$ take certain values.
More precisely, Lemma \ref{prop:non-degenerate} is the counterpart of Proposition \ref{prop:game-theory-degenerate}:

\begin{proposition}
\label{prop:non-degenerate}
In any normal noisy game $mG = \tuple{G^0, G^r, G^c}$, the probability that $G^x$ is degenerate (for $x \in \{r,c\}$) is $0$.
\end{proposition}

To formulate the counterpart of Proposition \ref{prop:game-theory-basics}, the following lemma will prove helpful:

\begin{lemma}
\label{lem:division}
Consider two independent random variables $X \sim \gndist{\mu_X, \devi_X}, Y \sim \gndist{\mu_Y, \devi_Y}$, with pdfs $f_X,f_Y$ respectively, and some $\Omega_1, \Omega_2 \in \mathbb{R} \cup \{-\infty\}$ such that $-\infty \leq \Omega_1 < \Omega_2 \leq 0$. Then:

\begin{equation*}
\begin{split}
& \prob{\Omega_1 \leq \frac{X}{Y} \leq \Omega_2, X < 0, Y > 0} = \int_{0}^{+\infty} \left(\int_{\Omega_1 y}^{\Omega_2 y} \! f_X(x) \, \mathrm{d}x \right) \frac{f_{Y}(y)}{y} \, \mathrm{d}y \\
& \prob{\Omega_1 \leq \frac{X}{Y} \leq \Omega_2, X > 0, Y < 0} =  \int_{-\infty}^{0} \left(\int_{\Omega_1 y}^{\Omega_2 y} \! f_X(x) \, \mathrm{d}x \right) \frac{f_{Y}(y)}{y} \, \mathrm{d}y \\
\end{split}
\end{equation*}
\end{lemma}

\input{Tables/prob-formulas-NE}

The next proposition determines the probability that $NE_x(G^x)$ will have each of its possible values (see also Table \ref{tab:prob-formulas-NE}):

\begin{proposition}
\label{prop:prob-NExG}
Consider a normal noisy game $mG \sim G^0 + \gndist{M, \Devi}$, and some $x \in \{r,c\}$.
Then, the probabilities $\prob{\OP{G^x}{x}{1}}$, $\prob{\OP{G^x}{x}{2}}$, $\prob{\ROM{G^x}{x}{\omega_1}{\omega_2}}$, $\prob{\RPM{G^x}{x}{\omega_1}{\omega_2}}$ and $\prob{\IN{G^x}{x}}$ are as shown in Table \ref{tab:prob-formulas-NE}.
\end{proposition}

\input{Tables/prob-emis-invemis}

\begin{ex1}[Running example]\label{ex:mot-def-computeProb-formulas-NE}
Continuing Example \ref{ex:mot-def}, we can now compute the probabilities that each of the robotic agents will believe that they play a game with pure, mixed or pure and mixed strategies. 
We start with the computation of the relevant $F_{\ugr yxi}(0)$ quantities for $x,y \in \{r,c\}$, $i \in \{1,2\}$, which are based on the respective pdf/cdf that were computed in Example \ref{ex:mot-def-computeCDFPDF}:
\begin{itemize}
    \item For the $r$ agent:
    \begin{itemize}
        \item[] $F_{\ugr{r}{r}{1}}(0) = 0.017$,  $F_{\ugr{r}{r}{2}}(0) = 0.921$,  $F_{\ugr{r}{c}{1}}(0) = 0.078$,  $F_{\ugr{r}{c}{2}}(0) = 0.983$
    \end{itemize}
    \item For the $c$ agent:
    \begin{itemize}
        \item[] $F_{\ugr{c}{c}{1}}(0) = 0.078$,  $F_{\ugr{c}{c}{2}}(0) = 0.983$,  $F_{\ugr{c}{r}{1}}(0) = 0.921$,  $F_{\ugr{c}{r}{2}}(0) = 0.016$
    \end{itemize}
\end{itemize}
Regarding the two double integrals in the third and fourth formulas in Table~\ref{tab:prob-formulas-NE} our computations yield the values $0.001$ and $0.229$ for $r$ respectively. Similarly, for $c$ we take the values $0.001$ and $0.189$.

Using the above results it is now easy to compute the following quantities, using the formulas of Table \ref{tab:prob-formulas-NE}:

\begin{itemize}
    \item For the $r$ agent:
    \begin{itemize}
        \item[] $\prob{\OP {G^r}r1} = 0.091$, $\prob{\OP {G^r}r2} = 0.085$, $\prob{\ROM {G^r}r {0.5}{0.7}} = 0.001$, $\prob{\RPM {G^r}r {0.5}{0.7}} = 0.207$
    \end{itemize}
    
    \item For the $c$ agent:
    \begin{itemize}
        \item[] $\prob{\OP {G^c}c1} = 0.091$, $\prob{\OP {G^c}c2} = 0.085$, $\prob{\ROM {G^c}c {0.3}{0.5}} = 0.001$, $\prob{\RPM {G^c}c {0.3}{0.5}} = 0.171$
    \end{itemize}
\end{itemize}
Not unexpectedly, the largest probability is that the agents retain the behaviour predicted by the original game (i.e., playing pure and mixed), but not significantly so.
\end{ex1}

Proposition \ref{prop:prob-NExG} (and the respective Table \ref{tab:prob-formulas-NE}), combined with Proposition \ref{prop:e-mis-no-prob} (and the respective Table \ref{tab:misinformed-cases}) easily leads to the following theorems (summarized in Table \ref{tab:prob-emis-invemis}):

\begin{theorem}
\label{thm:main-emis}
Consider a normal noisy game $mG \sim G^0 + \gndist{M, \Devi}$.
Then: 
\begin{equation*}
    \prob{mG: \varepsilon \text{-misinformed}} = \mathcal{P}_r^{mis} \cdot \mathcal{P}_c^{mis}
\end{equation*}
where, for $x \in \{r,c\}$, $\mathcal{P}_x^{mis}$ is determined by the second column of Table \ref{tab:prob-emis-invemis}.
\end{theorem}

\begin{theorem}
\label{thm:main-invemis}
Consider a normal noisy game $mG \sim G^0 + \gndist{M, \Devi}$.
Then: 
\begin{equation*}
    \prob{mG: \text{inverse-} \varepsilon \text{-misinformed}} = \mathcal{P}_r^{inv} \cdot \mathcal{P}_c^{inv}
\end{equation*}
where, for $x \in \{r,c\}$, $\mathcal{P}_x^{inv}$ is determined by the third column of Table \ref{tab:prob-emis-invemis}.
\end{theorem}

\begin{ex1}[Running example]\label{ex:mot-def-compute-prob-emis-invemis}
Returning to our running example (Example \ref{ex:mot-def}), let us now compute the probability for the respective noisy game to be (inverse-)$\varepsilon$- misinformed, for $\varepsilon = 0.1$. 
To do so, we plug in the formulas from Table~\ref{tab:prob-emis-invemis} into Theorems~\ref{thm:main-emis}-\ref{thm:main-invemis}, and, using the previously computed probabilities from Example \ref{ex:mot-def-computeProb-formulas-NE}, we take:
\begin{itemize}
    \item[] $\mathcal{P}_r^{mis} = 0.386 $, \hspace{.2em} $\mathcal{P}_r^{inv} = 0.207$,  \hspace{.2em} $\mathcal{P}_c^{mis} = 0.349$,  \hspace{.2em} $\mathcal{P}_c^{inv} = 0.171$
\end{itemize}
Thus, 
\begin{itemize}
    \item[] $\prob{mG: \varepsilon \text{-misinformed}} = 0.135$ 
    \item[] $\prob{mG: \text{inverse-} \varepsilon \text{-misinformed}} = 0.035$.
\end{itemize}

Thus, the conclusion of this analysis is that, the ``Battle of the Sexes'' (BoS) game, when receiving noise that follows the standard normal distribution ($\gndist{0,1}$) in each of its payoffs and for each player, will be $0.1$-misinformed with probability $13.5\%$ and inverse-$0.1$-misinformed with probability $3.5\%$.

Note that the original BoS game has 3 Nash equilibria, two pure and one mixed. Thus, the above results imply that, by defining closeness using $\varepsilon = 0.1$:
\begin{itemize}
    \item With probability $13.5\%$, all (misinformed) equilibrium points of the noisy game will be close to one of the expected equilibria (in BoS). Thus, under this probability, the agents will have one or more natural misinformed equilibria, all of which will be close to one of the BoS' Nash equilibria. This means that, with probability $13.5\%$, the agents' behaviour (no matter which of their equilibrium points they choose) will be close to the expected one.
    
    \item With probability $3.5\%$, for each of the three equilibria of BoS, there will be at least one (misinformed) equilibrium point that is close to it. Thus, under this probability, there will be at least 3 natural misinformed equilibria, although the agents may also have other (misinformed) equilibria as well that are not close to any Nash equilibrium of BoS. This means that, with probability $3.5\%$, all equilibria of BoS are within the valid options (modulo the closeness assumption) for the agents.
\end{itemize}
\end{ex1}

\section{Results for noisy games}
\label{sec:results_noisy}

The results of Section \ref{sec:noisy_probabilities} provide the formulas to compute the probability of a given normal noisy game to be (inverse-)$\varepsilon$-misinformed (i.e., behaviourally consistent). In this section, we explore the properties of these formulas, to understand better their behaviour. 

To do so, we first observe that the probability of a normal noisy game $mG \sim G^0 + \gndist{M, \Devi}$ being behaviourally consistent is essentially a function of:
\begin{itemize}
    \item The tolerance $\varepsilon$.
    \item The payoff matrix of the actual game of $mG$. This affects the probabilities in two ways: first, because it determines the equilibria of $G^0$, and, thus, the case to consider in Table \ref{tab:prob-emis-invemis}; second, because it affects $\mu_{\ugr{y}{x}{i}}$ (see Table \ref{tab:prob-formulas-U}). 
    \item The noise pattern, determined by the matrices $M, \Devi$.
\end{itemize}
In the following subsections, we study the effect of each of these parameters on the probability of $mG$ being (inverse-)$\varepsilon$-misinformed.

\subsection{Effect of modifying tolerance (\texorpdfstring{$\varepsilon$}{})}
\label{subsec:mod-epsilon}

With regards to tolerance ($\varepsilon$), we expect that larger values of tolerance would translate to higher probability of behavioural consistency. Although this is true, we also observe that there are several cases where increasing tolerance does not affect the probability of behavioural consistency.
The following proposition clarifies the situation:

\input{Tables/properties-e-monotonic-misinvmis}

\begin{proposition}
\label{prop:properties-e-monotonic-misinvmis}
Consider some $mG \sim G^0 + \gndist{M,\Devi}$ and $\varepsilon_1, \varepsilon_2$, such that $0 \leq \varepsilon_1 < \varepsilon_2$. Then:
\begin{enumerate}
    
    \item If $NE(G^0)$ contains a single pure strategy, then:
    \begin{itemize}
        \item $\prob{mG: \varepsilon_1 \text{-misinformed}} =  \prob{mG: \varepsilon_2 \text{-misinformed}}$
        \item $\prob{mG: \text{inverse-} \varepsilon_1 \text{-misinformed}} =  \prob{mG: \text{inverse-} \varepsilon_2 \text{-misinformed}}$  
    \end{itemize}
    
    \item If $NE(G^0)$ is finite and $((p^0,1-p^0),(q^0,1-q^0)) \in NE(G^0)$ for some $0 < p^0 < 1$, $0 < q^0 < 1$, then:
    \begin{enumerate}
        \item If $\max\{p^0, q^0, 1-p^0, 1-q^0\} \leq \varepsilon_1$, then: 
        \begin{itemize}
            \item $\prob{mG: \varepsilon_1 \text{-misinformed}} =   \prob{mG: \varepsilon_2 \text{-misinformed}}$
            \item $\prob{mG: \text{inverse-} \varepsilon_1 \text{-misinformed}} =  \prob{mG: \text{inverse-} \varepsilon_2 \text{-misinformed}}$  
        \end{itemize}
    
        \item If $\max\{p^0, q^0, 1-p^0, 1-q^0\} > \varepsilon_1$, then: 
        \begin{itemize}
            \item $\prob{mG: \varepsilon_1 \text{-misinformed}} <  \prob{mG: \varepsilon_2 \text{-misinformed}}$
            \item $\prob{mG: \text{inverse-} \varepsilon_1 \text{-misinformed}} <  \prob{mG: \text{inverse-} \varepsilon_2 \text{-misinformed}}$  
        \end{itemize}
    \end{enumerate}
    
    \item If $NE(G^0)$ is infinite, then:
    \begin{enumerate}
        \item If $\varepsilon_1 \geq 1$ or $\varepsilon_2 \leq 0.5$, then:
        \begin{itemize}
            \item $\prob{mG: \varepsilon_1 \text{-misinformed}} =   \prob{mG: \varepsilon_2 \text{-misinformed}}$
            \item $\prob{mG: \text{inverse-} \varepsilon_1 \text{-misinformed}} =   \prob{mG: \text{inverse-} \varepsilon_2 \text{-misinformed}}$  
        \end{itemize}
    
        \item If $\varepsilon_1 < 1$ and $\varepsilon_2 > 0.5$, then:
        \begin{itemize}
            \item $\prob{mG: \varepsilon_1 \text{-misinformed}} =   \prob{mG: \varepsilon_2 \text{-misinformed}}$
            \item $\prob{mG: \text{inverse-} \varepsilon_1 \text{-misinformed}} <  \prob{mG: \text{inverse-} \varepsilon_2 \text{-misinformed}}$  
        \end{itemize}
    \end{enumerate}
\end{enumerate}
\end{proposition}

\input{Tables/properties-e-minmax}

Proposition \ref{prop:properties-e-monotonic-misinvmis} has several interesting consequences. First, we note that the probability for a given $mG$ to be (inverse-)$\varepsilon$-misinformed is non-decreasing with respect to $\varepsilon$.
When there is a pure Nash equilibrium, the choice of $\varepsilon$ is irrelevant to the value of these probabilities.
When there is a mixed Nash equilibrium (case 2 of the proposition), there is a limit above which $\varepsilon$ does not affect the value of the related probability; this limit depends on the actual mixed equilibrium, but it is always equal to, or larger than $0.5$, and smaller than $1$.
Finally, in the case where there is an infinite number of equilibria, $\varepsilon$ affects the probabilities only for certain values (between $0.5$ and $1$, and only for the inverse-$\varepsilon$-misinformed case), as detailed in case 3 of Proposition \ref{prop:properties-e-monotonic-misinvmis}. These are summarised in Table \ref{tab:properties-e-monotonic-misinvmis}.

Our results (and Table \ref{tab:properties-e-monotonic-misinvmis}) indicate that the minimal value for the probability of $mG$ being (inverse-)$\varepsilon$-misinformed is given for $\varepsilon = 0$. Its maximal value is taken for an appropriate $\varepsilon$ (depending on the case); in all cases $\varepsilon = 1$ would also give that maximal value. 
These maximal/minimal values can be easily deduced by Table \ref{tab:prob-emis-invemis} for the above choices of $\varepsilon$, and are given in Table \ref{tab:properties-e-minmax} for convenience. Note that the actual result for the minimal/maximal values results by multiplying 
$\mathcal{P}_r^{mis}$ with $\mathcal{P}_r^{mis}$, and
$\mathcal{P}_r^{inv}$ with $\mathcal{P}_r^{inv}$ for $\varepsilon$-misinformed and inverse-$\varepsilon$-misinformed respectively.

Another important result (albeit relatively obvious) is that the probability of $mG$ being (inverse-)$\varepsilon$-misinformed, viewed as a function of $\varepsilon$, is continuous. This is a direct consequence of the results in Tables \ref{tab:prob-formulas-U}, \ref{tab:prob-formulas-NE}, \ref{tab:prob-emis-invemis}. An important consequence of this fact, by well-known results of calculus, is that, for any given target value for the probabilities of (inverse-)$\varepsilon$-misinformed (within the bounds shown in Table \ref{tab:properties-e-minmax}), there exists some $\varepsilon$ whose application would result to that value for the respective probability.

\subsection{Effect of changing the game (\texorpdfstring{$G^0$}{}) and the mean (\texorpdfstring{$M$}{})}
\label{subsec:mod-G0-M}

Consider a misinformation game $mG \sim G^0 + \gndist{M,\Devi}$, and let us informally ponder on the effect of bias in the noise of a game. A biased noise is noise whose mean $M$ is non-zero, i.e., $M \neq \tbl{0}$. Let us consider only player $r$, for simplicity. In such a scenario, we know that $G^r \sim G^0 + \gndist{M^r, \Devi^r}$. Observe that this is the same as writing $G^r \sim (G^0 + M^r) + \gndist{\tbl{0},\Devi^r}$. Using this simple reasoning, the computation of the probabilities of behavioural consistency for $mG$ for biased noise can be reduced to computations related to some $\overline{mG}$ with unbiased noise ($M = \tbl{0}$), whose actual game will be the sum of $G^0$ and $M^x$. 

However, there are two caveats here.
First, since $M^r$ may be different than $M^c$, our original misinformation game is essentially reduced to two different misinformation games (say $\overline{mG_r}, \overline{mG_c}$), i.e., one per player.
Second, in the case where the equilibria of $G^0$ are different than the equilibria of $G^0 + M^x$, care should be taken to consult the proper line in Table \ref{tab:prob-emis-invemis} while computing the probability of $mG$ being (inverse-)$\varepsilon$-misinformed. In particular, the line to consider should be the one related to the equilibria of $G^0$, not $G^0 + M^x$. This means that the probability of $mG$ being (inverse-)$\varepsilon$-misinformed may not be the same as the respective probability for $\overline{mG_r}, \overline{mG_c}$.

To prove the above ideas formally, we start with the following proposition:

\begin{proposition}
\label{prop:effect-G0-M-mod}
Consider two noisy games $mG \sim G^0 + \gndist{M,\Devi}$, $\overline{mG} \sim \overline{G^0} + \gndist{\overline{M},\Devi}$. Suppose that there exists $a \in \mathbb{R}$, $x \in \{r,c\}$ such that $G^0 + M^x = \overline{G^0} + \overline{M^x} + \tbl{a}$. Then:
\begin{itemize}
    
    \item For any $i \in \{1,2\}$, 
    \begin{equation*}
        \prob{\OP{G^x}{x}{i}} = \prob{\OP{\overline{G^x}}{x}{i}}
    \end{equation*}
    
    \item For any $0 \leq \omega_1 \leq \omega_2 \leq 1$, 
    \begin{equation*}
        \prob{\ROM{G^x}{x}{\omega_1}{\omega_2}} = \prob{\ROM{\overline{G^x}}{x}{\omega_1}{\omega_2}}
    \end{equation*}
    
    \item For any $0 \leq \omega_1 \leq \omega_2 \leq 1$, 
    \begin{equation*}
        \prob{\RPM{G^x}{x}{\omega_1}{\omega_2}} = \prob{\RPM{\overline{G^x}}{x}{\omega_1}{\omega_2}}
    \end{equation*}

\end{itemize}
\end{proposition}

Proposition \ref{prop:effect-G0-M-mod} implies that, given a noisy game $mG \sim G^0 + \gndist{M,\Devi}$ and a player $x \in \{r,c\}$, we can generate some other noisy game (say $\overline{mG}$), whose probabilities related to the various outcomes (equilibria) of the game $\overline{G^x}$ of $\overline{mG}$ are identical to the respective ones for $G^x$ (in $mG$). As a matter of fact, there is an infinite number of noisy games that satisfy this property: for any given $\overline{G^0}$ we can find an infinite number of $\overline{M}$ that do this, and for any given $\overline{M}$ we can find an infinite number of $\overline{G^0}$ that do this. 
This observation motivates us to consider some interesting special cases, formalised as corollaries below.

The first interesting case is when $\overline{M} = \tbl{0}$. Given a noisy game $mG$, the following corollary shows that the probabilities related to the various outcomes (equilibria) of the game $G^x$ in $mG$ can be predicted by looking at a properly defined noisy game $\overline{mG}$ where the noise is unbiased (i.e., $\overline{M} = \tbl{0}$). 
Formally:

\begin{corollary}
\label{cor:unbiased-noise}
Consider a noisy game $mG \sim G^0 + \gndist{M,\Devi}$, and some $x \in \{r,c\}$.
Set $\overline{G^0} = G^0 + M^x$, and $\overline{mG} \sim \overline{G^0} + \gndist{\tbl{0},\Devi}$. Then:
\begin{itemize}
    
    \item For any $i \in \{1,2\}$, 
    \begin{equation*}
        \prob{\OP{G^x}{x}{i}} = \prob{\OP{\overline{G^x}}{x}{i}}
    \end{equation*}
    
    \item For any $0 \leq \omega_1 \leq \omega_2 \leq 1$, 
    \begin{equation*}
        \prob{\ROM{G^x}{x}{\omega_1}{\omega_2}} = \prob{\ROM{\overline{G^x}}{x}{\omega_1}{\omega_2}}
    \end{equation*}
    
    \item For any $0 \leq \omega_1 \leq \omega_2 \leq 1$, 
    \begin{equation*}
        \prob{\RPM{G^x}{x}{\omega_1}{\omega_2}} = \prob{\RPM{\overline{G^x}}{x}{\omega_1}{\omega_2}}
    \end{equation*}

\end{itemize}
\end{corollary}

Combining Corollary \ref{cor:unbiased-noise} with Theorems \ref{thm:main-emis}, \ref{thm:main-invemis}, it is easy to compute the probability that $mG$ is (inverse-)$\varepsilon$-misinformed, using the respective probabilities for $\overline{mG}$. This is one of the main results of this subsection, as it allows us to restrict our study to noisy games with unbiased noise only.

An interesting observation is that Corollary \ref{cor:unbiased-noise} applies for some $x \in \{r,c\}$. Thus, we need to define two different $\overline{mG}$ (one for each player $x \in \{r,c\}$) in order to compute the probability that $mG$ is (inverse-)$\varepsilon$-misinformed.
The following corollary holds for both $x \in \{r,c\}$ (and thus foregoes this need), but applies only when $M^r = M^c$, i.e., when the noise received by the two players has the same bias:

\begin{corollary}
\label{cor:unbiased-noise-Mr-equals-Mc}
Consider a noisy game $mG \sim G^0 + \gndist{M,\Devi}$, where $M = (M^*;M^*)$.
Set $\overline{G^0} = G^0 + M^*$, and $\overline{mG} \sim \overline{G^0} + \gndist{\tbl{0},\Devi}$. Then:
\begin{itemize}
    
    \item For any $i \in \{1,2\}$ and $x \in \{r,c\}$,
    \begin{equation*}
        \prob{\OP{G^x}{x}{i}} = \prob{\OP{\overline{G^x}}{x}{i}}
    \end{equation*}
    \item For any $0 \leq \omega_1 \leq \omega_2 \leq 1$ and $x \in \{r,c\}$, 
    \begin{equation*}
        \prob{\ROM{G^x}{x}{\omega_1}{\omega_2}} = \prob{\ROM{\overline{G^x}}{x}{\omega_1}{\omega_2}}
    \end{equation*}
    \item For any $0 \leq \omega_1 \leq \omega_2 \leq 1$ and $x \in \{r,c\}$,
    \begin{equation*}
        \prob{\RPM{G^x}{x}{\omega_1}{\omega_2}} = \prob{\RPM{\overline{G^x}}{x}{\omega_1}{\omega_2}}
    \end{equation*}

\end{itemize}
\end{corollary}

Proposition \ref{prop:effect-G0-M-mod} and Corollary \ref{cor:unbiased-noise} provide the probability of the different events to occur (e.g., the probability that $G^x$ has a certain equilibrium), but do not directly provide the probability for $mG$ being (inverse-)$\varepsilon$-misinformed. Indeed, since $G^0$ and $\overline{G^0}$ may have different equilibria, the computation of the probabilities for $mG$ and $\overline{mG}$ being (inverse-)$\varepsilon$-misinformed may use different rows in Table \ref{tab:prob-emis-invemis}. This is unnecessary only when the two games have the same equilibria:

\begin{corollary}
\label{cor:same-nash-effect-M-G0}
Consider a noisy game $mG \sim G^0 + \gndist{M,\Devi}$.
Set:

\begin{center}
    $\overline{G^0} = G^0 + M^r$, \hspace{.5em} $\widehat{G^0} = G^0 + M^c$, \hspace{.5em} $\overline{mG} \sim \overline{G^0} + \gndist{\tbl{0},\Devi}$, \hspace{.5em} $\widehat{mG} \sim \widehat{G^0} + \gndist{\tbl{0},\Devi}$
\end{center}

\noindent If $NE(G^0) = NE(\overline{G^0}) = NE(\widehat{G^0})$ then:
\begin{itemize}
    \item $\prob{mG: \varepsilon\text{-misinformed}} = 
    \overline{\mathcal{P}_r^{mis}} \cdot \widehat{\mathcal{P}_c^{mis}}$
    \item $\prob{mG: \text{inverse-}\varepsilon\text{-misinformed}} = \overline{\mathcal{P}_r^{inv}} \cdot \widehat{\mathcal{P}_c^{inv}}$
\end{itemize}
where 
$\overline{\mathcal{P}_r^{mis}}$,
$\overline{\mathcal{P}_r^{inv}}$,
$\widehat{\mathcal{P}_c^{mis}}$,
$\widehat{\mathcal{P}_c^{inv}}$ are the probabilities of Table \ref{tab:prob-emis-invemis} for $\overline{mG}$, $\widehat{mG}$ respectively.
\end{corollary}

Note that, in Corollary \ref{cor:same-nash-effect-M-G0}, the computation of the probability for $mG$ to be (inverse-)$\varepsilon$-misinformed, occurs via the combination of quantities from two different noisy games ($\overline{mG}, \widehat{mG}$). As with Corollary \ref{cor:unbiased-noise}, this can be avoided when the noise received by the two players has the same bias, in which case we get a direct computation of the related probability:

\begin{corollary}
\label{cor:same-nash-effect-M-G0-Mr-equals-Mc}
Consider the noisy game $mG \sim G^0 + \gndist{M,\Devi}$, where $M = (M^*;M^*)$.
Set $\overline{G^0} = G^0 + M^*$ and 
$\overline{mG} \sim \overline{G^0} + \gndist{\tbl{0},\Devi}$.
If $NE(G^0) = NE(\overline{G^0})$ then:
\begin{itemize}
    \item $\prob{mG: \varepsilon\text{-misinformed}} = \prob{\overline{mG}: \varepsilon\text{-misinformed}}$
    \item $\prob{mG: \text{inverse-}\varepsilon\text{-misinformed}} = \prob{\overline{mG}: \text{inverse-}\varepsilon\text{-misinformed}}$
\end{itemize}
\end{corollary}

Corollary \ref{cor:same-nash-effect-M-G0-Mr-equals-Mc} is the most specific result, as it gives us a method of computing the probabilities of a noisy game being (inverse-)$\varepsilon$-misinformed using the respective probabilities of another noisy game, under specific assumptions.

The last proposition of this subsection follows easily from Proposition \ref{prop:effect-G0-M-mod}, and shows an elegant, and expected, property of noisy games. In particular, changing the payoff matrix of a game by adding any fixed constant number to all payoffs, does not modify the probability of the respective noisy game to be (inverse-)$\varepsilon$-misinformed (for a fixed noise pattern). This is expected, because the addition of a fixed number in the payoffs does not change the structure of the game, and, thus, the two games are considered ``equivalent'' in standard game theory.
The proposition below proves a more complex version of this statement, showing that the same is true for the noise pattern: adding a fixed amount of bias across the board does not modify the respective probabilities. Formally:

\begin{proposition}
\label{prop:effect-of-sum-on-G-M}
Consider a noisy game $mG \sim G^0 + \gndist{M,\Devi}$, and constant numbers $a_G, a_r, a_c \in \mathbb{R}$.
Set $\overline{G^0} = G^0 + \tbl{a_G}$ and $\overline{M} = (\overline{M^r};\overline{M^c})$, where $\overline{M^x} = M^x + \tbl{a_x}$ for $x \in \{r,c\}$. 
Moreover, set $\overline{mG} \sim \overline{G^0} + \gndist{\overline{M},\Devi}$.
Then:
\begin{itemize}
    \item $\prob{mG: \varepsilon\text{-misinformed}} = \prob{\overline{mG}: \varepsilon\text{-misinformed}}$
    \item $\prob{mG: \text{inverse-}\varepsilon\text{-misinformed}} =   \prob{\overline{mG}: \text{inverse-}\varepsilon\text{-misinformed}}$
\end{itemize}
\end{proposition}

\subsection{Effect of modifying noise intensity (\texorpdfstring{$\Devi$}{})}
\label{subsec:mod-D}

Although adding a fixed constant number to the game's payoffs does not modify the respective probabilities (Proposition \ref{prop:effect-of-sum-on-G-M}), this is not the case when changing the ``scale'' of a game (by multiplying all its payoffs by a constant number, say $\lambda > 0$).
In particular, changing the scale of a game will affect its ``resilience'' to noise, without changing the game's properties and behaviour, because it increases the ``amount of noise'' necessary to change the sign of the various $\ug yxi$.
As a matter of fact, multiplying the payoffs by a sufficiently large number would minimize the effect of the noise, as its effects on the payoffs would be, comparatively smaller (analogously, using a sufficiently small positive number would maximize the effect of the noise).

In Proposition \ref{prop:mg-change-scale} (and especially in Corollary \ref{cor:mg-change-scale}), we quantify this effect, by showing that we need to multiply the noise intensity (standard deviation) by $\lambda^2$ in order for the noise to have the same effect on a game scaled by $\lambda$. Formally:

\begin{proposition}
\label{prop:mg-change-scale}
Consider a normal noisy game $mG \sim G^0 + \gndist{M,\Devi}$ and some $\lambda > 0$. Set:
$\overline {G^0} = \lambda G^0$, $\overline M = \lambda M$ and $\overline \Devi = \lambda^2 \Devi$, and consider the normal noisy game $\overline {mG} = \overline {G^0} + \gndist{\overline M, \overline \Devi}$.
Then:
\begin{itemize}
    \item $\prob{mG: \varepsilon\text{-misinformed}} = \prob{\overline{mG}: \varepsilon\text{-misinformed}}$
    \item $\prob{mG: \text{inverse-}\varepsilon\text{-misinformed}} = \prob{\overline{mG}: \text{inverse-}\varepsilon\text{-misinformed}}$
\end{itemize}
\end{proposition}

An obvious and interesting corollary of Propositions \ref{prop:effect-of-sum-on-G-M} and \ref{prop:mg-change-scale} is the following:

\begin{corollary}
\label{cor:mg-change-scale}
Consider a normal noisy game $mG \sim G^0 + \gndist{\tbl 0,\Devi}$ and some $\lambda > 0$, $k \in \mathbb R$. Set:
$\overline {G^0} = \lambda G^0 + k$ and $\overline \Devi = \lambda^2 \Devi$, and consider the normal noisy game $\overline {mG} = \overline {G^0} + \gndist{\tbl 0, \overline \Devi}$.
Then:
\begin{itemize}
    \item $\prob{mG: \varepsilon\text{-misinformed}} = \prob{\overline{mG}: \varepsilon\text{-misinformed}}$
    \item $\prob{mG: \text{inverse-}\varepsilon\text{-misinformed}} =  \prob{\overline{mG}: \text{inverse-}\varepsilon\text{-misinformed}}$
\end{itemize}
\end{corollary}

From the previous theoretical results, the effect of noise in the outcome of an abstract $2 \times 2$ bimatrix game has the following characteristics: for small values of the noise intensity (standard deviation), players almost surely have the same behaviour as in the actual game, whereas for large noise intensity, the behaviour of players cannot be predicted as their games will be almost random. Also, observe that the formulas giving the probabilities for (inverse)-$\varepsilon$-misinformed are continuous with respect to the standard deviation. Given the above, one would expect that, by increasing the standard deviation, we would monotonically transit from the first extreme to the second. However, this does not always hold, as the following counter-example shows.

\addtocounter{example}{1}
\begin{example}
Consider as actual game the classical Prisoner's Dilemma (see Figure~\ref{fig:Payoff_matrix_PD}), which has a pure Nash equilibrium with strategy profile $((0,1),(0,1))$. We produce a noisy game, in which the noise only affects the upper left elements of the actual payoff matrix, where we add noise according to a random variable following the normal distribution $\gndist{0,\devi^2}$.
From Theorems~\ref{thm:main-emis},~\ref{thm:main-invemis}, we can compute the probabilities for this game to be (inverse-)$\varepsilon$-misinformed. The result is shown in Figure \ref{fig:counter_example}, where we plot $\prob{mG: \varepsilon \text{-misinformed}}$ (blue line) and $\prob{mG: \text{inverse-} \varepsilon \text{-misinformed}}$ (orange line), for $\devi \in (0,10)$.
As is obvious by this figure, these functions are not monotonic with respect to \devi.

\begin{figure}
    \centering
    \includegraphics[scale = 0.3]{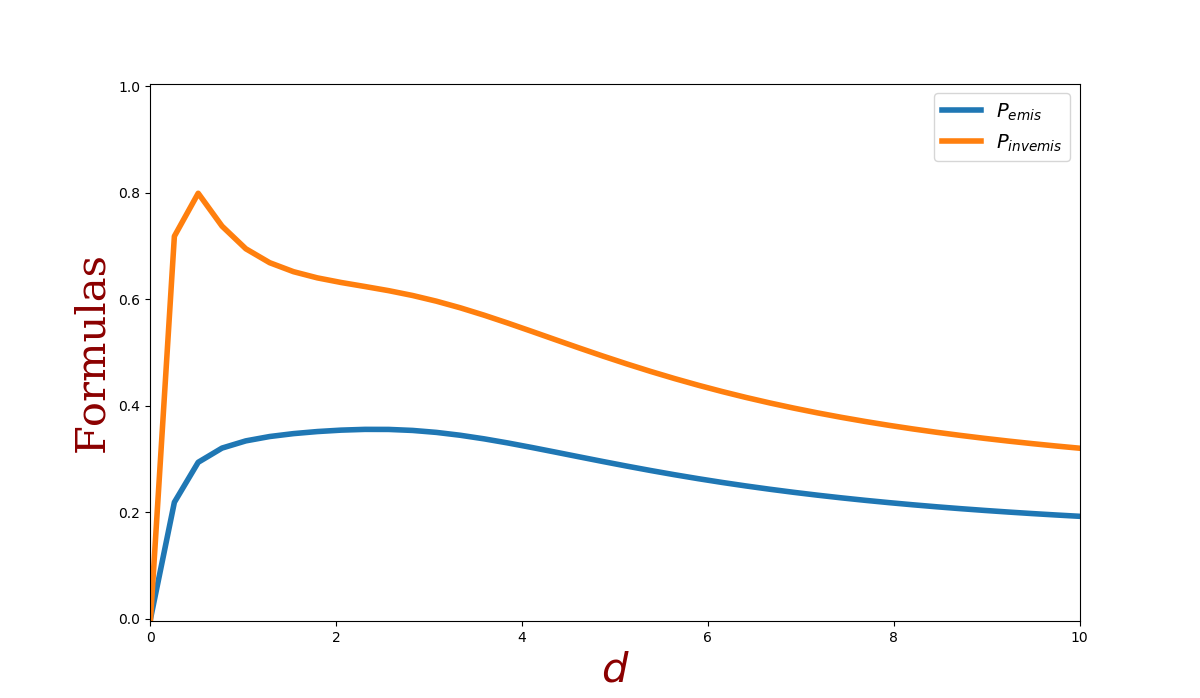}
    \caption{Example showing that $\prob{mG: \varepsilon \text{-misinformed}}$ (blue), $\prob{mG: \text{inverse-} \varepsilon \text{-misinformed}}$ (orange) (vertical axis) have non-monotonic dependence to noise (horizontal axis).}
    \label{fig:counter_example}
\end{figure}
\end{example}

\section{Discussion and experiments}\label{sec:discuss}

In this section we report on experiments that validate our basic results, and we investigate the effect of noise on the players' decisions, for the four $2 \times 2$ bimatrix games shown in Figure~\ref{fig:test cases}.
The games were chosen to capture the following cases: i) dominant equilibrium (Prisoner's Dilemma), ii) unique mixed Nash equilibrium (Matching Pennies), iii) multiple Nash equilibria (Battle of the Sexes), and iv) dominant equilibrium that coincides with the optimal outcome (Win-Win).

\begin{figure}
        \centering
        \begin{subfigure}[b]{0.25\textwidth}
            \centering
             \[\left(\begin{array}{cc}
    (2, 2) & (0, 3) \\
    (3, 0) & (1, 1)
    \end{array}\right)\]
            \caption[]%
            {{\footnotesize Prisoners' Dilemma}}    
            \label{fig:Payoff_matrix_PD}
        \end{subfigure}
        \medskip
        \begin{subfigure}[b]{0.24\textwidth}  
            \centering 
             \[\left(\begin{array}{cc}
    (1, -1) & (-1, 1) \\
    (-1, 1) & (1, -1)
    \end{array}\right)\]
            \caption[]%
            {{\footnotesize Matching Pennies}}    
            \label{fig:Payoff_matrix_MP}
        \end{subfigure}
        \medskip
        \begin{subfigure}[b]{0.24\textwidth}   
            \centering 
             \[\left(\begin{array}{cc}
    (2, 1) & (0, 0) \\
    (0, 0) & (1, 2)
    \end{array}\right)\]
            \caption[]%
            {{\footnotesize Battle of the Sexes}}    
            \label{fig:Payoff_matrix_BoS}
        \end{subfigure}
        \medskip
        \begin{subfigure}[b]{0.23\textwidth}   
            \centering 
             \[\left(\begin{array}{cc}
    (3, 2) & (4, 4) \\
    (1, 1) & (2, 3)
    \end{array}\right)\]
            \caption[]%
            {{\footnotesize Win-Win}}    
            \label{fig:Payoff_matrix_WW}
        \end{subfigure}
        \caption{Games to be used for the experimental evaluation.}
        \label{fig:test cases}
    \end{figure}

\subsection{Theoretical and Experimental Computation of the Probability that a Game is (Inverse-)$\varepsilon$-misinformed}

We consider that the actual game undergoes an additive noise that follows the normal distribution $\mathcal{N}(\tbl 0, \tbl{\devi^2})$ where $\devi \in \{ 0.001, 0.5, 1, \ldots, 10\}$.

\begin{figure*}
    \centering
    \begin{adjustbox}{minipage=\linewidth}
    \begin{subfigure}[b]{0.485\textwidth}
        \centering
        \includegraphics[width=\textwidth]{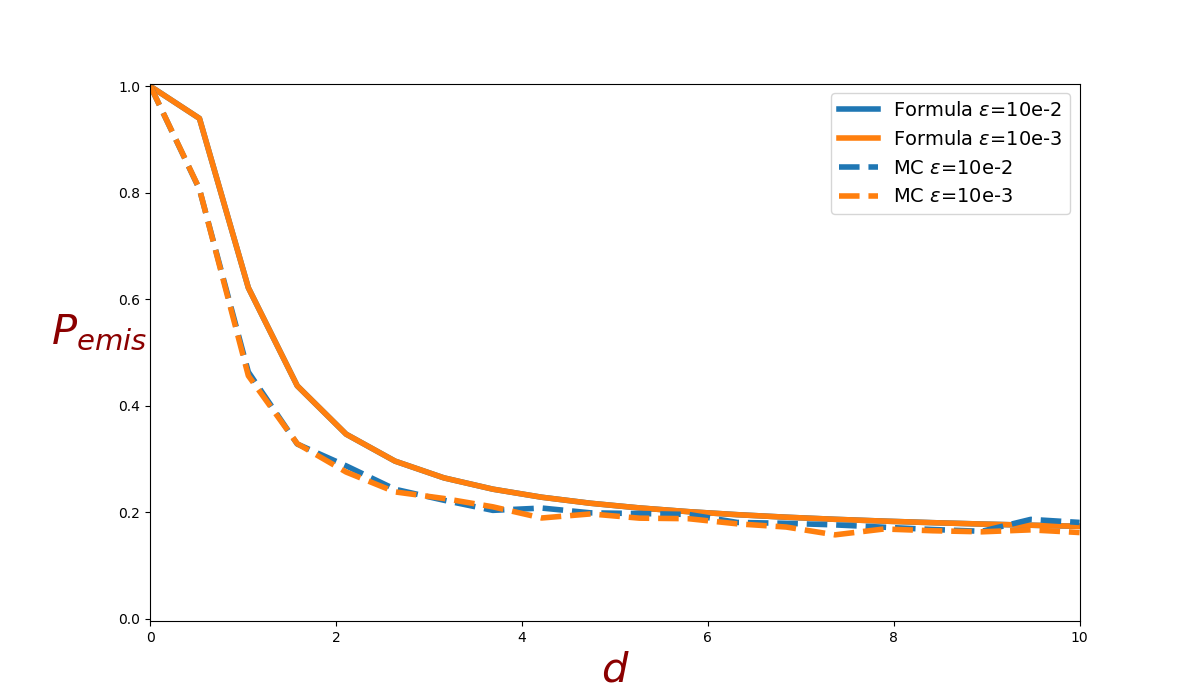}
        \caption[]%
        {{\small $\varepsilon$ misinformed.}}    
        \label{fig:e_mis_PD}
    \end{subfigure}
    \hfill
    \begin{subfigure}[b]{0.485\textwidth}  
        \centering 
        \includegraphics[width=\textwidth]{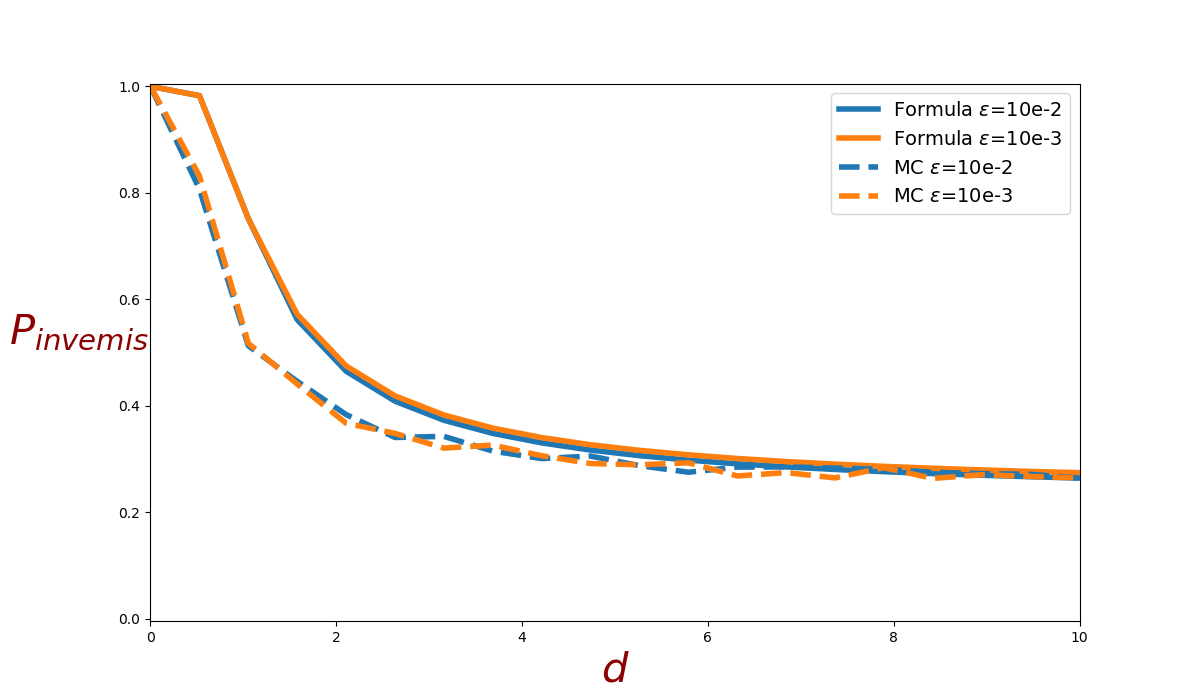}
        \caption[]%
        {{\small Inverse $\varepsilon$ misinformed.}}    
        \label{fig:inv_e_mis_PD}
    \end{subfigure}
    \caption{Monte Carlo (MC) simulation and probabilistic formulas for Prisoner's Dilemma and Win-Win games. Vertical axis: \eqref{fig:e_mis_PD} $\prob{mG: \varepsilon\text{-misinformed}}$, \eqref{fig:inv_e_mis_PD} $\prob{mG: \text{inverse-}\varepsilon\text{-misinformed}}$. Horizontal axis: noise intensity $d$.}
    \label{fig:Prisoner's Dilemma}
    \end{adjustbox}
\end{figure*}

\begin{figure*}
    \centering
    \begin{adjustbox}{minipage=\linewidth}
    \begin{subfigure}[b]{0.485\textwidth}
        \centering
        \includegraphics[width=\textwidth]{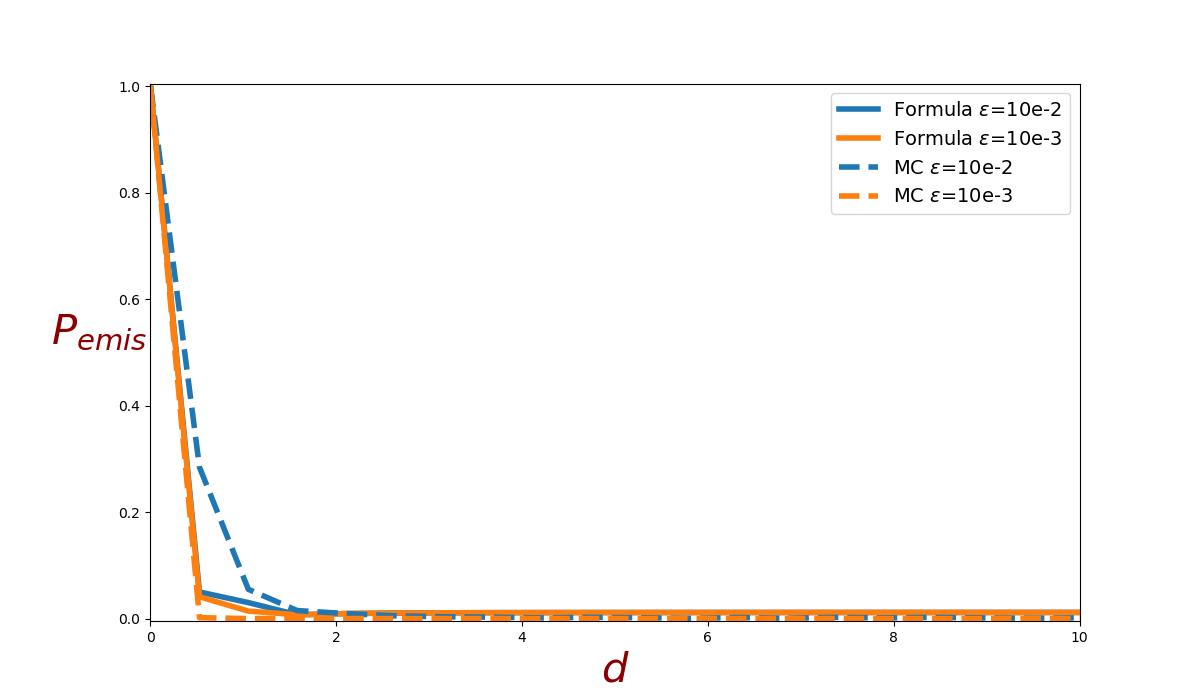}
        \caption[]%
        {{\small $\varepsilon$ misinformed.}}    
        \label{fig:e_mis_MP}
    \end{subfigure}
    \hfill
    \begin{subfigure}[b]{0.485\textwidth}  
        \centering 
        \includegraphics[width=\textwidth]{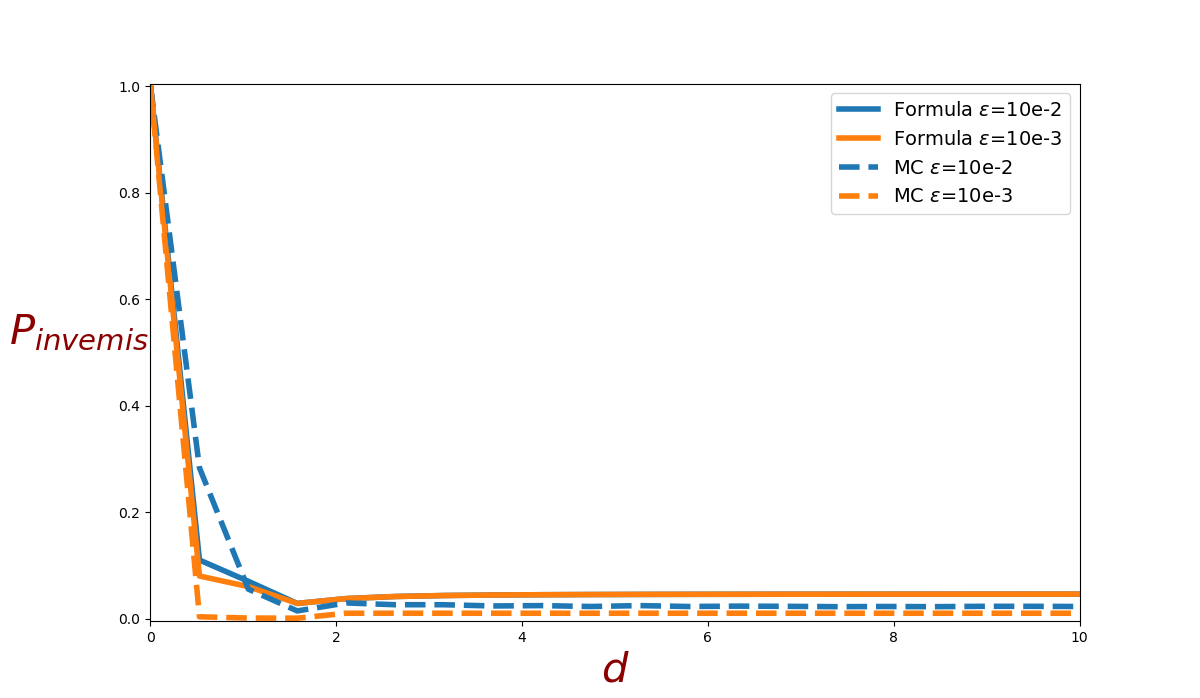}
        \caption[]%
        {{\small Inverse $\varepsilon$ misinformed.}}    
        \label{fig:inv_e_mis_MP}
    \end{subfigure}
    \caption{Monte Carlo (MC) simulation and probabilistic formulas for Matching Pennies. Vertical axis: \eqref{fig:e_mis_MP} $\prob{mG: \varepsilon\text{-misinformed}}$, \eqref{fig:inv_e_mis_MP} $\prob{mG: \text{inverse-}\varepsilon\text{-misinformed}}$. Horizontal axis: noise intensity $d$.} 
    \label{fig:Matching Pennies}
    \end{adjustbox}
\end{figure*}

\begin{figure*}
    \centering
    \begin{adjustbox}{minipage=\linewidth}
    \begin{subfigure}[b]{0.485\textwidth}
        \centering
        \includegraphics[width=\textwidth]{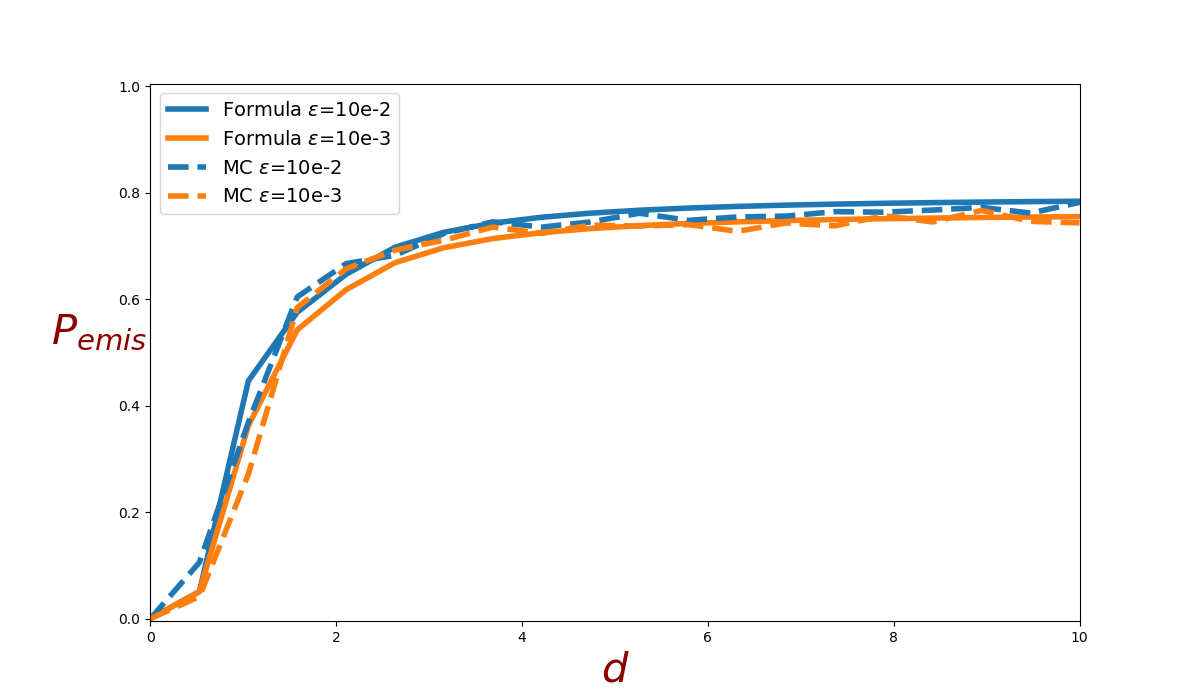}
        \caption[]%
        {{\small $\varepsilon$ misinformed.}}    
        \label{fig:e Mis BoS}
    \end{subfigure}
    \hfill
    \begin{subfigure}[b]{0.485\textwidth}  
        \centering 
        \includegraphics[width=\textwidth]{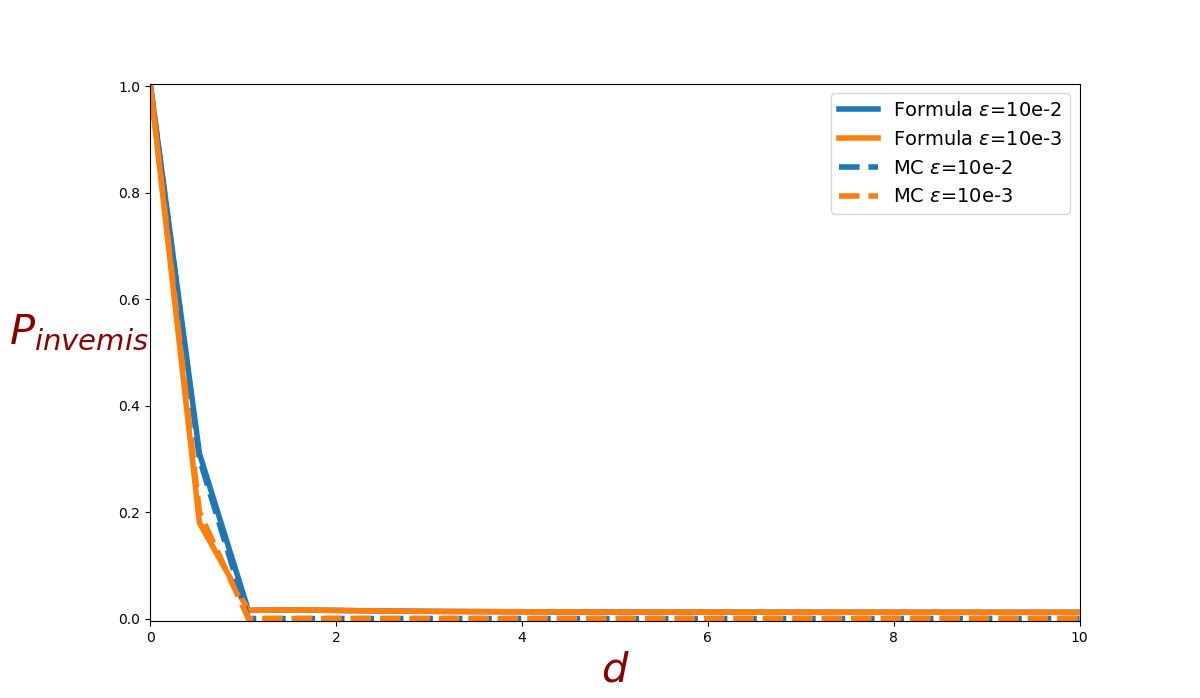}
        \caption[]%
        {{\small Inverse $\varepsilon$ misinformed.}}    
        \label{fig:inv e Mis BoS}
    \end{subfigure}
    \caption{Monte Carlo (MC) simulation and probabilistic formulas for Battle of the Sexes. Vertical axis: \eqref{fig:e Mis BoS} $\prob{mG: \varepsilon\text{-misinformed}}$, \eqref{fig:inv e Mis BoS} $\prob{mG: \text{inverse-}\varepsilon\text{-misinformed}}$. Horizontal axis: noise intensity $d$.}
    \label{fig:Battle of the Sexes}
    \end{adjustbox}
\end{figure*}

We compare the theoretical values of probabilities that we get from Theorems~\ref{thm:main-emis}, \ref{thm:main-invemis}, with the respective values calculated through Monte Carlo simulations.
The Monte Carlo simulations were conducted as follows: we generate a game $G^0$, which can be one of the four games shown in Figure \ref{fig:test cases}. Then, for each of the above values for $\devi$, we create the respective noisy game $mG=G^0+\gndist{\tbl 0, \tbl{\devi^2}}$. To be more precise, we generate a misinformation game, where the misinformation stems from the incorporation of additive noise stemming from one random experiment that follows the above distribution ($\gndist{\tbl 0, \tbl{\devi^2}}$).
We derive the natural misinformed equilibrium and check about $\varepsilon$-closeness. We perform $3,000$ repetitions of the above process and calculate:
\begin{enumerate} 
    \item [a)] the percentage of games that are $\varepsilon$-misinformed (i.e., all nmes of $mG$ are $\varepsilon$-close to one Nash equilibrium of $G^0$, according to the first bullet of Definition~\ref{def:epsilon-misinformed}),
    \item [b)] the percentage of games that are inverse $\varepsilon$-misinformed (i.e., all Nash equilibria of $G^0$ are $\varepsilon$-close to one nme of $mG$, according to
    the second bullet of Definition~\ref{def:epsilon-misinformed}).
\end{enumerate}
We repeat the simulations for two different values of $\varepsilon$ ($\varepsilon \in \{10^{-2}, 10^{-3}\}$). 
The results are shown in Figures~\ref{fig:Prisoner's Dilemma}-\ref{fig:Battle of the Sexes}.
As the Prisoner's Dilemma and the Win-Win games both have a unique pure Nash equilibrium, their behavioural consistency is similar. Hence, Figure~\ref{fig:Prisoner's Dilemma} shows both cases.
In all subplots, we have plots of two colours. The blue ones depict the computations for $\varepsilon = 10^{-2}$, whereas the red ones depict the computations for $\varepsilon = 10^{-3}$. 

In part (a) of the figures, the horizontal axis depicts the different values for the standard deviation $\devi$ of the noise, and the vertical axis depicts the probability of a game being $\varepsilon$-misinformed according to Theorem~\ref{thm:main-emis} (solid line) or the probability of a game being $\varepsilon$-misinformed according to the Monte Carlo simulations (dotted lines). 
The same hold for part (b) of the figures, but for the inverse-$\varepsilon$-misinformed case (Theorem~\ref{thm:main-invemis}).
In both subfigures, a high value of the probability calculated in the vertical axis implies a small effect of noise on players' decisions.
As expected, the theoretical results are very close to the experimental ones.

These figures give rise to various remarks concerning the influence of noise to players' strategic choices. Although some general patterns emerge, the effect of noise in the behavioural consistency of the game greatly depends on the type and number of Nash equilibria that it has, so we split our analysis in 3 different cases.

\CASE{Case 1: Unique Pure Nash equilibrium}

The case of a unique pure Nash equilibrium appears in the Prisoner's Dilemma and Win-Win games, whose behaviour is depicted in Figure~\ref{fig:Prisoner's Dilemma}.

For Prisoner's Dilemma, we observe that, for small values of the standard deviation ($\devi \ll 1$), the nme of $mG$ will usually be the same as the NE of the original game ($G^0$). Thus, both probabilities $\prob[emis]{mG; \varepsilon}$ and $\prob[invemis]{mG; \varepsilon}$ will have values close to $1$. As $\devi$ increases, noise will produce misinformation games $G^r, G^c$ with different Nash equilibria than that of the actual game (different means non-close, by definition, in this case) with an increasing probability, thereby reducing the probability for behavioural consistency.

As \devi\ increases further, each of the different possible sets of equilibria will appear with almost equal probability in $G^r, G^c$, leading to a convergence in the plots of Figure \ref{fig:Prisoner's Dilemma}.
In particular, $\prob[emis]{mG; \varepsilon}$ converges to approximately 14\%, whereas $\prob[invemis]{mG; \varepsilon}$ converges to approximately 25\%. 
This can be theoretically predicted by observing Table \ref{tab:game-theory-basics-depiction}. For a large enough noise, the original orderings among the elements of the payoff matrix become increasingly irrelevant, and the actual orderings in each of $G^r, G^c$ become totally random. As a result, the equilibrium strategy for $r$ in $G^r$ will be a pure one with a probability of 6/8 (3/8 for each strategy), a mixed one with 1/8 probability, and pure-and-mixed with 1/8 probability. The same is true of course for $c$ in $G^c$. Combining these observations with Table \ref{tab:prob-emis-invemis} and Theorems \ref{thm:main-emis}, \ref{thm:main-invemis}, we get the above numbers for the convergence of $\prob[emis]{mG; \varepsilon}$, $\prob[invemis]{mG; \varepsilon}$.

Similar remarks hold for the Win-Win game that has one pure Nash equilibrium strategy profile (namely, $((1, 0), (0, 1))$).

\CASE{Case 2: Unique Mixed Nash equilibrium}

The case of a unique mixed Nash equilibrium appears in the Matching Pennies game, which has one Nash equilibrium strategy profile $((1/2, 1/2), (1/2, 1/2))$, and whose behaviour is depicted in Figure~\ref{fig:Matching Pennies}.

As in case 1, we observe that for small values of the standard deviation ($\devi \ll 1$), $mG$ will have the same $nme$ as the NE in $G^0$. Thus, both probabilities $\prob[emis]{mG; \varepsilon}$ and $\prob[invemis]{mG; \varepsilon}$ will have values close to $1$. As $\devi$ increases, noise will produce games $G^r, G^c$ with different Nash equilibria than that of the actual game $G^0$, and the respective probabilities fall sharply (much faster compared to the Prisoner's Dilemma case), converging to a value close to $0$ for large values of the standard deviation. 
This is explained by the fact that, although a mixed nme is achieved in some of the produced games, this is often not close to the actual mixed one, leading to games that are (usually) not (inverse-)$\varepsilon$-misinformed. For example, for $\varepsilon = 10^{-2}$, the function $\prob[emis]{mG; \varepsilon}$ convergences at around $0.03\%$.

\CASE{Case 3: Multiple Nash equilibria}

The case of two pure and one mixed Nash equilibrium appears in the Battle of the Sexes game, whose behaviour is depicted in Figure~\ref{fig:Battle of the Sexes}.
The Nash equilibrium strategy profiles of Battle of the Sexes are: $\{ ((1,0), (1,0)), ((0,1), (0,1)), \\ ((2/3,1/3), (1/3,2/3)) \}$.

Unlike other games, we observe that the Battle of the Sexes has zero probability of being $\varepsilon$-misinformed for small values of \devi. This is explained by the fact that, for small values of \devi, $G^r, G^c$ will be very similar to $G^0$, each giving 3 equilibrium strategies (for the respective player). Thus, there are 9 nmes, one for each combination of equilibrium strategies (see Definition \ref{def:natural_misinformed_eq}), so some of them will not be $\varepsilon$-close to one of the three equilibria of $G^0$. 
By Definition \ref{def:epsilon-misinformed} this means that the respective game is not $\varepsilon$-misinformed, so $\prob[emis]{mG; \varepsilon}$ will be close to $0$.

As \devi\ increases, and the games $G^r, G^c$ become less and less predictable, the probability of being $\varepsilon$-misinformed becomes larger, reaching a plateau at around 72\%. The explanation here is analogous to the one given for the other two cases: in order for a misinformation game to \emph{not} be $\varepsilon$-misinformed, it should either have one pure equilibrium (but not one of the two that are in the equilibria of $G^0$), or it should have one mixed equilibrium (but not $\varepsilon$-close to the one of $G^0$). Based on the analysis of the Prisoner's Dilemma game, the probability of the former is around 28\%; based on the analysis of the Matching Pennies game, the probability of the latter is close to $0$; combining these observations, we conclude that a plateau at around 72\% is reasonable.

For the inverse-$\varepsilon$-misinformed case (part (b) of Figure~\ref{fig:Battle of the Sexes}), small values of \devi\ result to high values for $\prob[invemis]{mG; \varepsilon}$, as expected. As \devi\ increases, the probability decreases at a rate even faster than the one observed for Matching Pennies, eventually converging at a value close to $0$.
This is explained by the fact that, in order for the game to be inverse-$\varepsilon$-misinformed, it should have, among other things, also a mixed equilibrium that is close to the respective mixed of $G^0$. As we established in Case 2 above, this has a very low probability for large values of \devi.


\subsection{Optimal strategy profiles in terms of efficiency}

In this subsection, we report on experiments that investigate whether the misinformation game $mG$ that results from a given actual game $G^0$ has natural misinformed equilibria that are best or worst in terms of efficiency (social welfare). We then evaluate the effect of noise on each of the four games under consideration.

We performed Monte Carlo simulations as in the previous section and calculated:
\begin{enumerate} 
    \item [a)] the percentage $P_{best}$ of misinformation games that have a natural misinformed equilibrium that maximizes social welfare (best nme),
    \item [b)] the percentage $P_{worst}$ of misinformation games that have a natural misinformed equilibrium that minimizes social welfare (worst nme).
\end{enumerate}

\begin{figure}
    \centering
    \begin{adjustbox}{minipage=\linewidth}
    \begin{subfigure}[b]{0.485\textwidth}
        \centering
        \includegraphics[width=\textwidth,height=0.2\textheight]{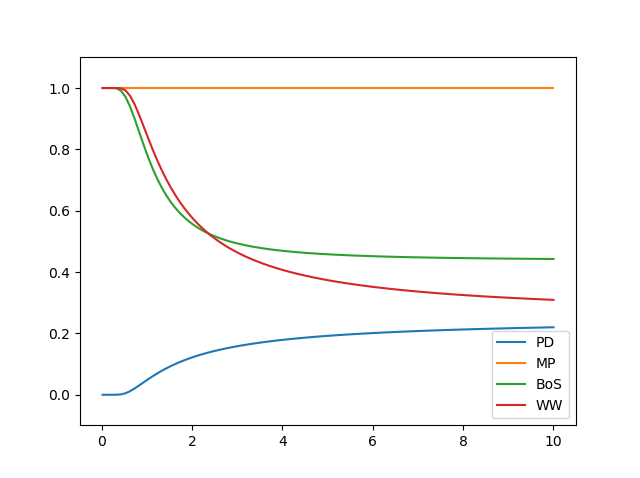}
    \caption{Percentage of misinformation games that result in the best nme.}
    \label{fig:best_sp}
    \end{subfigure}
    \hfill
    \begin{subfigure}[b]{0.485\textwidth}  
        \centering 
        \includegraphics[width=\textwidth,height=0.2\textheight]{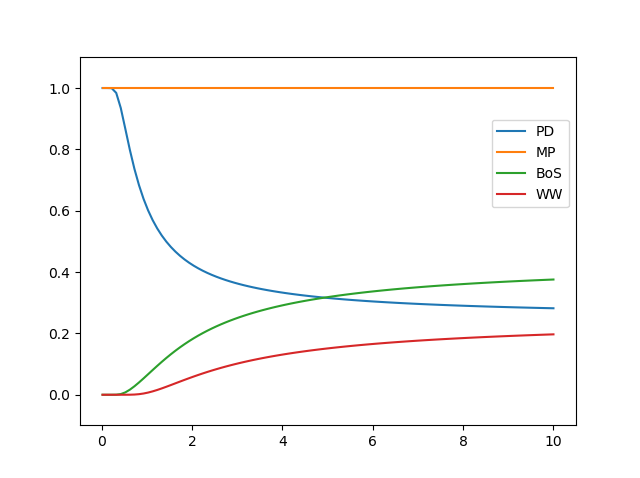}
    \caption{Percentage of misinformation games that result in the worst nme.}
    \label{fig:worst_sp}
    \end{subfigure}
    \caption{Strategy profiles in terms of efficiency.}
    \label{fig:efficiency}
    \end{adjustbox}
\end{figure}

We repeat the simulations for all values of $\devi$ in $\{0.02, 0.04, \ldots, 10\}$ and for $\varepsilon = 10^{-2}$. 
The results are shown in Figures~\ref{fig:best_sp} and \ref{fig:worst_sp}.

In Matching Pennies, as it is a constant-sum game, all strategy profiles provide the same level of social welfare, so the respective line is flat, regardless of the value of $\devi$ (see Figures~\ref{fig:best_sp} and \ref{fig:worst_sp}). 
In other words, the noise has no effect with respect to the optimal outcome. 

In Prisoners' Dilemma, the best strategy profile is $((1, 0), (1, 0))$ and the worst one is $((0, 1), (0, 1))$ which coincides with the pure NE of the actual game $G^0$. We observe that, for small values of $\devi$, only a few repetitions provide the best nme (Figure~\ref{fig:best_sp}), while most of them provide the worst nme (Figure~\ref{fig:worst_sp}); this is in line with the results given in the previous subsection.
As $\devi$ increases, the percentage of games resulting in the best strategy increases too, implying that noise has a positive effect on Prisoners' Dilemma.

In the Battle of the Sexes, the best strategy profiles are $((1, 0), (1, 0))$ and $((0, 1), (0, 1))$ (these are also the pure Nash equilibria of the actual game),
and the worst strategy profiles are  $((1, 0), (0, 1))$ and $((0, 1), (1, 0))$. We observe that, for small values of $\devi$, most of the misinformation games result in one of the best strategy profiles (Figure~\ref{fig:best_sp}). As $\devi$ increases, this percentage decreases, implying that noise has a negative effect on the Battle of the Sexes: players are not forced to choose better strategies. 

In the Win-Win game, the best strategy profile is $((1, 0), (0, 1))$ and the worst one is $((0, 1), (1, 0))$. The same observations as in the Battle of the Sexes hold for the Win-Win game.

To summarize, as the percentage $P_{best}$ increases (or $P_{worst}$ decreases) with respect to $\devi$, noise is beneficial. This is the case for Prisoners' Dilemma. 
On the contrary, noise deteriorates the efficiency of the system if the percentage $P_{best}$ decreases (or $P_{worst}$ increases) with respect to $\devi$ as in Win-Win and Battle of the Sexes games.  Finally, the efficiency of the system is independent of the noise in the Matching Pennies game.

Given the above, as expected, noise deteriorates the social welfare in games where the original Nash equilibrium is already ``good'' for the social welfare (Battle of the Sexes, Win-Win), as it induces a more ``random'' behaviour. On the contrary, it improves the situation in games where the original equilibrium is ``bad'' (e.g., Prisoner's Dilemma). In constant sum games (e.g., Matching Pennies), noise has no effect with regards to the social welfare.


\subsection{PoM vs PoA}

In this subsection, we compare the price of anarchy $PoA$ with the price of misinformation $PoM$ for the four games of interest. Both metrics measure social welfare, with or without misinformation respectively, and take values that are higher than or equal to $1$. 

\begin{figure}
    \centering
    \begin{adjustbox}{minipage=\linewidth}
    \begin{subfigure}[b]{0.485\textwidth}
        \centering
        \includegraphics[width=\textwidth,height=0.3\textheight]{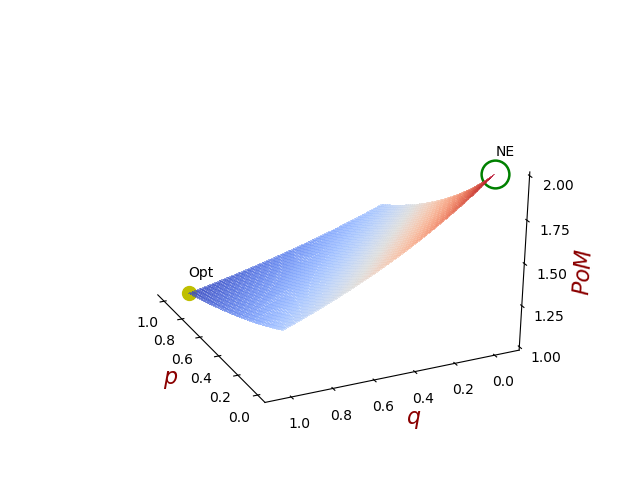}
    \caption{\footnotesize PoM plane for the Prisoner's Dilemma.}
    \label{fig:PoM_PD} 
    \end{subfigure}
    \hfill
    \begin{subfigure}[b]{0.485\textwidth}  
        \centering 
        \includegraphics[width=\textwidth,height=0.3\textheight]{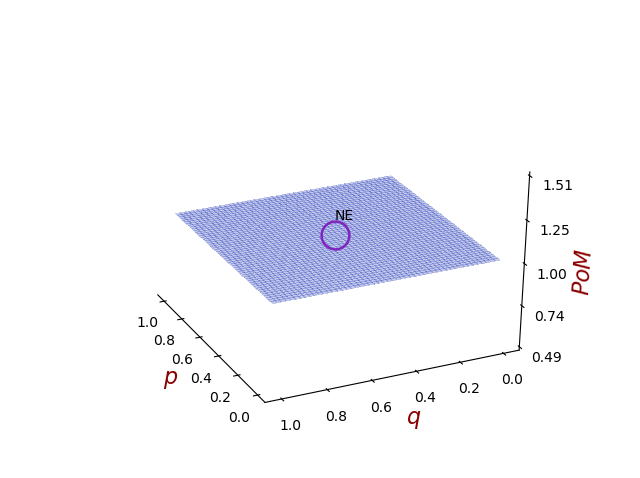}
    \caption{\small PoM plane for the Matching Pennies.}
    \label{fig:PoM_MP}
    \end{subfigure}
    \vskip\baselineskip
        \begin{subfigure}[b]{0.485\textwidth}
        \centering
        \includegraphics[width=\textwidth,height=0.3\textheight]{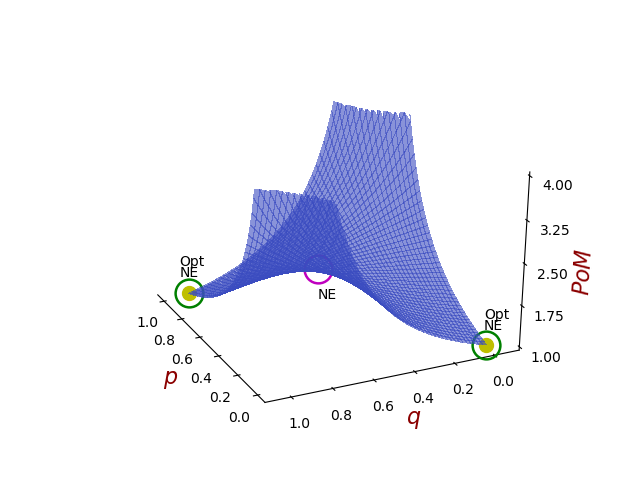}
    \caption{\footnotesize PoM plane for the Battle of the Sexes.}
    \label{fig:PoM_BoS} 
    \end{subfigure}
    \hfill
    \begin{subfigure}[b]{0.485\textwidth}  
        \centering 
        \includegraphics[width=\textwidth,height=0.3\textheight]{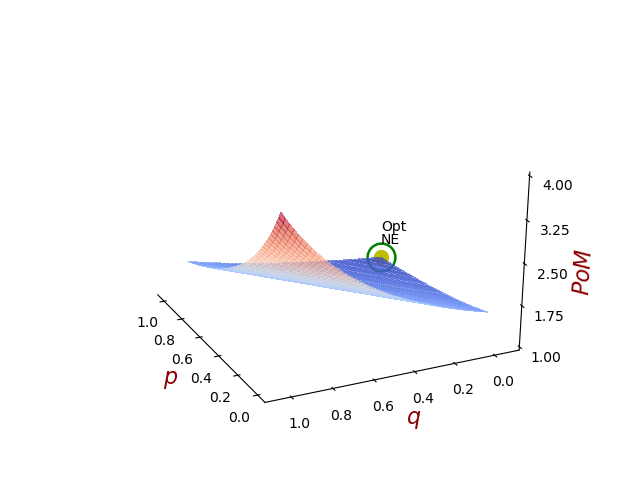}
    \caption{\footnotesize PoM plane for the Win-Win.}
    \label{fig:PoM_WW}
    \end{subfigure}
    \caption{\footnotesize PoM plane for games in Table~\ref{fig:test cases}.}
    \label{fig:PoMplane}
    \end{adjustbox}
\end{figure}

Given a bimatrix game $G$ with payoff matrix $P = (P_r; P_c)$ we use Definition~\ref{def:PoM} to compute $PoM$ for all values of pairs $(p,q)$, where $p,q \in [0,1]$. The values of $p,q$ are non other than the values in the joint strategy profile $\sigma = (\vec {\mathbf p}, \vec {\mathbf q})$ $= ((p,1-p),(q, 1-q))$. In formula~\ref{eq:PoM_max}, the quantities in the fraction are given by the formula $SW(\sigma) = {\bf p}^T(P_r + P_c){\bf q}$. 
The respective graphs are shown in Figures~\ref{fig:PoM_PD}-\ref{fig:PoM_WW}.

We can make the following observations on social welfare planes of Figures~\ref{fig:PoM_PD}-\ref{fig:PoM_WW} that present the range of values of $PoM$:
\begin{enumerate}
    \item In Prisoner's Dilemma we note that the social welfare plane is monotonic (see Figure~\ref{fig:PoM_PD}). The minimum value is in the bottom left corner (``bluest'') and the maximum value is in the upper right corner (``redest'').
    We know that the $PoA$ in this game is $2$, which is equal to the minimum social welfare, so any distortion in the payoff matrices of the game does not deteriorate the efficiency of the game, and $PoM \leq PoA$, for every level of noise.
    \item In Matching Pennies we observe that the social welfare plane is constant (Figure~\ref{fig:PoM_MP}). That is, $PoM$ remains constant as any combination of the values of the payoff matrix results in the same social welfare value. Thus, noise may affect the strategic behaviour of players, but keeps the social welfare constant. Note that, in zero-sum games such as Matching Pennies, the value of $PoM$ and $PoA$ cannot be calculated (the denominator of the respective formulas takes the value of zero). To mitigate this inconvenience we add proper values to each element of the payoff matrices and produce a constant-sum game, without affecting the strategic behaviour of players.
    \item In Battle of the Sexes we observe that the two pure Nash equilibria of the game are the optimal strategic behaviours (Figure~\ref{fig:PoM_BoS}). Thus, $PoA$ depends on the mixed Nash equilibrium, and noise could improve or degrade the efficiency of the system.
    \item In Win-Win, the unique Nash equilibrium coincides with the optimal one, thus $PoA = 1$ (Figure~\ref{fig:PoM_WW}). Therefore, any misinformation cannot improve the outcome of this game, and $PoM \geq PoA$.
\end{enumerate}

\section{Conclusion and Future Work}
\label{sec:concl}

In this paper we studied a novel game-theoretic setting, where players receive the information regarding the game's payoffs with a distortion that affects the elements of the payoff matrix. This distortion was assumed to be due to additive noise that follows a normal distribution, and could be due to communication errors that may appear when the game's parameters are communicated through a noisy channel, or when some malfunction in the sender or receiver distorts this information.
In such noisy settings, it is possible that each player knows a different game compared to her opponent and compared to the actual (originally communicated) one. 

We model this situation using misinformation games, an appropriate theoretical setting introduced previously in \cite{VFBM}, and define a subclass of misinformation games called \emph{noisy games} (see Section~\ref{sec:noisy_definitions}). 
The main problem considered in this setting is the computation of the probability for behavioural consistency, i.e., the probability that the agents' behaviour will be ``close'' (under some formal definition of closeness) to the one expected according to the original game, despite the noise. 
Towards this, two alternative formal definitions of behavioural consistency are given (Subsection \ref{subsec:distance-closeness}) and the respective probabilities are computed in Section~\ref{sec:noisy_probabilities}. 
Note that, due to the complexity of the formulas, we restricted ourselves to 2-player bimatrix games with 2 strategies per player.

We elaborate on those formulas and prove a number of related results (Section \ref{sec:results_noisy}), which help understand their properties. Such properties include the effect of the definition of closeness and/or the noise structure in the respective probabilities for behavioural consistency, as well as a study of how different interventions and modifications on the original game would affect these probabilities.

Moreover, we perform several numerical experiments using four well-known bimatrix games as benchmarks (see Figure~\ref{fig:test cases}). Initially, we compare the probabilistic formulas with Monte Carlo simulation to verify their correctness. Then, we derive general remarks as to the efficiency of the system regarding the additive noise, in terms of social welfare. To do so, we use the \emph{Price of Misinformation} metric, which is inspired by the well-known Price of Anarchy metric and quantifies how benevolent/malevolent is the misinformation caused by the noise with regards to game performance (related to social welfare).

Undeniably, the 2 players' bimatrix games with 2 strategies per player is a very restricting setting. Unsurprisingly however, even in this simple setting our analysis highlighted the richness, intricacy and interdependence of the probabilistic events, mathematical objects and techniques that were involved, leading to complex mathematical computations and stiff formulations as regards the end results. 
Having said that, we plan to consider more complex settings in the future, i.e., scenarios with more than two players and/or scenarios where each player may have more than two strategies. Further, we could consider deriving analogous probabilistic formulas for other classes of noise distributions (e.g., poisson or laplacian).

Moreover, an immediate future step is to provide tools to quantify the \emph{sensitivity} of a game to random noise, i.e., determine ``how much noise'' the game can withstand so that the behaviour of the players remains close (under the sense of behavioural consistency) to the expected ones, with a certain probability. A related research question is how sensitivity is affected by inconsequential changes in the game specification (e.g., change of scale); in this direction, results like Proposition \ref{prop:mg-change-scale} can help.
This analysis could be used as a tool for game designers to improve their designs and make them more robust to unexpected circumstances and noise in the communication channels.

\newpage

\appendix


\section{Proofs for the Results Appearing in the Paper}
\label{appendix:proofs}

\subsection{Normal Form Games}
\label{appendix:normal form games}

\begin{proofOf}{\Cref{prop:game-theory-degenerate}} 
Let's consider $G= \tuple{N,S,P}$, for $P=(P_c; P_c)$. 
Suppose that $G$ is degenerate. By definition, there is a pure strategy (say $s_i$, by player $x \in \{r,c\}$) that has two pure best responses. Suppose that $x = r, i = 1$. Then, since $s_1,s_2$ are equally preferred by $c$, it follows that $P_c[1,1] = P_c[1,2]$, i.e., $\ug{c}{1} = 0$. The other cases (i.e., when $x = c$ and/or $i = 2$) are analogous.\\ 
For the opposite, suppose that $\ug{r}{1} = 0$. Then $P_c[1,1] = P_c[1,2]$, so $c$ has two pure best responses for the strategy $s_1$ of $r$, which means that $G$ is degenerate. The proof is analogous for the other cases.
\end{proofOf}

\vspace{4pt}

\begin{proofOf}{\Cref{prop:game-theory-basics-mixed-value}}
Suppose that $x=r$.
From classical game theoretic results (e.g., see \cite{Algorithmic_Game_Theory_book}, \cite{Osborne_Rubinstein_book}), and our assumptions, we get that $p$ will satisfy the following equation:
\begin{equation*}
    p \cdot P_c[1,1] + (1-p) \cdot P_c[2,1] = 
p \cdot P_c[1,2] + (1-p) \cdot P_c[2,2]
\end{equation*}
The result now follows trivially by solving this equation and applying the definition of \ug{c}{i}.\\
Analogously, for the case where $x = c$, we get the following equation:
\begin{equation*}
    p \cdot P_r[1,1] + (1-p) \cdot P_r[1,2] = 
p \cdot P_r[2,1] + (1-p) \cdot P_r[2,2]
\end{equation*}
Solving it, as above, will give the required result.
\end{proofOf} 

\vspace{4pt}

\begin{proofOf}{\Cref{prop:game-theory-basics}}
By Proposition \ref{prop:game-theory-degenerate}, we conclude that $\ug{x}{i} \neq 0$ for all $x \in \{r,c\}, i \in \{1,2\}$. This means that the different (mutually exclusive) cases of the formulation of the proposition cover all possible cases for a non-degenerate game (see also Table \ref{tab:game-theory-basics-depiction}). Thus, it suffices to show the ``only if'' part for each different case.\\
For (1a), note that player $x$ will play $(1,0)$ (i.e., $s_1$) regardless of the choice of $\bar x$, so $NE_x(G) = \{(1,0)\}$ and \OP{}{x}{1} is true.\\
For (1b), note that the only Nash equilibrium of $G$ is $((1,0),(1,0))$, which proves the result.\\
Next, (1c) is analogous to (1b).\\
The cases (2a), (2b), (2c) are analogous to (1a), (1b), (1c) respectively.\\
With regards to (3a), it can be easily shown that the game can have no pure Nash equilibrium. Thus, it must have a mixed one (by the result of Nash \cite{Nash51}). Moreover, it cannot have more than one mixed, as this would render it degenerate\footnote{Immediate consequence of Corollary 3.7~\cite{Algorithmic_Game_Theory_book}.} (see \cite{Algorithmic_Game_Theory_book},\cite{Avis2010EnumerationON},\cite{Osborne_Rubinstein_book}). 

Thus, $NE_x(G) = \{(p,1-p)\}$, for some $0 < p < 1$. 
By Proposition \ref{prop:game-theory-basics-mixed-value}, it follows that $p = \frac{\ug{\bar x}{2}}{\ug{\bar x}{2} - \ug{\bar x}{1}}$, which shows the result.\\
The case (3b) is analogous.\\
For (4a), we observe that the values of \ug{x}{i} imply that the game has exactly two pure Nash equilibria, namely: $((1,0),(1,0))$ and $((0,1),(0,1))$. By \cite{Algorithmic_Game_Theory_book}, \cite{Osborne_Rubinstein_book}, it must also have one (unique) mixed equilibrium, as we examine a non-degenerate case.\\ 
Thus, $NE_x(G) = \{(1,0), (0,1), (p,1-p)\}$ for some $0 < p < 1$.
Again, using Proposition \ref{prop:game-theory-basics-mixed-value}, it follows that $p = \frac{\ug{\bar x}{2}}{\ug{\bar x}{2} - \ug{\bar x}{1}}$, which shows the result.\\
For (4b) the proof is analogous, except that here the pure Nash equilibria of $G$ are: $((1,0),(0,1))$ and $((0,1),(1,0))$.
\end{proofOf}

\subsection{Misinformation Games and Noisy games}
\label{appendix:misinformation and noisy games}

\begin{proofOf}{\Cref{prop:e-mis-no-prob}}
By definition, $mG$ is $\varepsilon$-misinformed if and only if for all $\sigma^* = (\sigma_r^*, \sigma_c^*) \in NME(mG)$ there exists $\sigma^0 = (\sigma_r^0,\sigma_c^0) \in NE(G^0)$ such that $\sigma^*, \sigma^0$ are $\varepsilon$-close. More formally:

\begin{equation*}
\begin{split}
    & mG:\varepsilon \text{ misinformed } \\
    & \hspace{.1cm}\Leftrightarrow \forall \sigma^* = (\sigma_r^*, \sigma_c^*) \in NME(mG) \;
    \exists \sigma^0 = (\sigma_r^0,\sigma_c^0) \in NE(G^0): \sigma^* \in \close{\sigma^0} \\
    & \hspace{.1cm}\Leftrightarrow \forall \sigma^* = (\sigma_r^*, \sigma_c^*) \in NME(mG), 
    \sigma^* \in \close{NE(G^0)} \\
     & \hspace{.1cm} \Leftrightarrow \forall \sigma^* = (\sigma_r^*, \sigma_c^*) \in NME(mG) \left(
    \sigma_r^* \in \close{NE_r(G^0)} \wedge \sigma_c^* \in \close{NE_c(G^0)} \right)  \\
    & \hspace{.1cm} \Leftrightarrow \forall \sigma_r^* \in NE_r(G^r), \sigma_r^* \in \close{NE_r(G^0)} \wedge  \forall \sigma_c^* \in NE_c(G^c), \sigma_c^* \in \close{NE_c(G^0)} \\
    & \hspace{.1cm} \Leftrightarrow \forall x \in \{r,c\} \enspace \forall \sigma_x^* \in NE_x(G^x), \sigma_x^* \in \close{NE_x(G^0)} \\
\end{split}
\end{equation*}
Now let us fix some $x$ and consider the different cases with regards to $NE_x(G^0)$:
\begin{itemize}
    \item If $NE_x(G^0)$ contains a single pure strategy, i.e., \OP{G^0}{x}{i} is true for some $i \in \{1,2\}$, then the expression $\forall \sigma_x^* \in NE_x(G^x), \sigma_x^* \in \close{NE_x(G^0)}$ is true if and only if $NE_x(G^x)$ contains the same pure strategy, and no other, i.e., if and only if    $\OP{G^x}{x}{i}$ is true.

    \item If $NE_x(G^0)$ contains a single mixed strategy, i.e., \OM{G^0}{x}{p^0} is true for some $0 < p^0 < 1$, then the expression $\forall \sigma_x^* \in NE_x(G^x), \sigma_x^* \in \close{NE_x(G^0)}$ is true if and only if $NE_x(G^x)$ contains a single mixed strategy that is $\varepsilon$-close to $(p^0,1-p^0)$, i.e., $\ROM{G^x}{x}{\omega_1}{\omega_2}$ is true, where 
    $\omega_1 = \max\{0, p^0 - \varepsilon\}$,
    $\omega_2 = \min\{1, p^0 + \varepsilon\}$.
    Note that the $\max, \min$ are necessary to cater for the case where $p^0 - \varepsilon$, $p^0 + \varepsilon$ are smaller than $0$ or greater than $1$, respectively.
    
    \item If $NE_x(G^0)$ contains two pure and one mixed strategies, i.e., \PM{G^0}{x}{p^0} is true for some $0 < p^0 < 1$, then the expression $\forall \sigma_x^* \in NE_x(G^x), \sigma_x^* \in \close{NE_x(G^0)}$ is true if and only if $NE_x(G^x)$ contains either a pure or a mixed strategy that is $\varepsilon$-close to $(p^0,1-p^0)$. This is expressed by the expression in bullet \#3 of the proposition.
    
    \item If $NE_x(G^0) = \Sigma_x$, i.e., \IN{G^0}{x} is true, then, no matter the contents of $NE_x(G^x)$, the expression $\forall \sigma_x^* \in NE_x(G^x), \sigma_x^* \in \close{NE_x(G^0)}$ is true.
\end{itemize}
This, combined with the fact that these are the only cases with regards to the value of $NE_x(G^0)$, conclude the proof.
\end{proofOf} 

\vspace{4pt}

\begin{proofOf}{ \Cref{prop:inv-e-mis-no-prob}}
By definition, $mG$ is inverse-$\varepsilon$-misinformed if and only if for all $\sigma^0 = (\sigma_r^0, \sigma_c^*) \in NE(G^0)$ there exists $\sigma^* = (\sigma_r^*,\sigma_c^*) \in NME(mG)$ such that $\sigma^*, \sigma^0$ are $\varepsilon$-close. More formally:

\begin{equation*}
\begin{split}
    & mG:\text{ inverse-} \varepsilon \text{-misinformed } \\
    & \hspace{.1cm} \Leftrightarrow \forall \sigma^0 = (\sigma_r^0, \sigma_c^0) \in NE(G^0) \;
    \exists \sigma^* = (\sigma_r^*, \sigma_c^*) \in NME(mG): \sigma^* \in \close{\sigma^0} \\
    & \hspace{.1cm} \Leftrightarrow \left( \forall \sigma_r^0 \in NE_r(G^0) \;
    \exists \sigma_r^* \in NE_r(G^r): \sigma_r^* \in \close{NE_r(\sigma_r^0)} \right)\\
    & \hspace{1cm} \wedge \left( \forall \sigma_c^0 \in NE_c(G^0) \;  \exists \sigma_c^* \in NE_c(G^c): 
    \sigma_c^* \in \close{NE_c(\sigma_c^0)} \right) \\
    & \hspace{.1cm} \Leftrightarrow \forall x \in \{r,c\} \forall \sigma_x^0 \in NE_x(G^0) \;
    \exists \sigma_x^* \in NE_x(G^x): \sigma_x^* \in \close{NE_x(\sigma_x^0)} \\
\end{split}
\end{equation*}
Now let us fix some $x$ and consider the different cases with regards to $NE_x(G^0)$:
\begin{itemize}
    \item If $NE_x(G^0)$ contains a single pure strategy, i.e., \OP{G^0}{x}{i} is true for some $i \in \{1,2\}$, then the expression $\forall \sigma_x^0 \in NE_x(G^0) \;
    \exists \sigma_x^* \in NE_x(G^x): 
    \sigma_x^* \in \close{NE_x(\sigma_x^0)}$ is true if and only if $NE_x(G^x)$ contains the same pure strategy, possibly in addition to others, i.e., (given that $G^x$ is non-degenerate) if and only if $\OP{G^x}{x}{i} \bigvee \RPM{G^x}{x}{0}{1}$ is true.

    \item If $NE_x(G^0)$ contains a single mixed strategy, i.e., \OM{G^0}{x}{p^0} is true for some $0 < p^0 < 1$, then the expression $\forall \sigma_x^0 \in NE_x(G^0) \;
    \exists \sigma_x^* \in NE_x(G^x): 
    \sigma_x^* \in \close{NE_x(\sigma_x^0)}$ is true if and only if $NE_x(G^x)$ contains a mixed strategy that is $\varepsilon$-close to $(p^0,1-p^0)$, possibly in addition to others, i.e., (given that $G^x$ is non-degenerate) $\ROM{G^x}{x}{\omega_1}{\omega_2}$ is true, where 
    $\omega_1 = \max\{0, p^0 - \varepsilon\}$,
    $\omega_2 = \min\{1, p^0 + \varepsilon\}$.
    Note that the $\max, \min$ are necessary to cater for the case where $p^0 - \varepsilon$, $p^0 + \varepsilon$ are smaller than $0$ or greater than $1$, respectively.
    
    \item If $NE_x(G^0)$ contains two pure and one mixed strategies, i.e., \PM{G^0}{x}{p^0} is true for some $0 < p^0 < 1$, then the expression $\forall \sigma_x^0 \in NE_x(G^0) \;
    \exists \sigma_x^* \in NE_x(G^x): 
    \sigma_x^* \in \close{NE_x(\sigma_x^0)}$ is true if and only if $NE_x(G^x)$ contains two pure and a mixed strategy that is $\varepsilon$-close to $(p^0,1-p^0)$, i.e., (given that $G^x$ is non-degenerate) $\RPM{G^x}{x}{\omega_1}{\omega_2}$ is true, where 
    $\omega_1 = \max\{0, p^0 - \varepsilon\}$,
    $\omega_2 = \min\{1, p^0 + \varepsilon\}$.

    \item If $NE_x(G^0) = \Sigma_x$, i.e., \IN{G^0}{x} is true, then, $\forall \sigma_x^0 \in NE_x(G^0) \;
    \exists \sigma_x^* \in NE_x(G^x): 
    \sigma_x^* \in \close{NE_x(\sigma_x^0)}$ is true if and only if at least one of the strategies in $NE_x(G^x)$ is $\varepsilon$-close to each strategy in $NE_x(G^0)$. Given that $NE_x(G^x)$ is finite (because $G^x$ is non-degenerate), this can only hold if \PM{G^x}{x}{p^x} for some $p^x$ such that $(p,1-p) \in \close{(p, 1-p)}$ for all $0 < p < 1$. From the latter, we conclude that $\varepsilon \geq 0.5$ and $\max\{0, p^0 - \varepsilon\} < p^x < \min\{1, p^0 + \varepsilon\}$, which leads to the requirement in bullet \#4 of the proposition.
\end{itemize}
This, combined with the fact that these are the only cases with regards to the value of $NE_x(G^0)$, conclude the proof.
\end{proofOf}

\subsection{Probabilities}
\label{appendix:probabilities}

\begin{proofOf}{ \Cref{prop:non-degenerate}}
The result is direct from Proposition \ref{prop:game-theory-degenerate} and the fact that $\prob{\ugr{x}{y}{i} = 0} = 0$ for any $x,y \in \{r,c\}$, $i \in \{1,2\}$.
\end{proofOf}

\vspace{4pt}

\begin{proofOf}{ \Cref{lem:division}}
For the first result, we observe that, since $\Omega_1 < \Omega_2 \leq 0$:
\begin{equation*}
    \begin{split}
        \Omega_1 \leq \frac{X}{Y} \leq \Omega_2 \bigwedge X < 0 \bigwedge Y > 0 \Leftrightarrow  \Omega_1 \leq \frac{X}{Y} \leq \Omega_2 \bigwedge Y > 0
    \end{split}
\end{equation*}

Thus, it suffices to compute the probability of the latter (simpler) event.\\
Now, set $Z = \frac{X}{Y}$. Then $f_{Z \vert Y}(z \vert y) = f_X(z y)$, so:
\begin{equation*}
    f_{ZY}(z,y) = f_{Z \vert Y}(z \vert y) \cdot f_Y(y) = f_X(z y) \cdot f_Y(y)
\end{equation*}
Therefore:

\begin{equation*}
\begin{split}
& \prob{\Omega_1 \leq \tfrac{X}{Y} \leq \Omega_2, X < 0, Y > 0}  = \prob{\Omega_1 \leq Z \leq \Omega_2, Y > 0}  \\
& \hspace{.5cm} = \int_{0}^{+\infty} \int_{\Omega_1}^{\Omega_2} \! f_{ZY}(z,y) \, \mathrm{d}z \, \mathrm{d}y  = \int_{0}^{+\infty} \int_{\Omega_1}^{\Omega_2} \! f_X(zy) \cdot f_Y(y) \, \mathrm{d}z \, \mathrm{d}y  \\
& \hspace{.5cm} = \int_{0}^{+\infty} \left(\int_{\Omega_1}^{\Omega_2} \! f_X(zy) \, \mathrm{d}z \right) f_{Y}(y) \, \mathrm{d}y  = \int_{0}^{+\infty} \left(\int_{\Omega_1 y}^{\Omega_2 y} \! \frac{1}{y} f_X(x) \, \mathrm{d}x \right) f_{Y}(y) \, \mathrm{d}y  \\
& \hspace{.5cm} = \int_{0}^{+\infty} \left(\int_{\Omega_1 y}^{\Omega_2 y} \! f_X(x) \, \mathrm{d}x \right) \frac{f_{Y}(y)}{y} \, \mathrm{d}y \\
\end{split}
\end{equation*}
The proof of the second result is completely analogous.
\end{proofOf}

\vspace{4pt}

\begin{proofOf}{ \Cref{prop:prob-NExG}}
The results on $\OP{G^x}{x}{i}$ ($i\in \{1,2\}$) are direct consequences of Proposition \ref{prop:game-theory-basics}, the fact that $\ugr{y}{x}{i}$ are normal random variables as described in Table \ref{tab:prob-formulas-U}, and the independence/mutual exclusiveness of the involved random variables (which allow us to use the restricted disjunction/conjunction formulas from formula (\ref{eq:prob-conj-disj}), Subsection \ref{subsec:probabilities}).\\
For the case of $\ROM{G^x}{x}{\omega_1}{\omega_2}$, applying Corollary \ref{cor:game-theory-basics}, we get that $\ROM{G^x}{x}{\omega_1}{\omega_2}$ is true if and only if:
$$
\begin{aligned}
& (\ug[G^x]{x}{1} > 0) \bigwedge (\ug[G^x]{x}{2} < 0) \bigwedge (\ug[G^x]{\bar x}{1} < 0) \\ 
& \bigwedge (\ug[G^x]{\bar x}{2} > 0) \bigwedge \left( \omega_1 < \frac{\ug[G^x]{\bar x}{2}}{\ug[G^x]{\bar x}{2} - \ug[G^x]{\bar x}{1}} < \omega_2 \right) \\
\bigvee & \\ 
& (\ug[G^x]{x}{1} < 0) \bigwedge (\ug[G^x]{x}{2} > 0) \bigwedge  (\ug[G^x]{\bar x}{1} > 0) \bigwedge \\
& (\ug[G^x]{\bar x}{2} < 0) \bigwedge \left(\omega_1 < \frac{\ug[G^x]{\bar x}{2}}{\ug[G^x]{\bar x}{2} - \ug[G^x]{\bar x}{1}} < \omega_2 \right) \\
\end{aligned}
$$
Obviously, the above disjunction contains mutually exclusive events, so the probability $\prob{\ROM{G^x}{x}{\omega_1}{\omega_2}}$ is the sum of the probability of each disjunct (by the restricted disjunctive formula -- see formula (\ref{eq:prob-conj-disj}), Subsection \ref{subsec:probabilities}). So, let us compute the probability of the first disjunct.\\
We observe that the events $\ug[G^x]{x}{1}$, $\ug[G^x]{x}{2}$ are independent to each other and also independent to the other conjuncts. Moreover:
\begin{equation*}
\begin{aligned}
\omega_1 < \frac{\ug[G^x]{\bar x}{2}}{\ug[G^x]{\bar x}{2} - \ug[G^x]{\bar x}{1}} < \omega_2 & \Leftrightarrow \frac{\omega_1-1}{\omega_1} < \frac{\ug[G^x]{\bar x}{1}}{\ug[G^x]{\bar x}{2}} < \frac{\omega_2-1}{\omega_2} & \\
\end{aligned}
\end{equation*}
Thus, we can apply Lemma \ref{lem:division} for the last three conjuncts (for $\Omega_1 = \frac{\omega_1-1}{\omega_1}$, $\Omega_2 = \frac{\omega_2-1}{\omega_2}$), getting that the probability of the first conjunction is equal to:
\begin{equation*}
    \begin{split}
        & (1 - F_{\ugr{x}{x}{1}}(0)) \cdot F_{\ugr{x}{x}{2}}(0) \cdot \int_{0}^{+\infty} \left(\int_{\Omega_1 u_2}^{\Omega_2 u_2} \! f_{\ugr{x}{\bar x}{1}}(u_1) \, \mathrm{d}u_1 \right) \frac{f_{\ugr{x}{\bar x}{2}}(u_2)}{u_2} \, \mathrm{d}u_2 \\
    \end{split}
\end{equation*}
Working analogously for the second disjunct, and summing the resulting probability with the one above, we get the result.\\
For $\RPM{G^x}{x}{\omega_1}{\omega_2}$, we work analogously, applying the second bullet of Corollary \ref{cor:game-theory-basics} as above.\\
For $\IN{G^x}{x}$, we observe that if $\IN{G^x}{x}$ is true, then $G^x$ is degenerate, which has probability $0$.
\end{proofOf}

\vspace{4pt}

\begin{proofOf}{ \Cref{thm:main-emis}}
The proof is direct from Proposition \ref{prop:e-mis-no-prob} (and the respective Table \ref{tab:misinformed-cases}), combined with the fact that the different cases in the disjunction are mutually exclusive, so we can use the restricted disjunction formula of (\ref{eq:prob-conj-disj}) in Subsection \ref{subsec:probabilities}.
\end{proofOf}

\vspace{4pt}

\begin{proofOf}{ \Cref{thm:main-invemis}]}
The proof is direct from Proposition \ref{prop:e-mis-no-prob} (and the respective Table \ref{tab:misinformed-cases}), combined with the fact that the different cases in the disjunction are mutually exclusive, so we can use the restricted disjunction formula of (\ref{eq:prob-conj-disj}) in Subsection \ref{subsec:probabilities}.
\end{proofOf}

\subsection{Effect of modifying tolerance (\texorpdfstring{$\varepsilon$}{})}
\label{appendix:mod-epsilon}

\begin{proofOf}{ \Cref{prop:properties-e-monotonic-misinvmis}}
We first observe that, for any $x\in \{r,c\}$ and any $a,b,c$ such that:
$0 \leq a \leq b \leq c \leq 1$, we have that:

\begin{align}
    & \prob{\ROM{G^x}{x}{a}{c}} = \prob{\ROM{G^x}{x}{a}{b}} + \prob{\ROM{G^x}{x}{b}{c}}  \tag{ROM1} \\ %
    & \prob{\ROM{G^x}{x}{a}{c}} = 0 \Leftrightarrow a = c \tag{ROM2} \\
    & \prob{\RPM{G^x}{x}{a}{c}} = \prob{\RPM{G^x}{x}{a}{b}} + \prob{\RPM{G^x}{x}{b}{c}} \tag{RPM1} \\
    & \prob{\RPM{G^x}{x}{a}{c}} = 0 \Leftrightarrow a = c \tag{RPM2} 
\end{align}

\noindent From Theorem \ref{thm:main-emis}, and for $i = 1,2$:
\begin{equation*}
\prob{mG: \varepsilon_i \text{-misinformed}} = P_{r,i} \cdot P_{c,i},
\end{equation*}
where $P_{r,i}, P_{c,i}$ are determined by the second column of Table \ref{tab:prob-emis-invemis} for the respective $\varepsilon_i$.\\
Similarly, 
from Theorem \ref{thm:main-invemis}, and for $i = 1,2$:
\begin{equation*}
\prob{mG: \text{inverse-} \varepsilon_i \text{-misinformed}} = P_{r,i}' \cdot P_{c,i}',
\end{equation*}
where $P_{r,i}', P_{c,i}'$ are determined by the third column of Table \ref{tab:prob-emis-invemis} for the respective $\varepsilon_i$.\\
Now, let us focus on the first bullet of the proposition.
By Tables \ref{tab:prob-formulas-U}, \ref{tab:prob-formulas-NE}, \ref{tab:prob-emis-invemis}, it is easy to conclude that, for any $x \in \{r,c\}$, $i \in \{1,2\}$, the computation of $P_{x,i}, P_{x,i}'$ is not affected by the value of $\varepsilon_i$, and, thus:
$P_{x,1} = P_{x,2}$, $P_{x,1}' = P_{x,2}'$ for $x\in\{r,c\}$, which shows the result.\\
Now, let us focus on the second bullet, and let us consider $P_{r,i}, P_{r,i}'$ first.
Set:
\begin{equation*}
\begin{split}
& \omega_{1,1} = \max\{0,p^0-\varepsilon_1\}, \;\; 
\omega_{2,1} = \min\{1,p^0+\varepsilon_1\}, \\
& \omega_{1,2} = \max\{0,p^0-\varepsilon_2\}, \;\; 
\omega_{2,2} = \min\{1,p^0+\varepsilon_2\} \\
\end{split}
\end{equation*}
Since $0 \leq \varepsilon_1 < \varepsilon_2$, we get that:
$0 \leq \omega_{1,2} \leq \omega_{1,1} \leq \omega_{2,1} \leq \omega_{2,2} \leq 1$.
Moreover, since $\varepsilon_1 < \varepsilon_2$, it follows that:\\
\begin{equation*}
\begin{split}
 & \omega_{1,1} = \omega_{1,2} \Leftrightarrow \omega_{1,1} = \omega_{1,2} = 0 \Leftrightarrow 
 \left( \begin{array}{c}
      p^0 \leq \varepsilon_1  \\
      \text{and} \\
      p^0 \leq \varepsilon_2
  \end{array}\right) \Leftrightarrow p^0 \leq \varepsilon_1
\end{split}
\end{equation*}
\noindent Analogously:
\begin{equation*}
\begin{split}
 & \omega_{2,1} = \omega_{2,2} \Leftrightarrow \omega_{2,1} = \omega_{2,2} = 1 \Leftrightarrow \left( \begin{array}{c}
      1 - p^0 \leq \varepsilon_1  \\
      \text{and} \\
      1 - p^0 \leq \varepsilon_2 
  \end{array}\right) \Leftrightarrow 1 - p^0 \leq \varepsilon_1
\end{split}
\end{equation*}

\noindent Using the order among $\omega_{i,j}$, and by applying (ROM1) twice, we get that:

\begin{equation*}
    \begin{split}
        \prob{\ROM{G^r}{r}{\omega_{1,2}}{\omega_{2,2}}} & =  \prob{\ROM{G^r}{r}{\omega_{1,2}}{\omega_{1,1}}} + \prob{\ROM{G^r}{r}{\omega_{1,1}}{\omega_{2,1}}} \\
        & \hspace{3em} + \prob{\ROM{G^r}{r}{\omega_{2,1}}{\omega_{2,2}}} \\
    \end{split}
\end{equation*}
Now given the fact that probabilities are non-negative, and (ROM2), we have:
\begin{equation*}
    \begin{split}
    & \prob{\ROM{G^r}{r}{\omega_{1,1}}{\omega_{2,1}}} \leq  \prob{\ROM{G^r}{r}{\omega_{1,2}}{\omega_{2,2}}} \text { and: } \\
    & \prob{\ROM{G^r}{r}{\omega_{1,1}}{\omega_{2,1}}} = \prob{\ROM{G^r}{r}{\omega_{1,2}}{\omega_{2,2}}} \Leftrightarrow \\
    & \hspace{1.7cm} \left( \omega_{1,1} = \omega_{1,2} \text{ and } \omega_{2,1} = \omega_{2,2} \right)
    \end{split}
\end{equation*}
Using analogous reasoning we get:
\begin{equation*}
    \begin{split}
    &\prob{\RPM{G^r}{r}{\omega_{1,1}}{\omega_{2,1}}} \leq  \prob{\RPM{G^r}{r}{\omega_{1,2}}{\omega_{2,2}}} \text { and: } \\
    & \prob{\RPM{G^r}{r}{\omega_{1,1}}{\omega_{2,1}}} = \prob{\RPM{G^r}{r}{\omega_{1,2}}{\omega_{2,2}}} \Leftrightarrow \\
    & \hspace{1.7cm} \left( \omega_{1,1} = \omega_{1,2} \text{ and } \omega_{2,1} = \omega_{2,2} \right)
    \end{split}
\end{equation*}
Using the above, and Tables \ref{tab:prob-formulas-U}, \ref{tab:prob-formulas-NE}, \ref{tab:prob-emis-invemis}, we can easily conclude that $P_{r,1} \leq P_{r,2}$ and $P_{r,1}' \leq P_{r,2}'$. Moreover:
\begin{equation*}
    \begin{split}
        & P_{r,1} = P_{r,2} \Leftrightarrow \left( \begin{array}{c}
         \omega_{1,1} = \omega_{1,2}  \\
         \text{and} \\
         \omega_{2,1} = \omega_{2,2}
     \end{array}\right) \Leftrightarrow \left( \begin{array}{c}
         p^0 \leq \varepsilon_1  \\
         \text{and} \\
         1-p^0 \leq \varepsilon_1
     \end{array}\right)
    \end{split}
\end{equation*}
Analogously:
\begin{equation*}
P_{r,1}' = P_{r,2}' \Leftrightarrow \left( \begin{array}{c}
         p^0 \leq \varepsilon_1  \\
         \text{and} \\
         1 - p^0 \leq \varepsilon_1
     \end{array}\right)
\end{equation*}

\noindent Reasoning analogously for the case of $P_{c,i}, P_{c,i}'$, we get:
\begin{equation*}
\text{For } P_{c,1} \leq P_{c,2} \text{ : } P_{c,1} = P_{c,2} \Leftrightarrow \left( \begin{array}{c}
         q^0 \leq \varepsilon_1   \\
         \text{and} \\
         1 - q^0 \leq \varepsilon_1
     \end{array}\right)
\end{equation*}

\begin{equation*}
\text{For } P_{c,1}' \leq P_{c,2}' \text{ : } P_{c,1}' = P_{c,2}' \Leftrightarrow \left( \begin{array}{c}
         q^0 \leq \varepsilon_1   \\
         \text{and} \\
         1 - q^0 \leq \varepsilon_1
     \end{array}\right)
\end{equation*}

By the hypothesis of the second bullet with regards to $NE(G^0)$, Tables \ref{tab:prob-formulas-U}, \ref{tab:prob-formulas-NE}, \ref{tab:prob-emis-invemis}, and the above relations, the cases (2a), (2b) of the Theorem follow easily.\\
Now let us focus on the third bullet. First, we observe that, by Table \ref{tab:prob-emis-invemis}, the result is obvious for the case of $\varepsilon$-misinformed, so let us focus on the case of inverse-$\varepsilon$-misinformed. 
If $\varepsilon_2 \leq 0.5$, then $\varepsilon_1 \leq 0.5$, so the result is again obvious by Table \ref{tab:prob-emis-invemis}.
So let us focus on the scenario where $\varepsilon_2 > 0.5$.\\
To show the result for this case, we use an approach similar to the one employed for the second bullet. In particular, we consider $P_{r,i}'$ first.
Set:

\begin{equation*}
\begin{split}
& \omega_{1,1}' = \max\{0,1-\varepsilon_1\}, \;\; 
\omega_{2,1}' = \min\{1,\varepsilon_1\}, \\
& \omega_{1,2}' = \max\{0,1-\varepsilon_2\}, \;\; 
\omega_{2,2}' = \min\{1,\varepsilon_2\} \\
\end{split}
\end{equation*}
Using an analogous procedure (as in the second bullet), and the fact that $0 \leq \varepsilon_1 < \varepsilon_2$, we conclude that:
$$ 
\begin{aligned}
0 \leq \omega_{1,2}' \leq \omega_{1,1}' \leq \omega_{2,1}' \leq \omega_{2,2}' \leq 1 \\
\omega_{1,1}' = \omega_{1,2}' \Leftrightarrow \varepsilon_1 \leq 1 \\
\omega_{2,1}' = \omega_{2,2}' \Leftrightarrow \varepsilon_1 \leq 1 \\
\end{aligned}
$$
Also, using (RPM1), (RPM2), and the fact that probabilities are non-negative, we get, as in the second bullet:
\begin{equation*}
\begin{split}
& \prob{\RPM{G^r}{r}{\omega_{1,1}'}{\omega_{2,1}'}} \leq  \prob{\RPM{G^r}{r}{\omega_{1,2}'}{\omega_{2,2}'}} \\
& \prob{\RPM{G^r}{r}{\omega_{1,1}'}{\omega_{2,1}'}} = \prob{\RPM{G^r}{r}{\omega_{1,2}'}{\omega_{2,2}'}}  \\
& \hspace{1.2cm} \Leftrightarrow \left(\omega_{1,1}' = \omega_{1,2}' \text{ and } \omega_{2,1}' = \omega_{2,2}' \right)\\
\end{split}
\end{equation*}

\noindent Therefore, given that $\varepsilon_1 > \varepsilon_2 > 0.5$:

\begin{equation*}
    \left(P_{r,1}' \leq P_{r,2}' \text{ and } P_{r,1}' = P_{r,2}' \right) \Leftrightarrow \varepsilon_1 \leq 1 \\
\end{equation*}

\noindent Working analogously for $P_{c,i}'$, we get: 
\begin{equation*}
    \left( P_{c,1}' \leq P_{c,2}' \text{ and } P_{c,1}' = P_{c,2}' \right) \Leftrightarrow \varepsilon_1 \leq 1 \\
\end{equation*}
By the hypothesis of the second bullet with regards to $NE(G^0)$, Tables \ref{tab:prob-formulas-U}, \ref{tab:prob-formulas-NE}, \ref{tab:prob-emis-invemis}, and the above relations, the remaining subcases of (3a), (3b) of the Theorem follow easily.
\end{proofOf}

\subsection{Effect of changing the game (\texorpdfstring{$G^0$}{}) and the mean (\texorpdfstring{$M$}{})}
\label{appendix:mod-G0-M}

\begin{proofOf}{ \Cref{prop:effect-G0-M-mod}}
From Table \ref{tab:prob-formulas-U}, we observe that, for the given $x$, and for any $i \in \{1,2\}$:
\begin{equation*}
\begin{split}
\mu_{\ugr{x}{r}{i}} & = (P^0_r[1,i] + M^x_r[1,i]) - (P^0_r[2,i] + M^x_r[2,i]) \\
& = (\overline{P^0_r}[1,i] + \overline{M^x_r}[1,i] + a) - (\overline{P^0_r}[2,i] + \overline{M^x_r}[2,i] + a) \\
& = \overline{\mu_{\ugr{x}{r}{i}}}.
\end{split}
\end{equation*}
Analogously, we can show that 
$\mu_{\ugr{x}{c}{i}} = \overline{\mu_{\ugr{x}{c}{i}}}$ for any $i \in \{1,2\}$.
Also, it is clear that $\devi_{\ugr{x}{y}{i}} = \overline{\devi_{\ugr{x}{y}{i}}}$ for any $y \in \{r,c\}$, $i \in \{1,2\}$.
Combining these two facts, the results are obvious.
\end{proofOf}

\vspace{4pt}

\begin{proofOf}{ \Cref{prop:effect-of-sum-on-G-M}}
Take any $x\in \{r,c\}$. Set $b_x = - a_G - a_x$.
We observe that $G^0 + M^x = \overline{G^0} + \overline{M^x} + \tbl{b_x}$.
Thus, by Proposition \ref{prop:effect-G0-M-mod}, we get, for $x \in \{r,c\}$:
\begin{itemize}
    
    \item For any $i \in \{1,2\}$, 
    \begin{equation*}
        \prob{\OP{G^x}{x}{i}} = \prob{\OP{\overline{G^x}}{x}{i}}
    \end{equation*}
    
    \item For any $0 \leq \omega_1 \leq \omega_2 \leq 1$, 
    \begin{equation*}
        \prob{\ROM{G^x}{x}{\omega_1}{\omega_2}} = \prob{\ROM{\overline{G^x}}{x}{\omega_1}{\omega_2}}
    \end{equation*}
    
    \item For any $0 \leq \omega_1 \leq \omega_2 \leq 1$,
    \begin{equation*}
        \prob{\RPM{G^x}{x}{\omega_1}{\omega_2}} = \prob{\RPM{\overline{G^x}}{x}{\omega_1}{\omega_2}}
    \end{equation*}

\end{itemize}
In addition, game theoretic results tell us that $NE(G^0) = NE(\overline{G^0})$.
Combining the above with Theorems \ref{thm:main-emis}, \ref{thm:main-invemis} and Table \ref{tab:prob-emis-invemis}, the result follows directly.
\end{proofOf}

\subsection{Effect of modifying noise intensity (\texorpdfstring{$\Devi$}{})}
\label{appendix:mod-D}

\begin{proofOf}{\Cref{prop:mg-change-scale}}
Consider the family of random variables $\overline{\ugr y x i}$ for $\overline {mG}$. By the definition of $\overline M, \overline \Devi$ and by Table \ref{tab:prob-formulas-U}, it follows that, for any $x,y \in \{r,c\}$, and for any $i \in \{1,2\}$, it holds that:
\begin{equation*}
    \mu_{\overline{\ugr y x i}} = \lambda \mu_{\ugr y x i} \quad \text{ and } \quad  
\devi_{\overline{\ugr y x i}} = \lambda^2 \devi_{\ugr y x i}
\end{equation*}
Therefore $\overline{\ugr yxi} = \lambda \ugr yxi$.\\
Using the latter relationship, we get, for any $x,y \in \{r,c\}$, $i\in \{1,2\}$:
\begin{equation*}
\begin{split}
    F_{\overline{\ugr y x i}}(0) & = \prob{\overline{\ugr y x i} \leq 0} = \prob{\lambda \cdot \ugr y x i \leq 0}  = \prob{\ugr y x i \leq 0} =  F_{\ugr y x i}(0)
\end{split}
\end{equation*}
To simplify the equations in the following, let us set, for any $x \in \{r,c\}$, $i \in \{1,2\}$:\\
$U_i = \ugr {x}{\bar x} {1}$, 
$\overline U_i = \overline{\ugr {x}{\bar x} {1}}$, and let $f_i, \overline f_i$ the respective cdfs for $U_i, \overline U_i$. Then, using Lemma \ref{lem:division}, and the above notation, for any $\omega_1,\omega_2 \in \mathbb R$ such that $0 \leq \omega_1 < \omega_2 \leq 1$, it holds that:
\begin{equation*}
\begin{split}
     & \int_{0}^{+\infty} \left(\int_{\frac{\omega_1 -  1}{\omega_1} u_2}^{\frac{\omega_2 - 1}{\omega_2} u_2} \! f_{\overline {\ugr{x}{\bar x}{1}}}(u_1) \, \mathrm{d}u_1 \right) \frac{f_{\overline{\ugr{x}{\bar x}{2}}}(u_2)}{u_2} \, \mathrm{d}u_2  \\
    & \hspace{2cm} = \int_{0}^{+\infty} \left(\int_{\frac{\omega_1 -  1}{\omega_1} u_2}^{\frac{\omega_2 - 1}{\omega_2} u_2} \! \overline f_1(u_1) \, \mathrm{d}u_1 \right) \frac{\overline f_2(u_2)}{u_2} \, \mathrm{d}u_2 \\
    & \hspace{2cm} = \prob{
    \frac{\omega_1-1}{\omega_1} \leq 
    \frac  { \overline U_1  }   {   \overline U_2   } \leq
    \frac{\omega_1-1}{\omega_1}, 
    \overline U_1 < 0,
    \overline U_2 > 0
    }  \\
    & \hspace{2cm} = \prob{
    \frac{\omega_1-1}{\omega_1} \leq 
    \frac  { \lambda \cdot U_1}
           { \lambda \cdot U_2} \leq
    \frac{\omega_1-1}{\omega_1}, 
    \lambda \cdot U_1 < 0,
    \lambda \cdot U_2 > 0
    } \\
    & \hspace{2cm} = \prob{
    \frac{\omega_1-1}{\omega_1} \leq 
    \frac  {U_1} {U_2} \leq
    \frac{\omega_1-1}{\omega_1}, 
    U_1 < 0,
    U_2 > 0
    } \\
    & \hspace{2cm} = \int_{0}^{+\infty} \left(\int_{\frac{\omega_1 -  1}{\omega_1} u_2}^{\frac{\omega_2 - 1}{\omega_2} u_2} \! f_1(u_1) \, \mathrm{d}u_1 \right) \frac{f_2(u_2)}{u_2} \, \mathrm{d}u_2 \\
    & \hspace{2cm} = \int_{0}^{+\infty} \left(\int_{\frac{\omega_1 -  1}{\omega_1} u_2}^{\frac{\omega_2 - 1}{\omega_2} u_2} \! f_{\ugr{x}{\bar x}{1}}(u_1) \, \mathrm{d}u_1 \right) \frac{f_{\ugr{x}{\bar x}{2}}(u_2)}{u_2} \, \mathrm{d}u_2
\end{split}
\end{equation*}
Analogously, it can be shown that:
\begin{equation*}
\begin{split}
    & \int_{-\infty}^{0} \left(\int_{\frac{\omega_1 -  1}{\omega_1} u_2}^{\frac{\omega_2 -  1}{\omega_2} u_2} \! f_{\ugr{x}{\bar x}{1}}(u_1) \, \mathrm{d}u_1 \right) \frac{f_{\ugr{x}{\bar x}{2}}(u_2)}{u_2} \, \mathrm{d}u_2 = \\
    & \hspace{3cm} \int_{-\infty}^{0} \left(\int_{\frac{\omega_1 -  1}{\omega_1} u_2}^{\frac{\omega_2 -  1}{\omega_2} u_2} \! f_{\overline {\ugr{x}{\bar x}{1}}}(u_1) \, \mathrm{d}u_1 \right) \frac{f_{\overline{\ugr{x}{\bar x}{2}}}(u_2)}{u_2} \, \mathrm{d}u_2
\end{split}
\end{equation*}
From the above equations, it is obvious that the respective probabilities in Table \ref{tab:prob-formulas-NE} for $mG$ and $\overline{mG}$ are equal. Moreover, since $\overline{G^0} = \lambda \cdot G^0$, it follows that the Nash equilibria of $G^0$ and $\overline{G^0}$ are the same. Combining these facts with Table \ref{tab:prob-emis-invemis} and Theorems \ref{thm:main-emis}, \ref{thm:main-invemis}, the result follows.
\end{proofOf}


\bibliographystyle{plain}
\bibliography{ref}

\end{document}

%% file: Tables/game-theory-basics-depiction.tex
\begin{table}
    \caption{Visualising the cases of Proposition \ref{prop:game-theory-basics}, for $x = r$.}
    \label{tab:game-theory-basics-depiction}
\centering
\adjustbox{max width=\textwidth}{%
    \begin{tabular}{|| c || M{0.9cm}|M{0.9cm}|M{0.9cm}|M{0.9cm}|| M{4cm} || M{2.4cm}||}
    \hline
    
    \textbf{Case (from} &
    \multicolumn{4}{c||}{\textbf{Value of }\ug{x}{i}\textbf{, for:}} &
    \textbf{Nash equilibrium} &
    \textbf{Schematic} \\ 
    
    \textbf{Proposition} &
    $x=r$ & 
    $x=r$ & 
    $x=c$ & 
    $x=c$ & 
    \textbf{strategies for $x=r$} & 
    \textbf{depiction} \\
    
    \textbf{\ref{prop:game-theory-basics})} &
    $i=1$ &
    $i=2$ &
    $i=1$ &
    $i=2$ &
    \textbf{($NE_r(G)$)} & 
    \textbf{of the case}
    \\ \hline \hline

    \addLineTable   {>} {>} {<>} {<>} {(1,0)} {\ } {1a}
    \addLineTable   {>} {<} {>} {>} {(1,0)} { } {1b}
    \addLineTable   {<} {>} {<} {<} {(1,0)} {} {1c}
    
    \addLineTable   {<} {<} {<>} {<>} {(0,1)} {} {2a}  
    \addLineTable   {<} {>} {>} {>} {(0,1)} {} {2b}  
    \addLineTable   {>} {<} {<} {<} {(0,1)} {} {2c}  
    
    \addLineTable   {>} {<} {<} {>} {(p,1-p)} {p=\frac{\ug{c}{2}}{\ug{c}{2} - \ug{c}{1}}} {3a}  
    \addLineTable   {<} {>} {>} {<} {(p,1-p)} {p=\frac{\ug{c}{2}}{\ug{c}{2} - \ug{c}{1}}} {3b}
    
    \addLineTable   {>} {<} {>} {<} {(1,0),(0,1),(p,1-p)} {p=\frac{\ug{c}{2}}{\ug{c}{2} - \ug{c}{1}}} {4a}  
    \addLineTable   {<} {>} {<} {>} {(1,0),(0,1),(p,1-p)} {p=\frac{\ug{c}{2}}{\ug{c}{2} - \ug{c}{1}}} {4b}
    
    \end{tabular}}
\end{table}

%% file: Tables/misinformed-cases.tex
\begin{table}
    \caption{Scenarios for $\varepsilon$-misinformed and inverse-$\varepsilon$-misinformed}
    \label{tab:misinformed-cases}
    \centering
    \begin{tabular}{|p{2cm}||p{4cm}|p{4.6cm}|}
    \hline
    
    \textbf{Condition on $G^0$} &  
    \multicolumn{2}{c|}{\textbf{Condition on $G^x$}} \\ 
    
    &
    \textbf{For $\varepsilon$-misinformed} &   
    \textbf{For inverse-$\varepsilon$-misinformed} \\ \hline \hline
    
    \OP{G^0}{x}{i}  &
    \OP{G^x}{x}{i}  &
    \OP{G^x}{x}{i} $\bigvee$ \newline \RPM{G^x}{x}{0}{1} \\  \hline
    
    \OM{G^0}{x}{p^0}  &
    \ROM{G^x}{x}{\omega_1}{\omega_2}  &
    \ROM{G^x}{x}{\omega_1}{\omega_2} $\bigvee$ \newline \RPM{G^x}{x}{\omega_1}{\omega_2} \\ [3.5ex] \hline
    
    \PM{G^0}{x}{p^0}  &
    \RPM{G^x}{x}{\omega_1}{\omega_2} $\bigvee$ \newline 
    \ROM{G^x}{x}{\omega_1}{\omega_2} $\bigvee$ \newline 
    \OP{G^x}{x}{1} $\bigvee$ \newline 
    \OP{G^x}{x}{2}    &
    \RPM{G^x}{x}{\omega_1}{\omega_2} \\ [5.5ex] \hline
    
    \IN{G^0}{x} &
    Always true &
    If $\varepsilon < 0.5$: always false \newline
    If $\varepsilon \geq 0.5$: \RPM{G^x}{x}{\omega_1'}{\omega_2'} \\ [3.5ex]\hline
    
    \multicolumn{3}{p{11cm}}{
    In all the above: \newline 
    $0 < p^0 < 1$, \newline
    $\omega_1 = \max\{0, p^0 - \varepsilon\}$,
    $\omega_2 = \min\{1, p^0 + \varepsilon\}$, \newline
    $\omega_1' = \max\{0, 1 - \varepsilon\}$, 
    $\omega_2' = \min\{1, \varepsilon\}$\newline 
    \ 
    }
    \end{tabular}
\end{table}

%% file: Tables/prob-formulas-U.tex
\begin{table}
\caption{Formulas related to $\ugr{y}{x}{i}$ for a given $mG \sim G^0 + \gndist{M, \Devi}$.}
\label{tab:prob-formulas-U}
\centering   
\begin{tabular}{|M{.8\textwidth}|}
\hline

\textbf{Results related to $\ugr{y}{x}{i}$} \\
\hline
$$
\begin{aligned}
\mu_{\ugr{y}{r}{i}} = & (P^0_r[1,i] + M^y_r[1,i]) - (P^0_r[2,i] + M^y_r[2,i]) \\
\devi_{\ugr{y}{r}{i}} = & \sqrt{(\Devi^y_r[1,i])^2 + (\Devi^y_r[2,i])^2} \\
\mu_{\ugr{y}{c}{i}} = & (P^0_c[i,1] + M^y_c[i,1]) - (P^0_c[i,2] + M^y_c[i,2]) \\
\devi_{\ugr{y}{c}{i}} = & \sqrt{(\Devi^y_c[i,1])^2 + (\Devi^y_c[i,2])^2} \\
f_{\ugr{y}{x}{i}}(u) = & \frac{1}{\devi_{\ugr{y}{x}{i}}} \phi \left(\frac{u-\mu_{\ugr{y}{x}{i}}}{\devi_{\ugr{y}{x}{i}}}\right)  \\
F_{\ugr{y}{x}{i}}(u) = & \Phi\left(\frac{u-\mu_{\ugr{y}{x}{i}}}{\devi_{\ugr{y}{x}{i}}}\right) \\
\prob{\ugr{y}{x}{i} < 0} = & F_{\ugr{y}{x}{i}}(0) \\
\prob{\ugr{y}{x}{i} > 0} = & 1 - F_{\ugr{y}{x}{i}}(0) \\
\prob{\ugr{y}{x}{i} = 0} = & 0 \\
\end{aligned}
$$
\\ \hline
\end{tabular}
\end{table}

%% file: Tables/prob-formulas-NE.tex
\begin{table}
\caption{Various probabilities pertaining to a given $mG \sim G^0 + \gndist{M, \Devi}$ (see also Proposition \ref{prop:prob-NExG}).}
\label{tab:prob-formulas-NE}
\centering   
\begin{tabular}{|M{\textwidth}|}
\hline

\textbf{Results related to $NE_x(G^x)$ (Proposition \ref{prop:prob-NExG})} 
\\\hline 
    $$
    \begin{aligned}
    \prob{\OP{G^x}{x}{1}} = 
    & (1 - F_{\ugr{x}{x}{1}}(0)) \cdot (1 - F_{\ugr{x}{x}{2}}(0)) + \\
    & \hspace{-.5em} (1 - F_{\ugr{x}{x}{1}}(0)) \cdot F_{\ugr{x}{x}{2}}(0) \cdot (1 - F_{\ugr{x}{\bar x}{1}}(0)) \cdot (1 - F_{\ugr{x}{\bar x}{2}}(0)) + \\
    & \hspace{-.5em} F_{\ugr{x}{x}{1}}(0) \cdot (1 - F_{\ugr{x}{x}{2}}(0)) \cdot F_{\ugr{x}{\bar x}{1}}(0) \cdot F_{\ugr{x}{\bar x}{2}}(0) \\
    \prob{\OP{G^x}{x}{2}} = 
    & F_{\ugr{x}{x}{1}}(0) \cdot F_{\ugr{x}{x}{2}}(0) + \\
    & \hspace{-.5em} F_{\ugr{x}{x}{1}}(0) \cdot (1 - F_{\ugr{x}{x}{2}}(0)) \cdot  (1 - F_{\ugr{x}{\bar x}{1}}(0)) \cdot (1 - F_{\ugr{x}{\bar x}{2}}(0)) + \\
    & \hspace{-.5em} (1 - F_{\ugr{x}{x}{1}}(0)) \cdot F_{\ugr{x}{x}{2}}(0) \cdot F_{\ugr{x}{\bar x}{1}}(0) \cdot F_{\ugr{x}{\bar x}{2}}(0) \\
    \prob{\ROM{G^x}{x}{\omega_1}{\omega_2}} = \\
    & \hspace{-2.5cm} (1 - F_{\ugr{x}{x}{1}}(0)) \cdot F_{\ugr{x}{x}{2}}(0) \cdot  \int_{0}^{+\infty} \left(\int_{\frac{\omega_1 -  1}{\omega_1} u_2}^{\frac{\omega_2 -  1}{\omega_2} u_2} \! f_{\ugr{x}{\bar x}{1}}(u_1) \, \mathrm{d}u_1 \right) \frac{f_{\ugr{x}{\bar x}{2}}(u_2)}{u_2} \, \mathrm{d}u_2 + \\
    & \hspace{-2.5cm} F_{\ugr{x}{x}{1}}(0) \cdot (1 - F_{\ugr{x}{x}{2}}(0)) \cdot \int_{-\infty}^{0} \left(\int_{\frac{\omega_1 -  1}{\omega_1} u_2}^{\frac{\omega_2 -  1}{\omega_2} u_2} \! f_{\ugr{x}{\bar x}{1}}(u_1) \, \mathrm{d}u_1 \right) \frac{f_{\ugr{x}{\bar x}{2}}(u_2)}{u_2} \, \mathrm{d}u_2 \\
    \prob{\RPM{G^x}{x}{\omega_1}{\omega_2}} = \\
    & \hspace{-2.5cm} (1 - F_{\ugr{x}{x}{1}}(0)) \cdot F_{\ugr{x}{x}{2}}(0) \cdot  \int_{-\infty}^{0} \left(\int_{\frac{\omega_1 -  1}{\omega_1} u_2}^{\frac{\omega_2 -  1}{\omega_2} u_2} \!  f_{\ugr{x}{\bar x}{1}}(u_1) \, \mathrm{d}u_1 \right) \frac{f_{\ugr{x}{\bar x}{2}}(u_2)}{u_2} \, \mathrm{d}u_2 \; + \\
    & \hspace{-2.5cm} F_{\ugr{x}{x}{1}}(0) \cdot (1 - F_{\ugr{x}{x}{2}}(0)) \cdot  \int_{0}^{+\infty} \left(\int_{\frac{\omega_1 -  1}{\omega_1} u_2}^{\frac{\omega_2 -  1}{\omega_2} u_2} \! f_{\ugr{x}{\bar x}{1}}(u_1) \, \mathrm{d}u_1 \right) \frac{f_{\ugr{x}{\bar x}{2}}(u_2)}{u_2} \, \mathrm{d}u_2 \\
    \prob{\IN{G^x}{x}} = & 0 \\
    \end{aligned}
    $$
    \\ \hline
\end{tabular}
\end{table}

%% file: Tables/prob-emis-invemis.tex
\begin{table}
\caption{Probabilities for $\varepsilon$-misinformed and inverse-$\varepsilon$-misinformed ($\mathcal{P}_r^{mis} \cdot \mathcal{P}_c^{mis}$ and $\mathcal{P}_r^{inv} \cdot \mathcal{P}_c^{inv}$ respectively -- see also Theorems \ref{thm:main-emis}, \ref{thm:main-invemis}).}
    \label{tab:prob-emis-invemis}
    \centering
    \begin{tabular}{|p{2cm}||p{4.2cm}|p{4.2cm}|}
    \hline
    
    \textbf{Condition on $G^0$} \newline 
    \textbf{(value of $NE_x(G^0)$)}
    &  
    \textbf{Probability $\mathcal{P}_x^{mis}$ ($x\in \{r,c\}$) for $\varepsilon$-misinformed} \newline 
    \textbf{(Theorem \ref{thm:main-emis})}
    &
    \textbf{Probability $\mathcal{P}_x^{inv}$ ($x\in \{r,c\}$) for inverse-$\varepsilon$-misinformed} \newline 
    \textbf{(Theorem \ref{thm:main-invemis})}
    \\ \hline \hline
    
    \OP{G^0}{x}{i}  &
    \prob{\OP{G^x}{x}{i}}  &
    \prob{\OP{G^x}{x}{i}} $+$ \prob{\RPM{G^x}{x}{0}{1}} \\ [2ex] \hline
    
    \OM{G^0}{x}{p^0}  &
    \prob{\ROM{G^x}{x}{\omega_1}{\omega_2}}  &
    \prob{\ROM{G^x}{x}{\omega_1}{\omega_2}} $+$ \newline \prob{\RPM{G^x}{x}{\omega_1}{\omega_2}} \\ [3.5ex] \hline
    
    \PM{G^0}{x}{p^0}  &
    \prob{\RPM{G^x}{x}{\omega_1}{\omega_2}} $+$ \newline 
    \prob{\ROM{G^x}{x}{\omega_1}{\omega_2}} $+$ \newline 
    \prob{\OP{G^x}{x}{1}} $+$ \newline 
    \prob{\OP{G^x}{x}{2}}    &
    \prob{\RPM{G^x}{x}{\omega_1}{\omega_2}} \\ [5.5ex]\hline
    
    $\IN{G^0}{x}$ &
    1 &
    If $\varepsilon \leq 0.5$: 0 \newline
    If $\varepsilon > 0.5$: \prob{\RPM{G^x}{x}{\omega_1'}{\omega_2'}} \\ [3.5ex] \hline
    
    \multicolumn{3}{p{11cm}}{
    In all the above: \newline 
    $i \in \{1,2\}$, $0 < p^0 < 1$, \newline
    $\omega_1 = \max\{0, p^0 - \varepsilon\}$,
    $\omega_2 = \min\{1, p^0 + \varepsilon\}$, \newline
    $\omega_1' = \max\{0, 1 - \varepsilon\}$, 
    $\omega_2' = \min\{1, \varepsilon\}$
    }
    \end{tabular}
\end{table}

%% file: Tables/properties-e-monotonic-misinvmis.tex
\begin{table}
    \caption{Effect of tolerance on behavioural consistency (monotonicity).
\label{tab:properties-e-monotonic-misinvmis}}
    \centering
    \begin{tabular} { | p{3cm} || *{2}{M{3.7cm}|} }
    
    \hline
    \textbf{Condition on $G^0$} &
    \multicolumn{2}{c|}{\textbf{Monotonicity properties}} \\
    
    \textbf{(value of $NE(G^0)$)} &
    \textbf{For $\varepsilon$-misinformed} & 
    \textbf{For inverse-$\varepsilon$-misinformed} \\ \hline \hline

    $\OP{G^0}{r}{i} \wedge \OP{G^0}{c}{j}$ for some \newline
    $i,j \in \{1.2\}$ & 
    \multicolumn{2}{M{8.4cm}|}
    {Constant for all $\varepsilon \geq 0$} \\ \hline

    $\OM{G^0}{r}{p^0} \wedge \OM{G^0}{c}{q^0}$ for some \newline
    $0 < p^0 < 1, 0 < q^0 < 1$ & 
    \multicolumn{2}{M{8.4cm}|}
    {Strictly increasing for $0 \leq \varepsilon \leq \max \{p^0, q^0, 1-p^0, 1-q^0\}$, \newline constant otherwise} 
    \\ \hline

    $\PM{G^0}{r}{p^0} \wedge \PM{G^0}{c}{q^0}$ for some \newline
    $0 < p^0 < 1, 0 < q^0 < 1$ & 
    \multicolumn{2}{M{8.4cm}|}
    {Strictly increasing for $0 \leq \varepsilon \leq \max \{p^0, q^0, 1-p^0, 1-q^0\}$, \newline constant otherwise} 
    \\ \hline

    $\IN{G^0}{r} \vee \IN{G^0}{c}$ & 
    Constant for all $\varepsilon \geq 0$ &
    Strictly increasing 
    for $0.5 \leq \varepsilon \leq 1$, 
    constant otherwise 
    \\ \hline

    \end{tabular}
\end{table}

%% file: Tables/properties-e-minmax.tex
\begin{table}
        \caption{Minimal and maximal values for the probabilities of $mG$ being (inverse-)$\varepsilon$-misinformed (resulting by multiplying $\mathcal{P}_r^{mis}$ with $\mathcal{P}_c^{mis}$ and $\mathcal{P}_r^{inv}$ with $\mathcal{P}_c^{inv}$ respectively)}
    \label{tab:properties-e-minmax}
    \centering
    \begin{subtable}{1\textwidth}
    \centering
    \begin{tabular}{|p{2cm}||p{3.7cm}|p{4.6cm}|}
    \hline
    
    \textbf{Condition on $G^0$} \newline 
    \textbf{(value of $NE_x(G^0)$)}
    &  
    \textbf{Minimal value for probability} \newline
    \textbf{$P_x$ ($x\in \{r,c\}$) for $\varepsilon$-misinformed}
    &
    \textbf{Minimal value for probability} \newline
    \textbf{$P_x$ ($x\in \{r,c\}$) for inverse-$\varepsilon$-misinformed}
    \\ \hline \hline
    
    \OP{G^0}{x}{i} for some \newline
    $i \in \{1,2\}$ &
    \prob{\OP{G^x}{x}{i}}  &
    \prob{\OP{G^x}{x}{i}} $+$ \prob{\RPM{G^x}{x}{0}{1}} \\ [3.5ex] \hline
    
    \OM{G^0}{x}{p^0} for some \newline
    $0 < p^0 < 1$ &
    $0$  &
    $0$ \\ [3.5ex] \hline
    
    \PM{G^0}{x}{p^0} for some \newline
    $0 < p^0 < 1$ &
    \prob{\OP{G^x}{x}{1}} $+$ \prob{\OP{G^x}{x}{2}}    &
    $0$ \\ [3.5ex]\hline
    
    $\IN{G^0}{x}$ &
    $1$ &
    $0$ \\ [2ex] \hline
    
    \end{tabular}
    \caption{Minimal values}
    \end{subtable}
    \centering
    \begin{subtable}{1\textwidth}
    \centering
    \begin{tabular}{|p{2cm}||p{3.7cm}|p{4.6cm}|}
    \hline
    
    \textbf{Condition on $G^0$} \newline 
    \textbf{(value of $NE_x(G^0)$)}
    &  
    \textbf{Maximal value for probability} \newline
    \textbf{$P_x$ ($x\in \{r,c\}$) for $\varepsilon$-misinformed}
    &
    \textbf{Maximal value for probability} \newline
    \textbf{$P_x$ ($x\in \{r,c\}$) for inverse-$\varepsilon$-misinformed}
    \\ \hline \hline
    
    \OP{G^0}{x}{i} for some \newline
    $i \in \{1,2\}$ &
    \prob{\OP{G^x}{x}{i}}  &
    \prob{\OP{G^x}{x}{i}} $+$ \prob{\RPM{G^x}{x}{0}{1}} \\ [3.5ex] \hline
    
    \OM{G^0}{x}{p^0} \newline
    for some \newline
    $0 < p^0 < 1$  &
    \prob{\ROM{G^x}{x}{0}{1}}  &
    \prob{\ROM{G^x}{x}{0}{1}} $+$ \prob{\RPM{G^x}{x}{0}{1}} \\ [3.5ex] \hline
    
    \PM{G^0}{x}{p^0} for some \newline
    $0 < p^0 < 1$  &
    $1$ &
    \prob{\RPM{G^x}{x}{0}{1}} \\ [3.5ex] \hline
    
    $\IN{G^0}{x}$ &
    $1$ &
    \prob{\RPM{G^x}{x}{0}{1}} \\ [2ex] \hline
    
    \end{tabular}
    \caption{Maximal values}
    \end{subtable}
    
\end{table}